\documentclass[useAMS,usenatbib,usegraphicx]{mn2e}

\title[Metals around QSOs and Galaxies]{Metal Absorption Systems in Spectra of Pairs of QSOs}
\author[D. Tytler \etal]{David Tytler\thanks{E-mail: tytler@ucsd.edu},
  Mark Gleed, Carl Melis, Angela Chapman,
  David Kirkman\newauthor
  Dan Lubin, Pascal Paschos and Tridivesh Jena \newauthor\\
  Center for Astrophysics and Space Sciences,
  University of California San Diego,
  La Jolla, CA, 92093-0424 \newauthor
  and Arlin P.S. Crotts\newauthor\\
  Department of Astronomy, Columbia University,
  550 West 120th Street, New York, NY 10027
}

\usepackage{color}
\usepackage{graphicx}
 
\newcommand{\lya}{\mbox{Ly$\alpha$}}
\newcommand{\lyb}{\mbox{Ly$\beta$}}

\newcommand{\kms}{\mbox{km s$^{-1}$}}
\newcommand{\cmm}{\mbox{cm$^{-2}$}}

\newcommand{\om}{\mbox{$\Omega_m$}}
\newcommand{\ol}{\mbox{$\Omega_{\Lambda } $}}

\newcommand{\zabs}{\mbox{$z_{\rm abs}$}}
\newcommand{\zem}{\mbox{$z_{\rm em}$}}

\newcommand{\nhi}{\mbox{N$_{\rm H I}$}}
\newcommand{\lnhi}{\mbox{log \nhi}}

\newcommand{\etal}{{\it et al.}}
\newcommand{\lyaf} {\lya\ forest}

\newcommand{\dz}{\mbox{$\Delta z$}}
\newcommand{\dzee}{\mbox{$\Delta z_{EE}$}}
\newcommand{\dzea}{\mbox{$\Delta z_{EA}$}}
\newcommand{\dzaa}{\mbox{$\Delta z_{AA}$}}

\newcommand{\dvea}{\mbox{$v_{EA}$}}
\newcommand{\dvaa}{\mbox{$v_{AA}$}}
\newcommand{\vabs}{\mbox{$v_{abs}$}}
\newcommand{\wrest}{\mbox{$W_{rest}$}}
\newcommand{\wpm}{\mbox{$W_{pm}$}}

\newif\ifdraftmodep
\draftmodepfalse

\newif\ifapjp
\apjpfalse

\begin{document}
\date{\today}

\maketitle

\begin{abstract}
  We present the first large sample of metal absorption systems in
  pairs of QSOs with sightlines separated by about 1~Mpc at $z = 2$.  We found 691
  absorption systems in the spectra of 310 QSOs in 170 pairings.  Most of the systems
  contain C~IV or MgII absorption.

We see 17 cases of absorption in one line-of-sight within 200~\kms\
(1~Mpc) of absorption in the paired line-of-sight.  When we see
absorption in one line-of-sight, the probability of also seeing
absorption within about 500~\kms\ in the partner line-of-sight is at
least $\approx 50$\% at $<100$~kpc, declining rapidly to 23\% at 100
-- 200~kpc and 0.7\% by 1 -- 2~Mpc.  Although we occasionally probe an
individual absorbing halo with two lines-of-sight, the
absorber-absorber correlation is primarily a probe of the large scale
distribution of metals around galaxies and galaxy clustering. 
With redshifts errors of $\sim 23$~\kms , we 
detect clustering on 0.5~Mpc scales and we see a hint of
the ``fingers of God'' redshift-space distortion. The distribution is
consistent with absorbers arising in galaxies at $z=2$ with a normal
correlation function, normal systematic infall velocities and
unusually low random pair-wise velocity differences, more consistent
with blue than with red galaxies.  Absorption in gas flowing out from
galaxies with a mean velocity of 250~\kms\ would produce vastly more
redshift elongation than we see. The UV absorption from fast winds
that \citet{adelberger05} see in spectra of LBGs is not representative
of UV absorption that we see. Either the winds are confined to the UV
luminous star forming regions of LBGs and account for under 1/3 of the
absorption systems, or if they are common to all galaxies, they can
not extend to 40~kpc with large velocities, while continuing to make
UV absorption that we can detect. This suggests that the metals were
in place in the IGM long before $z = 2$.

Separately, we examine the absorption seen when a sight line passes a
second QSO. We see 19 absorbers within $\pm 400$~\kms\ of the
redshifts of the partner QSO and 30 within 1000~\kms .  These
transverse absorbers are more tightly clustered about the QSO redshift
than are associated C~IV absorbers seen in individual QSO spectra.
The probability of seeing absorption when a sight line passes a QSO is
approximately constant for impact parameters 0.1 -- 1.5~Mpc.  Perhaps
we do not see a rapid rise in the probability at small impact
parameters, because the UV from QSOs destroys some absorbers near to
the QSOs.  The 3D distribution of 64 absorbers around 313 QSOs is to
first order isotropic, with just a hint of the anisotropy expected if
the QSO UV emission is beamed into coaxial cones of half apex angle
$\sim 20$ degrees towards and away from the Earth.  Alternatively,
QSOs might emit UV isotropically but for a surprisingly short time of
only 0.3~Myr.  Anisotropy is obscured because some \zem\ values have
large errors, we do not know the distance from the QSOs at which a
given absorber will be destroyed, and the axis of the cone of UV
radiation has an unknown angle to our line-of-sight.

\end{abstract}


\section{Introduction}

In this paper we discuss correlations between metal line absorption
systems seen in the spectra of pairs of QSOs with sight lines
separated by approximately 1~Mpc proper at $z=2$. These metal systems
probe two separate environments; metals in the outer regions of
galaxies that are far from both QSO, and separately, metals in the
immediate vicinity of one or both QSOs. We will maintain the
distinction between these two environments throughout the paper.

We use moderate resolution spectra (FWHM 100 -- 250~\kms ), with a
wide range of SNR and hence we are sensitive to a wide range of
absorption line equivalent widths.  The absorption lines that we see
typically come from the outer parts of galaxies, and the regions
around the QSOs themselves. We learn little about the distribution of
metals in the IGM far from galaxies because those metals tend to make
absorption lines that are weaker than most that we can detect.

In general information that we gain from paired sight lines
complements what we have learnt from individual sight lines over the
last 30 years.  However, it remains hard to decide what is causing the
absorption, especially because we get velocity and not pure distance
information from redshifts in spectra. When we come to examine small
scale differences in redshifts, both in single lines of sight, between
paired sight lines, or between QSO absorption and adjacent galaxies,
peculiar velocities can be comparable to the Hubble flow. Moreover, these
peculiar velocities can have systematic as well as random components
on small scales.  We expect to see net infall relative to Hubble
motion when the absorbing region is over-dense, while we might see
outflows if the gas is ejected from a galaxy in a wind.

In the remainder of this introduction we discuss metal absorption
systems far from the QSO and then those near to the QSO and finally we
describe the distance measures that we use.  In \S 2 we describe the
sample of 310 QSOs and their separations from their partners.  In \S 3
we present the observations then in \S 4 we describe the absorption
systems that we found.  In \S 5 we discuss the redshift differences
between pair of absorbers in different sight lines, and absorbers near
to QSOs. In \S 6 we see that these coincidences have the same ions and
\wrest\ values as normal systems.  In \S 7 we see that the
absorber-absorber correlation is strong and inconsistent with
absorption in fast winds. In \S 8 we see that the absorbers have a
uniform distribution around the QSOs, with a hint of a lack of
absorbers near to the line of sight. We mention BAL systems in \S 9
and discuss our conclusions in \S 10.

\subsection{Metals Absorption Far from QSOs}

In general the correlation of metal line absorption between sight
lines tells us about the size and structure of the regions that
produce metal absorption.

When QSO sight lines are separated by 1 -- 100 kpc we see metal
absorption lines in each that can have very similar line profiles,
indicating that the light is passing through the same gas clouds.
\citet{rauch02} have studied ions such as Mg~II, Ca~II, and Fe~II in a
quadruple gravitationally lensed QSO to derive the structure of metal
clouds in the interstellar medium.

For all except exceptionally close sight lines with arcsecond
separations, it is hard to distinguish between absorption by a single
large cloud and absorption by clouds which are too small to
individually cover both sight lines, but that are clustered such that
one cloud covers each sight line. Or in terms of a continuous
distribution of gas, the density, ionization and metal abundance are
changing significantly over the distance between the sight lines. This
ambiguity was first stressed by \citet{weymann81} and is more apparent
in spectra of lower spectral resolution.

The sizes of the metal absorbing regions, either single clouds and
clusters of clouds, have been estimated from various pairings on the
sky. Lensed images of the same QSO \citep{steidel91} probe the
smallest separations and give proper sizes of $\sim 15$ to about $\sim
80$ kpc ($H_0 = 71$ km s$^{-1}$ Mpc$^{-1}$). Pairs of physically
distinct QSOs close to each other on the sky, like those used here,
typically probe separations greater than 1~Mpc, but the occasional
pair is closer and gives sizes of roughly 100~kpc for strong C~IV
absorbers \citep{crotts94}.  \citet{chen01b} find CI~V is very often
seen out to impact parameters (radii) of 140~kpc around galaxies seen
in images, but only rarely beyond that distance.  \citet{churchill07a}
discuss one E/SO galaxy that shows no metals at 58~kpc, showing that
not all galaxies absorb at large distances.  \citet{adelberger05}
examined the absorption near 1044 foreground galaxies in the spectra
of 23 background QSOs and found C~IV absorption with column densities
$N_{CIV} >> 10^{14}$ is often seen to radii of 40~kpc, while $N_{CIV}
\approx 10^{14}$ is seen to 80~kpc.  A corresponding size for M~II
absorbers at low-redshifts ($z \la 0.5$) is about 70~kpc
\citep{lanzetta90}. We will see strong correlations arising from sight
lines that pass through the same halo, but in general out paired sight
lines are too far apart to both probe the same halo.

On larger scales the correlation in the redshifts of metal line
absorption in adjacent sight lines depend on the clustering of
baryons, modified by the spatial variation in both the ratio of metals
to baryons, and the level of ionization of the metal atoms.

We know from many large surveys of single QSO sight lines that metal
lines are very strongly clustered on scales of $<150$ \kms , with
detectable clustering out to beyond 600 \kms\
\citep{sargent88a,petitjean94a,rauch96a,pichon03a,scannapieco06a}, and
possibly out to 140~Mpc (for $q_0=0.5$)
\citep{quashnock96a}. Absorption systems with higher H~I or metal line
column densities tend to be more strongly clustered and confined,
while those with low column densities are more widespread
\citep{sargent80,tytler87b,cristiani97b}.

With some assumptions we can convert velocities along sight lines into
a prediction for the correlation between adjacent sight lines. We need
to specify the cosmological model, and the 3D density field of the
metal ions. The connection between line of sight and transverse
clustering is complicated by velocity field distortions.  Ignoring
these velocity distortions, we would expect the see metal lines weakly
correlated in sight lines separated by $< 600/H(z) \simeq 3$ Mpc at
$z=2$.  \citet{crotts97a} found that the clustering of C~IV in HIRES
spectra of the triplet near 1623+27 was weaker than along the line of
sight, they suggest perhaps because peculiar velocities make the
clustering appear over extended along the line of sight.

\citet{scannapieco05a} argues that we expect metals to be highly
clustered around their sources that are in high density
peaks. \citet{scannapieco06a} show that the metal line correlations
function along a line of sight can be modelled with most metals
confined to bubbles of radius 2~Mpc comoving.  On the other hand,
\citet{pieri06a} show that the incidence of weak C~IV and O~VI
absorption is similar both near to galaxies (marked by strong metal
lines) and far from galaxies, indicating that some metals (mostly
below our detection threshold) are present well beyond the immediate
surroundings of galaxies.

\citet{coppolani06} did not see any significant correlation in 139
C~IV systems towards 32 pairs of QSOs, except for an over-density of
C~IV in front of a group of 4 QSOs.  Since the mean separation of
their QSO pairs is $>2$ arcmin, they conclude that metal enriched
``bubbles'' should be smaller. Below we will show that we see strong
correlations between sight lines, but only when the separations are
$<0.6$~Mpc.  \citet{simcoe06} study galaxies and intergalactic gas
towards a single QSO at z=2.73 and conclude that the metal absorption
can arise from bubbles of radii $\approx 100$~kpc and thickness
$\approx 1$~kpc.

On the largest scales \citet{jakobsen86a} and \citet{sargent87a}
discuss evidence for a supercluster filament causing absorption at
similar velocities in sight lines separated by 17.9
arcminutes. \citet{romani91a} describe how absorption in pairs of QSOs
could be used to find superclusters at high redshifts, before any
galaxies were known at these redshifts.  \citet{jakobsen92a} used C~IV
absorption in 12 QSOs including Tol 1037-2703 to detect a sheet
spanning tens of Mpc. \citet{tytler93a} saw no sign of periodicity on
10 -- 210~Mpc scales in the 3D distribution of 268 Mg~II absorption
systems. \citet{loh01a} examined the 3D distribution of 345 C~IV
systems from 276 QSOs and found evidence for clustering on scales up
to 220~Mpc ($q_0=0.5$). \citet{vandenberk00} found four metal line
absorption systems towards QSOs near to the HDF, two of which have
redshifts that place them in the second most populated peak in the
galaxy redshift distribution.  \citet{williger96} compiled a
statistically complete sample of C~IV absorption at $1.5 < z < 2.8$ in
25 sight lines. They found evidence for structure on 2 -- 50 Mpc
proper scales but not for a smaller sample of 11 Mg~II absorbers.
Overall, this body of work shows that the distribution of absorption
systems clearly reveals large scale structure, especially the redshift
spikes seen in all narrow angle galaxy surveys
\citep{broadhurst90a}. The sample presented here is not well suited to
studying correlations on these largest scales because we have very low
sampling density, and we very rarely see even one metal system when we
pass a given group of galaxies.

\subsection{Metals Absorption near to QSOs}

We are especially interested in absorption by metals around QSOs
because these metals will help us understand the feedback of QSOs on
the IGM. Specifically, we should learn about the ejection from the
QSOs of hot gas that contains metals.  We might see absorption from
the emission line region, the ISM of the QSO host galaxy, and galaxies
in a group that might contain the QSO host. In all cases we expect to
see enhanced photoionization by the QSO UV. If the QSO UV is not
isotropic then we might see an anisotropic spatial distribution of the
absorbers about the QSOs. We also expect to see a higher density of
absorbers near to the QSOs because QSO hosts are in over-dense
environments.

The results in the literature on absorption near to QSOs are
diverse. We will distinguish between absorption seen along a single
line of sight (from a QSO to the Earth) from that seen in the spectrum
of a second background QSO (transverse to the line of sight). It also
helps to distinguish different ions, since they probe different gas
densities and respond differently to the QSO environment.  Note
however that if we can show that QSOs emit isotropically, or instead
are strongly beamed, using one line of evidence, then this result may
hold even if other lines of evidence seem to be contradictory. The
other evidence might be based on a small sample or the measurements
may be less sensitive to the effect in question.

A significant fraction of metal lines systems are intrinsic to QSOs
and not at the positions implied by their redshifts.
\citet{richards99} found that the numbers of C~IV systems in a
heterogeneous catalogue of absorption systems depended on the optical
luminosity and radio properties of the QSOs. If this were confirmed in
a homogeneous sample it would be evidence that perhaps 36\% of C~IV
systems are intrinsic.  Recently \citet{misawa07b} used doublet ratios
of metal lines in Keck HIRES spectra of 37 QSOs to conduct the first
large survey of the frequency of absorbers that do not cover the QSO
UV radiation source. They found 28 reliable cases of intrinsic
absorption, corresponding to 10 -- 17\% of narrow C~IV systems at
velocities of 5000 -- 70,000~\kms\ from their QSOs, and at least 50\%
of QSOs show intrinsic systems. \citet{ganguly07a} estimate that 60\%
of QSOs, with a wide range of luminosity, display outflows in
absorption that we can see as BAL or associated absorption, or
absorption with partial coverage, time variability, high
photoionization or high metallicity.

Intrinsic gas that is ejected and clumped in velocity can appear like
a supercluster of galaxies along a single sight line. The examination
of absorption systems in pairs of QSOs can clarify the situation. The
outflows mentioned above are generally believed to be confined to
within a few pc to a few ~kpc of the QSOs
\citep{narayanan04,wise04a,misawa05b}, too small a distance to
intercept the other line of sight.

\subsubsection{Absorption near to QSOs and along individual Lines of Sight (los)}

We see various behaviours in the number of absorption lines from
different ions at velocities near to the QSOs emission redshifts. The
amount of H~I decreases, probably because of the increased
photoionization. The number of Lyman limit Systems (LLS) is little
changed, while the number of C~IV lines \citep{young82a} and DLAs
\citep{russell06a} increases.  N~V lines in particular are found
mostly near to the QSOs \citep{petitjean94b}.  \citet{prochaska07a}
also study the incidence of DLAs within 3000~\kms\ of their QSOs. At
$z < 2.5$ and $z > 3.5$ they see no deviation from a flat intervening
distribution, but at intermediate $z$ they see twice the usual number
of DLAs.  They had expected 5 -- 10 times more DLAs near the QSOs
because of clustering of galaxies around the QSOs and interpret this
lack as due to the enhanced ionization from the QSOs.

It has long been known \citep{tytler82} that there is no strong excess
of LLS with \zabs $\simeq $ \zem\ along individual sight lines to
mostly high luminosity QSOs, and hence no sign of absorption in the
host galaxies. There are two obvious explanations.  First, the QSOs
might be in galaxies (elliptical or lenticular) that do not have
enough H~I columns to make LLS, \lnhi $>17.2$~\cmm .  While high
luminosity QSOs are in elliptical galaxies, low luminosity QSOs are
often disk galaxies \citep{hamilton02a}, hence we predict that lower
luminosity QSOs may show enhanced LLS at \zabs $\simeq $ \zem .
Second, the QSO UV radiation or relativistic jets may have ionized the
gas in the sight line.  In some models, feedback from optical
\citep{dimatteo05a} or radio loud AGN \citep{best07a} can effect the
host galaxy as a whole, tending to make it appear like an elliptical
galaxy, and hence making the first explanation a consequence of the
second: the host is elliptical because of the QSOs effect on the host.

When an absorption system has velocities similar to that of the QSO
emission lines, we must be aware that the gas could have significant
peculiar velocity, either from gas ejected from the QSO or the motions
in a group of galaxies containing the QSO host galaxy.
\citet{guimaraes07a} find that the H~I proximity effect in 45 high
QSOs at $z > 4$ is less than expected, by an amount that implies that
QSOs reside in regions with overdensities of 5 to 2 within 3 and 10
$h^{-1}$Mpc, with higher luminosity QSOs in higher over-densities.

\subsubsection{Absorption near to QSOs and seen in background spectra: Transverse}

The results of searches for a relative decrease in H~I absorption near
foreground QSOs, the classical ``transverse proximity effect'', are
complex and confusing.  In early work,
\citet{dobrzycki91a,dobrzycki91b} found a large transverse void in the
amount of \lya\ absorption that they concluded was probably not caused
by UV from a foreground QSO. \citet{srianand97a} discussed a second
void that was less than expected from the foreground QSO luminosity.
\citet{fernandez95} and \citet{liske01} claimed marginal detections of
the transverse proximity effect while
\citet{crotts89,moller92,crotts98,croft04a,schriber04a} saw no lack of
H~I absorption. Rather, in some cases we have reports of enhanced,
instead of decreased H~I absorption near foreground QSOs
\citep{crotts98,croft04a,schriber04a}.

Three types of explanation have been discussed for the non-detections
of the expected H~I transverse proximity effect:
\begin{enumerate}
\item The QSOs were less luminous in the past, by factors of 10 -- 100
  some 1 -- 30~Myr ago \citep{schirber04a}.
\item The QSO radiation is beamed so that their typical luminosity in
  transverse directions is factors of 10 -- 100 lower than in the line
  of sight (los) direction
  \citep{barthel89,crotts89,moller92,antonucci93,schirber04a}.
\item The H~I does feel the full QSO UV luminosity but the increased
  photoionization is partly cancelled by increased gas density, and
  the increased number of galaxies near the QSO
  \citep{loeb95,schriber04a,rollinde05a,guimaraes07a,faucher07a,kim07a}.
\end{enumerate}
The first two explanations both make the radiation around the QSOs
anisotropic, but both narrow beam opening angles and ultra short UV
life times are disfavoured by other evidence \citet{urry95}.  The
third is widely agreed to be a real effect that enhances the numbers
of many types of absorbers near to QSOs.

In contrast with H~I, there are claims that the He~II ionization is
changed by the radiation from foreground QSOs
\citep{jakobsen03a,worseck06a,worseck06b,worseck07a}.
\citet{worseck06a} claim that the hardness of the radiation ionising H
and He~II changes near to four foreground QSOs and that lifetimes are
at least 10 -- 30~Myr ago.

Only a few cases of metal line systems near to foreground QSOs have
been found
\citep{shaver82a,shaver83a,shaver85a,dodorico02,adelberger06a} because
few QSO pairs were known until recently.  \citet{williger96} found
marginal evidence of an association between C~IV absorbers and a
grouping of 25 QSOs.

\citet{bowen06a} found excess Mg~II absorption in background QSOs at
the emission redshift of foreground QSOs. They considered four pairs
of QSOs with separations of 3 -- 15 arcsec, or 26 -- 97~kpc.  Since we
do not expect to see associated Mg~II in 4 out of 4 sight lines, they
conclude that the absorbing gas is not isotropic, and they discuss
possible explanations (host galaxy, nearby galaxies, ejected gas),
none of which were compelling to them.

\citet{hennawi07} examine the incidence of a subset of all LLS, those
with \lnhi $> 19$~\cmm , in the spectra of 149 QSO pairs with the
nearer QSOs at $1.8<z<4.0$.  They find 17 such LLS in transverse
pairings which they argue is larger than the number seen along lines
of sight by factors of 4 -- 20 times. \citet{hennawi06a} note their
concern over the level of completeness and the false positive rate in
their samples, since it is hard to find \lya\ lines with \lnhi $>
18.3$ or 19.3~\cmm\ in low SNR moderate resolution spectra, and the
reality of an excess depends on the precise minimum \nhi\ values, and
the number of false high column systems in their sample.  Like
\citet{bowen06a}, they suggest that absorbers are an-isotropically
distributed. \citet{hennawi07} suggest that the LLS absorbers in the
line of sight are photo-evaporated while those in the transverse
direction are not evaporated because they do not see the full QSO UV
flux.

\subsection{Distances}

We will measure separations in various units.  The separations of
sight lines in the plane of the sky are known in arcseconds.  The
errors on these separations are probably less than one arcsecond, but
they will be larger when the QSOs lack modern position measurements,
or when the two QSOs were measured in different coordinate reference
frames.  We use a flat cosmological model with \ol = 0.73, \om = 0.27
and $H_0 = 71$ \kms Mpc$^{-1}$.

We convert from arcsec to Mpc in the plane of the sky, an impact
parameter, using
\begin{equation}
b = \frac{c ~ \delta \phi }{(1+z) H_0}  \int^z_0 {dz \over \sqrt{\Omega_m (1+z)^3 + \Omega_\Lambda}}
\end{equation}
where $b$ is the transverse proper distance in the plane of the sky
corresponding to the angular separation $\delta \phi $ in radians and
measured at redshift $z$. The $b$ parameter is the proper equivalent
of the comoving $r_p$ often seen in galaxy literature.  For our
cosmological model this gives
\begin{equation}
b = \frac{0.020471 ~ \delta \theta}{(1+z)}  \int^z_0 {dz \over \sqrt{0.27 (1+z)^3 + 0.73}}
\end{equation}
where $\delta \theta$ is now the angular separation in arcseconds, and
the integral is 0.44567, 0.78566, 1.24223, 1.75678 for $z=0.5$, 1, 2,
3.  \citet[Eqn 1,2]{jakobsen03a} and \citet{pen99} give analytic
approximations, and \citet{wright06a} gives a Java calculator.  We do
not use the formulae given by \citet{liske00a} that are required for
large angles or large redshift differences.

Distances along the lines of sight are most conveniently expressed as
differences of redshift.  When we discuss correlations in the
positions of absorbers, we are interested in scales of about 1 Mpc or
100 -- 400~km/s depending on the redshift.  These correlations will
include higher density regions that will be expanding less rapidly
than the Hubble flow, and some of them may be bound with constant
proper size.

We shall convert the differences in redshift into velocity intervals
using
\begin{equation}
{v \over c} =  { (1+ z_1 )^{2}-(1+z_2 )^{2} \over (1+z_1 )^{2}+(1+z_2 )^{2} },
\end{equation}
which for small intervals is approximately $v/c = H(z)d/c \simeq
\Delta z / (1+z)$.  We can convert these velocities into proper
distances $d$ using $v=H(z)d$ where $[H(z)/H_0]^2 = (1+z)^3 \Omega _m
+ \Omega _{\Lambda}$ for our flat model and $H(z) =$ 120.70, 201.069,
301.31, and 416.91 km/s at $z = $1, 2, 3 and 4.

For small distances conversion from velocity along the line of sight
using the Hubble constant and ignoring peculiar velocities will be
highly inaccurate because of the distortions caused by the systematic
peculiar velocities \citep{kaiser88,mcdonald03,kim07a}.  The effect of
peculiar velocities has been studied in detail for H~I \lya\
absorption from the IGM \citep{hui99a, mcdonald99a, rollinde03}.  The
goal is to measure the correlation in the H~I absorption between
adjacent sight lines, divided by the correlation along individual
sight lines.  The division removes most of the effects of evolution
and is used in a geometric cosmological test which \citet{alcock79}
showed is particularly sensitive to \ol . In this paper we assume that
we know \ol\ with negligible error. We discuss redshift-space
distortions in \S 7 and \S 8.

\section{QSO Targets}
\label{starg}

We use spectra of the 310 QSOs which include 140 pairs and 10
triplets.  We treat each triple as 3, non-independent pairs, giving
170 QSO-to-QSO pairings.  We selected these QSOs because their
separations and redshifts are suitable for studying the correlations
in the H~I absorption in the \lyaf . When choosing objects to observe
we initially observed all known pairs separated by under a few
arcminutes. Later, as more pairs were announced by 2dF Quasar Redshift
Survey \citep{boyle97a,shanks00a,miller04a} and SDSS \citep{york00},
we strongly biased our observing to the closest known suitable pairs,
typically those within 120~arcseconds.  We also strongly biased our
sample to pairs with similar emission redshifts, \zem , to maximise
the redshift of overlap in the region between \lya\ and \lyb\ in the
paired spectra. However, some of the pairs discussed here do have
widely differing \zem\ values.  We also biased the sample against
pairs where C~IV BAL absorption was at \zabs\ values that would put
\lya\ at rest wavelengths 1070 -- 1170~\AA , the key wavelengths for
correlations of \lya\ in the \lyaf .  Hence the sample contains less
than the normal number of BAL QSOs, though there remain BAL systems in
34 of the QSOs, especially weak BAL systems and systems restricted to
\zabs\ similar to \zem .  If we knew from SDSS spectra that a QSO in a
pair had strong and widespread BAL absorption then we typically did
not obtain new spectra of higher resolution or higher signal-to-noise
ratio (SNR) to cover the \lyaf\ at wavelengths $< 3900 $~\AA\ that
were not covered by the SDSS spectra.

In Table \ref{tabtarg}
we list the J2000 coordinates of the 310 QSOs their \zem\ values and
the separation from the partner QSO in arcseconds.  We give each QSO a
label comprising a `P', a number, and a letter. The number identifies
the pair and the letter `a', `b' or `c' the QSO in that pair or
triple.  We show the label in bold face when that QSO is part of a
triple.  The angles listed are from the current QSO to the next QSO in
the sequence b, c then a; hence QSO P8c is 217 arcseconds from P8a.
The order of the QSOs in this table is not random and mostly relates
in part to when we obtained spectra.

\begin{table}
\caption{QSO Targets. In order of QSO pair Label number, not RA.}
\label{tabtarg}
\begin{tabular}{llllr}
\hline
 RA (J2000)    &    Dec (J2000)   &  \zem\  & Label & Separation \\
\hline
00 44 34.08 & +00 19 03.5 & 1.878 & P1a & 88.64"  \\
00 44 39.33 & +00 18 22.8 & 1.866 & P1b &  \\
00 55 57.46 & $-$32 55 39.0 & 2.250 & P2a & 122.53" \\
00 56 05.33 & $-$32 56 51.1 & 2.125 & P2b & \\
02 09 54.8 & $-$10 02 23.0 & 1.970 & P3a  & 12.04" \\
02 10 00.1 & $-$10 03 54.0 & 1.976 & P3b  & \\
02 56 42.6 & $-$33 15 21.0 & 1.915 & P4a  & 55.78" \\
02 56 47.02 & $-$33 15 27.0 & 1.863 & P4b & \\
03 10 06.08 & $-$19 21 24.9 & 2.144 & P5a & 60.27" \\
03 10 09.05 & $-$19 22 08.1 & 2.122 & P5b & \\
03 10 36.47 & $-$30 51 08.4 & 2.554 & P6a & 72.04" \\
03 10 41.06 & $-$30 50 27.5 & 2.544 & P6b & \\
09 14 04.1 & +46 10 44.9 & 2.180 & P7a   & 64.59" \\
09 14 10.3 & +46 10 50.01 & 2.370 & P7b  &  \\
09 56 58.73 & +69 38 52.5 & 2.048 & {\bf P8a} & 130.92" \\
09 57 21.22 & +69 37 54.5 & 2.054 & {\bf P8b} & 108.83" \\
09 57 25.94 & +69 36 08.5 & 2.048 & {\bf P8c} & 217.08" \\
11 45 47.55 & $-$00 31 06.7 & 2.043 & P9a  & 149.34" \\
11 45 53.67 & $-$00 33 04.5 & 2.055 & P9b & \\
12 12 51.14 & $-$00 53 42.2 & 2.473 & P10a & 74.01" \\
12 12 56.06 & $-$00 53 36.5 & 2.459 & P10b & \\
13 06 34.19 & +29 24 43.1 & 1.960 & P11a  & 27.82" \\
13 06 35.41 & +29 25 05.9 & 1.926 & P11b  & \\
13 21 47.86 & +01 06 04.8 & 2.130 & P12a  & 107.86" \\
13 21 54.33 & +01 06 51.9 & 1.971 & P12b  & \\
13 39 39.0 & +00 10 22.0 & 2.122 &  P13a  & 101.69" \\
13 39 45.4 & +00 09 45.0 & 1.869 &  P13b  & \\
13 46 21.4 & $-$00 38 05.0 & 1.894 & P14a  & 119.62" \\
13 46 25.6 & $-$00 39 47.0 & 1.848 & P14b  & \\
14 12 24.51 & $-$01 56 34.0 & 1.916 & P15a & 112.60" \\
14 12 29.73 & $-$01 55 13.1 & 2.030 & P15b & \\
14 20 45.98 & $-$00 05 18.0 & 2.193 & P16a & 299.36" \\
14 20 55.61 & $-$00 09 40.0 & 2.193 & P16b & \\
16 12 37.9 & +23 57 09.0 & 2.014 & P17a  & 117.22" \\
16 12 45.6 & +23 58 00.0 & 2.005 & P17b  & \\
16 45 01.09 & +46 26 16.0 & 3.790 & P18a  & 195.32" \\
16 45 19.62 & +46 25 38.3 & 3.831 & P18b  & \\
17 27 56.45 & +58 21 55.7 & 2.368 & P19a   & 111.58" \\
17 28 06.77 & +58 20 39.19 & 2.011 & P19b  & \\
17 30 14.71 & +54 56 57.5 & 2.127 & P20a   & 244.99" \\
17 30 42.38 & +54 56 01.1 & 2.112 & P20b  & \\
17 36 26.73 & +55 27 20.7 & 1.822 & P21a  & 101.46" \\
17 36 35.51 & +55 28 29.4 & 1.988 & P21b  &  \\
22 39 41.75 & $-$29 49 55.2 & 2.101 & {\bf P22a} & 154.98" \\
22 39 48.64 & $-$29 47 48.7 & 2.068 & {\bf P22b} & 63.52" \\
22 39 51.82 & $-$29 48 37.0 & 2.121 & {\bf P22c} & 152.44" \\
23 09 11.88 & $-$27 32 27.1 & 1.930 & P23a & 49.52" \\
23 09 15.34 & $-$27 32 45.3 & 1.927 & P23b & \\
23 26 03.52 & $-$29 37 40.4 & 2.310 & P24a & 141.22" \\
23 26 14.26 & $-$29 37 22.3 & 2.387 & P24b & \\
09 09 23.13 & +00 02 03.9 & 1.889 & P25a  & 14.99" \\
09 09 24.01 & +00 02 11.0 & 1.866 & P25b  &  \\
11 07 25.70 & +00 33 53.6 & 1.883 & P26a  & 24.82" \\
11 07 27.08 & +00 34 07.3 & 1.882 & P26b  &  \\
14 35 06.42 & +00 09 01.5 & 2.378 & P27a  & 33.24" \\
14 35 08.32 & +00 08 44.4 & 2.378 & P27b  &  \\
15 48 40.77 & +53 37 08.59 & 2.165 & P28a  & 126.11" \\
15 48 50.17 & +53 38 43.0 & 2.188 & P28b  &  \\
23 01 12.42 & $-$31 43 45.0 & 1.977 & P29a & 67.84" \\
23 01 17.62 & $-$31 43 59.2 & 2.132 & P29b & \\
00 45 26.49 & $-$32 00 16.91 & 1.885 & P30a & 79.62" \\
00 45 27.54 & $-$32 01 35.4 & 1.988 & P30b & \\
01 24 56.45 & $-$28 51 21.0 & 1.992 & P31a & 100.69" \\
01 24 53.09 & $-$28 52 51.5 & 2.094 & P31b & \\
\end{tabular}
\end{table}

\addtocounter{table}{-1}

\begin{table}
\caption{Continued.}
\begin{tabular}{lllll}
\hline
 RA (J2000)    &    Dec (J2000)   &  \zem\  & Label & Separation \\
\hline
01 35 14.53 & $-$00 53 18.9 & 2.111 & P32a & 258.23" \\
01 35 21.00 & $-$00 57 18.2 & 2.075 & P32b & \\
02 48 25.59 & $-$28 03 55.4 & 2.139 & P34a & 193.83" \\
02 48 40.13 & $-$28 03 32.4 & 2.209 & P34b & \\
21 36 19.40 & +00 41 31.0 & 2.030 & P35a & 288.90" \\
21 36 38.60 & +00 41 54.0 & 1.941 & P35b & \\
23 53 10.02 & $-$27 26 14.09 & 1.968 & P36a & 40.35" \\
23 53 13.03 & $-$27 26 09.4 & 2.303 & P36b & \\
00 08 52.71 & $-$29 00 44.1 & 2.645 & P37a & 78.53" \\
00 08 57.73 & $-$29 01 26.9 & 2.610 & P37b & \\
02 18 21.44 & $-$29 53 40.9 & 2.070 & P38a & 21.97" \\
02 18 22.96 & $-$29 53 31.3 & 1.917 & P38b & \\
03 06 40.91 & $-$30 10 31.9 & 2.093 & P39a & 51.21" \\
03 06 43.75 & $-$30 11 07.49 & 2.129 & P39b & \\
03 13 24.40 & $-$31 41 44.9 & 1.954 & P40a & 17.00" \\
03 13 25.51 & $-$31 41 54.3 & 2.065 & P40b & \\
03 33 20.90 & $-$06 12 16.8 & 2.050 & P41a & 145.71" \\
03 33 24.83 & $-$06 10 03.4 & 2.139 & P41b & \\
21 48 34.95 & $-$29 41 09.9 & 1.807 & P42a & 26.74" \\
21 48 36.61 & $-$29 40 54.19 & 2.089 & P42b & \\
22 32 20.27 & $-$28 38 58.7 & 2.204 & P43a & 50.81" \\
22 32 23.45 & $-$28 38 29.9 & 2.065 & P43b & \\
23 59 44.12 & $-$00 57 38.16 & 1.778 & P44a & 46.15" \\
23 59 45.48 & $-$00 58 19.56 & 1.814 & P44b & \\
00 59 34.10 & $-$08 43 13.1 & 2.074 & P45a & 269.92" \\
00 59 51.67 & $-$08 44 23.8 & 2.142 & P45b & \\
01 06 57.94 & $-$08 55 00.1 & 2.354 & P46a & 181.93" \\
01 06 58.41 & $-$08 58 01.9 & 1.827 & P46b & \\
03 40 23.50 & +00 31 11.8 & 1.910 & P47a & 217.35" \\
03 40 27.31 & +00 34 41.5 & 1.874 & P47b & \\
08 54 06.10 & +42 38 10.0 & 2.387 & {\bf P49a} & 283.22" \\
08 54 15.40 & +42 42 34.0 & 2.174 & {\bf P49b} & 449.63" \\
08 54 25.00 & +42 35 17.0 & 1.850 & {\bf P49c} & 271.04" \\
10 05 38.50 & +57 07 44.0 & 1.866 & P50a & 122.15" \\
10 05 41.30 & +57 05 44.0 & 2.306 & P50b & \\
10 40 32.20 & $-$27 27 48.6 & 2.331 & {\bf P51a} & 291.12" \\
10 40 33.50 & $-$27 22 58.0 & 1.937 & {\bf P51b} & 133.92" \\
10 40 40.32 & $-$27 24 36.4 & 2.460 & {\bf P51c} & 220.51" \\
10 41 21.90 & +56 30 01.0 & 2.052 & P52a & 65.09" \\
10 41 29.30 & +56 30 23.0 & 2.267 & P52b & \\
10 43 30.46 & $-$02 30 12.7 & 2.246 & P53a & 258.44" \\
10 43 42.53 & $-$02 33 17.3 & 1.993 & P53b & \\
11 06 10.70 & +64 00 09.0 & 2.201 & P54a & 170.00" \\
11 06 26.60 & +63 57 55.0 & 1.960 & P54b & \\
11 11 12.30 & +01 22 01.6 & 2.417 & {\bf P55a} & 91.33" \\
11 11 14.11 & +01 20 34.4 & 2.150 & {\bf P55b} & 280.80" \\
11 11 31.30 & +01 22 25.0 & 2.010 & {\bf P55c} & 285.96" \\
11 19 22.40 & +60 48 51.0 & 2.014 & {\bf P56a} & 142.19" \\
11 19 28.90 & +60 46 37.0 & 2.293 & {\bf P56b} & 164.79" \\
11 19 31.10 & +60 49 21.0 & 2.645 & {\bf P56c} & 70.35" \\
12 24 27.80 & $-$11 20 50.0 & 2.495 & P57a & 253.03" \\
12 24 41.40 & $-$11 23 25.0 & 2.171 & P57b & \\
13 31 25.93 & +00 44 14.0 & 2.020 & {\bf P58a} & 219.75" \\
13 31 38.50 & +00 42 21.1 & 2.429 & {\bf P58b} & 252.96" \\
13 31 50.51 & +00 45 10.7 & 1.893 & {\bf P58c} & 247.41" \\
14 16 47.60 & +63 02 51.0 & 2.034 & P60a & 284.83" \\
14 16 50.80 & +63 07 35.0 & 1.961 & P60b & \\
14 26 05.80 & +50 04 26.0 & 2.242 & P61a & 235.17" \\
14 26 28.00 & +50 02 48.0 & 2.324 & P61b & \\
14 53 29.53 & +00 23 57.3 & 2.538 & P62a & 259.89" \\
14 53 37.99 & +00 20 10.5 & 1.859 & P62b & \\
14 58 38.04 & +00 24 18.0 & 1.888 & {\bf P63a} & 387.74" \\
14 59 01.28 & +00 21 23.7 & 1.988 & {\bf P63b} & 180.73" \\
14 59 07.19 & +00 21 01.2 & 3.012 & {\bf P63c} & 91.46" \\
\end{tabular}
\end{table}

\addtocounter{table}{-1}

\begin{table}
\caption{Continued.}
\begin{tabular}{lllll}
\hline
 RA (J2000)    &    Dec (J2000)   &  \zem\  & Label & Separation \\
\hline
15 22 43.99 & +03 27 19.8 & 1.998 & P64a & 249.33" \\
15 22 46.66 & +03 31 25.9 & 2.287 & P64b & \\
16 06 28.39 & +17 31 26.0 & 2.040 & P65a & 265.05" \\
16 06 37.60 & +17 35 16.0 & 2.323 & P65b & \\
16 32 52.30 & +37.47 47.99 & 1.888 & P66a & 156.19" \\
16 32 57.60 & +37 50 11.0 & 2.152 & P66b & \\
22 40 26.20 & +00 39 38.0 & 2.200 & P67a & 213.50" \\
22 40 40.10 & +00 40 24.0 & 2.200 & P67b & \\
23 31 32.84 & +01 06 20.9 & 2.641 & P68a & 153.99" \\
23 31 39.75 & +01 04 27.0 & 2.245 & P68b & \\
23 46 46.02 & +12 45 30.18 & 2.763 & P69a & 334.44" \\
23 46 28.21 & +12 48 59.89 & 2.525 & P69b & \\
20 45 33.15 & $-$06 21 54.3 & 2.014 & P70a & 182.84" \\
20 45 22.28 & $-$06 23 19.0 & 2.157 & P70b & \\
13 48 08.70 & +28 40 07.0 & 2.464 & P71a & 59.09" \\
13 48 04.40 & +28 40 24.0 & 2.464 & P71b & \\
09 00 06.90 & +03 33 07 & 1.872 & P72a & 212.87" \\
08 59 52.70 & +03 33 18 & 2.163 & P72b & \\
08 04 00.30 & +30 20 46 & 3.446 & P73a & 269.22" \\
08 03 42.00 & +30 22 54 & 2.031 & P73b & \\
15 45 44.20 & +51 13 07 & 2.242 & P74a & 98.27" \\
15 45 34.60 & +51 12 28 & 2.453 & P74b & \\
17 17 30.70 & +26 22 27 & 2.203 & P75a & 211.93" \\
17 17 15.20 & +26 21 48 & 1.934 & P75b & \\
17 29 43.36 & +60 21 54.20 & 1.928 & P76a & 370.01" \\
17 30 30.20 & +60 19 47.40 & 2.215 & P76b & \\
17 18 45.00 & +30 26 47 & 2.028 & P77a & 160.61" \\
17 18 37.20 & +30 28 52 & 2.028 & P77b & \\
17 28 40.02 & +56 39 57.74 & 1.984 & P78a & 148.43" \\
17 28 52.65 & +56 41 43.58 & 1.769 & P78b & \\
14 57 56.27 & +57 44 46.90 & 2.130 & P79a & 73.63" \\
14 57 47.55 & +57 44 23.50 & 2.016 & P79b & \\
15 34 12.70 & +50 34 05 & 2.118 & P80a & 280.61" \\
15 33 48.30 & +50 31 28 & 2.215 & P80b & \\
15 08 38.11 & +60 35 40.10 & 2.179 & P81a & 120.46" \\
15 08 27.67 & +60 34 07.40 & 1.893 & P81b & \\
16 50 51.10 & +34 43 10 & 2.002 & P82a & 169.84" \\
16 50 43.30 & +34 45 30 & 1.984 & P82b & \\
07 55 45.60 & +40 56 43.61 & 2.348 & P83a & 138.18" \\
07 55 35.61 & +40 58 02.90 & 2.418 & P83b & \\
11 09 52.30 & +55 42 24 & 3.177 & P84a & 221.68" \\
11 09 27.20 & +55 41 20 & 3.472 & P84b & \\
11 26 34.30 & $-$01 24 36 & 3.741 & P85a & 278.71" \\
11 26 17.40 & $-$01 26 32 & 3.607 & P85b & \\
12 19 33.26 & +00 32 26.40 & 2.871 & P86a & 260.71" \\
12 19 22.19 & +00 29 05.40 & 2.627 & P86b & \\
13 54 42.90 & +59 28 56 & 2.554 & P87a & 161.67" \\
13 54 38.40 & +59 31 34 & 2.992 & P87b & \\
14 19 19.50 & +57 45 13 & 3.339 & P88a & 248.30" \\
14 19 00.60 & +57 48 30 & 2.937 & P88b & \\
14 29 51.87 & +63 16 31.90 & 2.403 & P89a & 188.86" \\
14 29 33.01 & +63 14 12.40 & 2.749 & P89b & \\
14 35 00.27 & +03 54 03.50 & 2.491 & P90a & 224.84" \\
14 34 55.38 & +03 50 30.90 & 2.853 & P90b & \\
14 41 34.30 & +61 39 19 & 2.435 & P91a & 296.10" \\
14 40 52.90 & +61 38 52 & 2.898 & P91b & \\
15 59 22.70 & +52 00 27 & 3.101 & P92a & 145.47" \\
15 59 17.40 & +52 02 44 & 3.042 & P92b & \\
21 36 29.40 & +10 29 52 & 2.555 & P93a & 237.83" \\
21 36 15.40 & +10 27 54 & 2.966 & P93b & \\
08 25 50.20 & +35 48 03 & 3.203 & P94a & 259.91" \\
08 25 40.10 & +35 44 14 & 3.846 & P94b & \\
08 31 15.90 & +38 14 24 & 3.073 & {\bf P95a} & 216.00" \\
08 30 53.00 & +38 12 43 & 3.171 & {\bf P95b} & 288.08" \\
08 30 52.90 & +38 09 07 & 3.149 & {\bf P95c} & 417.03" \\
\end{tabular}
\end{table}

\addtocounter{table}{-1}

\begin{table}
\caption{Continued.}
\begin{tabular}{lllll}
\hline
 RA (J2000)    &    Dec (J2000)   &  \zem\  & Label & Separation \\
\hline
10 19 37.00 & +55 23 55 & 3.231 & P96a & 125.37" \\
10 19 22.90 & +55 24 31 & 3.720 & P96b & \\
11 08 19.15 & $-$00 58 24.00 & 4.564 & P97a & 113.03" \\
11 08 13.86 & $-$00 59 44.50 & 4.033 & P97b & \\
13 02 16.91 & $-$03 38 03.70 & 3.714 & P98a & 141.24" \\
13 02 08.17 & $-$03 37 10.50 & 3.718 & P98b & \\
13 48 08.79 & +00 37 23.20 & 3.626 & P99a & 236.72" \\
13 47 55.68 & +00 39 35.00 & 3.814 & P99b & \\
17 19 37.90 & +29 18 05 & 3.079 & P100a & 106.46" \\
17 19 32.90 & +29 19 29 & 3.294 & P100b & \\
20 53 03.70 & $-$01 04 42 & 3.115 & P101a & 137.52" \\
20 53 02.90 & $-$01 02 25 & 3.251 & P101b & \\
08 59 59.14 & +02 05 19.70 & 2.980 & P102a & 275.29" \\
08 59 56.83 & +02 09 52.80 & 2.233 & P102b & \\
13 37 57.87 & +02 18 20.90 & 3.332 & P103a & 172.33" \\
13 37 56.34 & +02 15 30.10 & 2.314 & P103b & \\
10 42 53.43 & $-$00 13 00.90 & 2.958 & P104a & 289.81" \\
10 42 43.12 & $-$00 17 06.00 & 1.969 & P104b & \\
09 45 08.00 & +50 40 57 & 3.736 & P105a & 182.57" \\
09 44 53.80 & +50 43 00 & 3.789 & P105b & \\
11 04 11.60 & +02 46 55 & 2.533 & P106a & 131.25" \\
11 04 03.00 & +02 47 20 & 2.374 & P106b & \\
13 12 13.84 & +00 00 03.00 & 2.680 & P107a & 148.43" \\
13 12 13.29 & +00 02 31.20 & 2.842 & P107b & \\
15 54 07.74 & +01 00 10.10 & 2.606 & P108a & 239.10" \\
15 53 59.96 & +00 56 41.40 & 2.630 & P108b & \\
01 06 16.06 & +00 15 24.00 & 3.043 & P109a & 243.02" \\
01 06 12.22 & +00 19 20.10 & 3.110 & P109b & \\
08 40 55.70 & +37 04 37 & 3.152 & P111a & 178.10" \\
08 40 42.20 & +37 05 52 & 2.905 & P111b & \\
13 24 11.60 & +03 20 50 & 3.670 & P112a & 154.19" \\
13 24 01.50 & +03 20 20 & 3.371 & P112b & \\
14 30 06.40 & $-$01 20 20 & 3.249 & P113a & 199.75" \\
14 29 57.10 & $-$01 17 57 & 3.111 & P113b & \\
15 00 58.70 & +61 45 06 & 2.587 & P114a & 287.78" \\
15 00 23.50 & +61 47 29 & 2.994 & P114b & \\
15 37 29.50 & +58 32 24 & 3.059 & P115a & 202.31" \\
15 37 15.70 & +58 29 33 & 2.590 & P115b & \\
01 02 51.85 & $-$27 53 03.30 & 1.768 & P116a & 81.11" \\
01 02 57.35 & $-$27 53 38.82 & 1.800 & P116b & \\
15 19 13.29 & +23 46 58.72 & 1.834 & P117a & 101.25" \\
15 19 19.40 & +23 46 02.00 & 1.903 & P117b & \\
16 25 48.00 & +26 44 32.50 & 2.490 & {\bf P118a} & 146.44" \\
16 25 48.75 & +26 46 58.60 & 2.526 & {\bf P118b} & 180.33" \\
16 25 57.66 & +26 44 43.40 & 2.605 & {\bf P118c} & 129.86" \\
21 42 25.90 & $-$44 20 17.00 & 3.230 & P119a & 61.52" \\
21 42 22.20 & $-$44 19 30.00 & 3.220 & P119b & \\
21 43 07.01 & $-$44 50 47.60 & 3.250 & P120a & 33.15" \\
21 43 04.09 & $-$44 50 36.00 & 3.060 & P120b & \\
09 45 05.93 & $-$00 46 45.00 & 2.299 & P121a & 261.94" \\
09 44 54.24 & $-$00 43 30.40 & 2.292 & P121b & \\
09 27 43.02 & +29 07 34.7 & 2.2530 & P122a & 57.60" \\
09 27 47.27 & +29 07 20.7 & 2.2920 & P122b & \\
11 51 38.05 & +02 06 10.30 & 2.258 & P123a & 260.18" \\
11 51 22.14 & +02 04 26.30 & 2.401 & P123b & \\
14 11 08 & +62 24 52 & 2.264 & P124a & 200.66" \\
14 11 30.7 & +62 22 48 & 2.305 & P124b & \\
09 46 42.43 & +33 07 54.8 & 2.477 & P125a & 194.21" \\
09 46 56.17 & +33 06 25.8 & 2.484 & P125b & \\
08 33 21.61 & +08 12 38.3 & 2.518 & P126a & 208.57" \\
08 33 26.82 & +08 15 52.0 & 2.572 & P126b & \\
10 00 52.21 & +45 00 11.0 & 2.567 & P127a & 199.33" \\
10 00 54.38 & +45 03 29.0 & 2.649 & P127b & \\
09 35 31.84 & +36 33 17.6 & 2.858 & P128a & 231.84 \\
09 35 48.51 & +36 31 21.9 & 2.977 & P128b & \\
\end{tabular}
\end{table}

\addtocounter{table}{-1}

\begin{table}
\caption{Continued.}
\begin{tabular}{lllll}
\hline
 RA (J2000)    &    Dec (J2000)   &  \zem\  & Label & Separation \\
\hline
14 22 49.19 & +42 02 46.2 & 3.071 & P129a & 106.87" \\
14 22 39.88 & +42 02 20.4 & 3.236 & P129b & \\
12 09 17.94 & +11 38 30.4 & 3.105 & P130a & 196.41" \\
12 09 10.71 & +11 35 45.2 & 3.122 & P130b & \\
08 52 32.18 & +26 35 26.2 & 3.208 & P131a & 170.86" \\
08 52 37.94 & +26 37 58.6 & 3.29 & P131b & \\
08 03 05.84 & +50 32 15.3 & 3.244 & P132a & 190.91" \\
08 03 21.24 & +50 34 17.4 & 3.245 & P132b & \\
12 38 15.04 & +44 30 26.2 & 3.254 & P133a & 232.24" \\
12 38 31.46 & +44 32 58.2 & 3.268 & P133b & \\
08 36 59.84 & +35 10 19.4 & 3.319 & P134a & 269.47" \\
08 37 00.83 & +35 05 50.2 & 3.311 & P134b & \\
14 21 49.99 & +46 59 38.6 & 3.678 & P135a & 201.85" \\
14 22 09.71 & +46 59 32.5 & 3.798 & P135b & \\
10 54 16.47 & +51 27 24.6 & 2.367 & P136a & 238.50" \\
10 54 16.51 & +51 23 26.1 & 2.341 & P136b & \\
12 13 03.26 & +12 08 39.2 & 3.384 & P137a & 137.99" \\
12 13 10.72 & +12 07 15.1 & 3.469 & P137b & \\
10 39 41.49 & +55 09 27.8 & 3.709 & P138a & 258.46" \\
10 39 49.28 & +55 13 37.5 & 3.851 & P138b & \\
08 07 44.89 & +23 48 25.7 & 3.745 & P140a & 225.90" \\
08 07 35.01 & +23 51 26.4 & 3.76 & P140b & \\
22 47 21.06 & $-$09 15 45.7 & 4.13 & P141a & 284.94" \\
22 47 40.17 & $-$09 15 11.8 & 4.167 & P141b & \\
23 38 15.45 & $-$10 19 17.2 & 2.279 & P142a & 281.75" \\
23 37 56.58 & $-$10 20 00.1 & 2.436 & P142b & \\
14 34 08.31 & +23 22 30.0 & 3.97 & P143a & 166.07" \\
14 33 56.26 & +23 22 22.8 & 4.16 & P143b & \\
10 16 05.84 & +40 40 05.8 & 2.99 & P144a & 68.24" \\
10 16 01.50 & +40 40 52.9 & 2.963 & P144b & \\
13 56 07.50 & +58 12 36.1 & 3.32  & P145a & 89.15" \\
13 55 57.54 & +58 13 18.0 & 3.371 & P145b & \\
11 52 10.43 & +45 18 25.8 & 2.282 & P146a & 113.40" \\
11 52 00.54 & +45 17 41.4 & 2.379 & P146b & \\
15 09 25.64 & +50 56 09.3 & 2.365 & P147a & 119.68" \\
15 09 32.23 & +50 57 51.5 & 2.343 & P147b & \\
16 43 30.13 & +30 55 41.8 & 2.703 & P148a & 244.49" \\
16 43 41.28 & +30 58 59.8 & 2.559 & P148b & \\
09 16 11.02 & +33 11 30.5 & 3.112 & P149a & 152.45" \\
09 16 03.40 & +33 09 31.8 & 3.153 & P149b & \\
11 06 11.11 & +13 56 00.0 & 3.908 & P150a & 101.72" \\
11 06 16.68 & +13 54 58.6 & 3.846 & P150b & \\
08 15 14.28 & +06 05 42.5 & 2.495 & P151a & 64.08" \\
08 15 18.32 & +06 06 04.3 & 2.529 & P151b & \\
16 23 23.68 & +33 12 32.6 & 2.411 & P152a & 103.69" \\
16 23 24.76 & +33 10 49.8 & 2.585 & P152b & \\
11 43 16.98 & +13 24 00.8 & 2.514 & P153a & 138.30" \\
11 43 23.44 & +13 25 42.0 & 2.515 & P153b & \\
10 40 04.02 & +32 21 50.6 & 2.609 & P154a & 191.02" \\
10 40 19.09 & +32 21 56.4 & 2.649 & P154b & \\
11 16 10.68 & +41 18 14.5 & 2.98 & P155a & 13.75" \\
11 16 11.73 & +41 18 21.5 & 2.971 & P155b & \\
\hline
\end{tabular}
\end{table}

\begin{figure}
  \includegraphics[width=84mm]{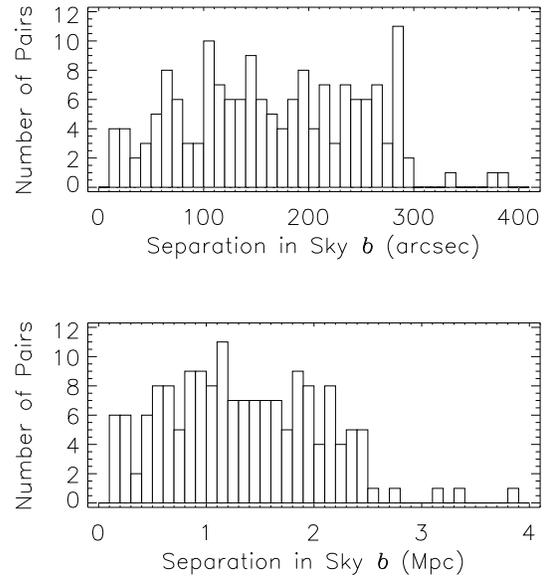}
  \caption{Separation of pairs of QSOs in arcseconds (upper, in bins of 10
  arcsec) and
proper Mpc in the plane of the sky, $b$ (lower, in bins of 0.1~Mpc).
We calculate the separation using the lower of the two \zem\ values.
In this and many other figures we move the
zero of the horizontal axis away from the vertical axis.
}
  \label{figmpcarcdist}
\end{figure}

In Fig. \ref{figmpcarcdist}
we show the distribution of the QSO pair separations in arcseconds and
in proper Mpc.  For this figure we use the lower of the two \zem\
values for each pair when we convert to Mpc since this is the highest
$z$ at which their sight lines overlap.

\begin{figure}
  \includegraphics[width=84mm]{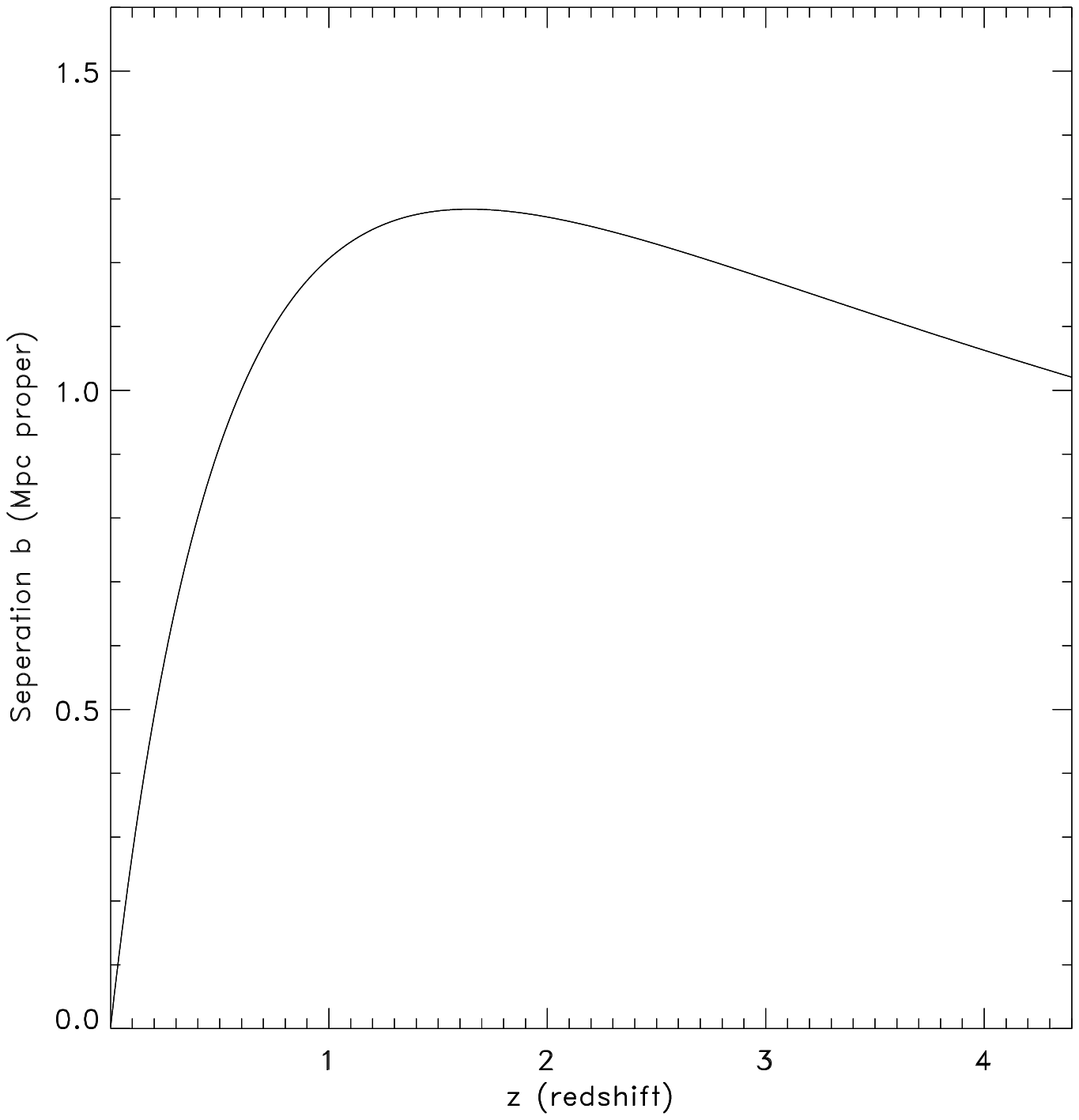}
  \caption{Proper distance between two sight lines separated by 150 arcsec
  as a function of redshift for the model we use in this paper with
  \ol = 0.73, \om = 0.27 and $H_0 = 71$ \kms Mpc$^{-1}$.
}
  \label{figtypsep}
\end{figure}

In Fig. \ref{figtypsep}
we show the proper distance between two sight lines
separated by 150 arcsec, the typical separation for our sample, as a function
of $z$. The sight lines reach a maximum separation at $z = 1.628$.
We will consider absorption systems at a wide range of \zabs\ values,
including low redshifts where the proper distances are $<< 1$ ~Mpc.

\subsection{Emission Redshifts}

The average emission redshift of the QSOs is $z = 2.470$, with a range
of 1.84 to 3.84.  We took emission redshift \zem\ values from the SDSS
when available and otherwise from the literature.  The SDSS redshifts
use effective rest frame wavelength for emission lines from
\citet{vandenberk01}, intended to give \zem\ values that approximate
the redshifts of the host galaxies of the QSOs, the so called
``systemic redshifts''. The SDSS redshifts are referenced to a
composite QSO spectrum with a zero point from the [O~III] emission
line \citep[\S 4.10.2.3]{stoughton02}. The redshift values are
obtained by the SDSS project either by cross-correlation with the
composite spectrum or using effective rest wavelengths for emission
lines from the composite.  We do not know how much these redshifts
differ from the systemic values, but we shall see in \S 5.1 that the
small dispersion in the difference between \zem\ values and \zabs\
values in partner QSOs suggests errors are $\sim 500$~\kms\ for some
QSOs. Although the errors may be many times larger for some QSOs,
depending on the emission lines used to obtain the \zem\ values
\citep{gaskell82,tytler92,vandenberk01,richards02a}, the number of
absorbers at negative velocities implies that \zem\ errors are
typically $<1000$~\kms .

\begin{figure}
  \includegraphics[width=84mm]{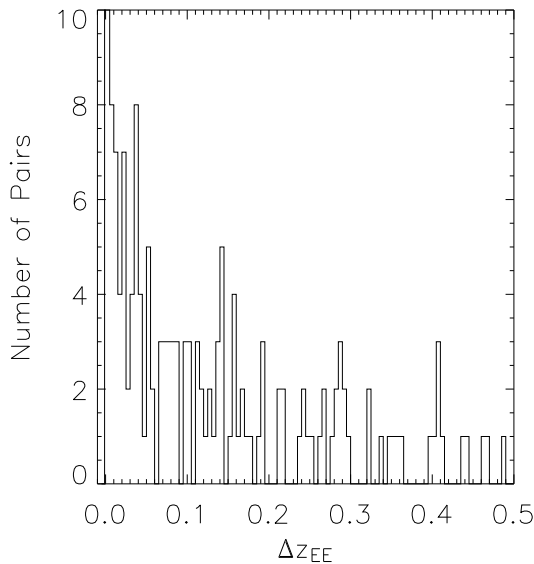}
  \caption{Distribution of the \dzee\ values, the difference between the
\zem\ values of the QSOs in each pair, in bins of 0.005.
}
  \label{figsepzem}
\end{figure}

In Fig. \ref{figsepzem} we show the distribution of the differences in the
\zem\ values, where we define for each pair of QSOs
\begin{equation}
\dzee\ = |z_{\rm em1}-z_{\rm em2}|.
\end{equation}
There is a strong tendency for QSOs in the sample to have very similar
\zem\ values, because most of the pairs with the smallest angular
separation are physical pairings \citep{hennawi06b,shen07a}, and
because we favoured pairs with the most similar \zem\ values when we
obtained spectra.  We see a wide range of \dzee\ values with a
pronounced excess at $< 0.04$ (4000~\kms ) approximately.  This excess
has a major effect on the correlation of \zabs\ values, because there
is also also a pronounced excess of absorbers with \zabs $\simeq $
\zem\ in the individual sight lines.

\section{Observations}

We use spectra that we obtained with LRIS on the Keck I telescope, the Kast
spectrograph on the Lick 3-m Shane telescope and from CTIO and KPNO.
We also use spectra from the SDSS data release 5 (DR5).

We obtained spectra with LRIS \citep{mccarthy98, oke95} from 2001 to
2004 September.  While we attempted to obtain spectra of a given pair
on a given night, there are many cases where the partner spectrum was
obtained on different nights or even in different years. The paired
spectra then often have different resolution and wavelength coverage
because LRIS changed.

In Table \ref{tabsetup}
we summarise the different grisms and gratings that we used with LRIS.
LRIS is a double spectrograph with independent blue and red dispersers
and cameras.  We used various grisms and grating to best match the
targets to changes that were made in the instrument.  We took blue
spectra using either the 400/3400 grism or the 1200/3400 grism.  The
high resolution 1200/3400 grism has the advantage of showing weak
absorption lines, but it sometimes had lower efficiency than the low
resolution grism and there is an unavoidable gap at wavelengths 3880
-- 4600 or 3770 -- 4600~\AA\ between the blue and red side spectrum.
We took red spectra using either the 600/5000 grating or the 900/5500
grating. For the red spectra LRIS-R was used with the 2048 $\times$
2048 Tektronix CCD with 24 $\mu $m pixels.  For blue spectra prior to
June 2002 LRIS-B was used with a SITe 2048 $\times$ 2048 engineering
grade CCD similar to the red CCD. After the mid 2002 LRIS upgrade,
blue spectra were obtained with a mosaic of two Marconi 2048 $\times$
4096 CCDs with 15 $\mu $m pixels.

\begin{table}
\caption{LRIS CCDs, Grisms and Gratings}
\label{tabsetup}
\begin{tabular}{lll}
\hline
 & \multicolumn{2}{c}{CCD Used} \\
\hline
 A  & \multicolumn{2}{l}{Marconi mosaic} \\
 B  & \multicolumn{2}{l}{old blue SITe} \\
 C  & \multicolumn{2}{l}{red} \\
\hline
    & Blue Grism\footnotemark[1] & Red Grating\footnotemark[1] \\
\hline
 1  & 1200/3400 & \\
 2  & 400/3400 & \\
 3  &  & 600/5000  \\
 4  &  & 900/5500  \\
 5  & 600/4000 & \\
\hline
\end{tabular}
 \\
1 Grooves per mm/Blaze Wavelength (\AA ) \\
\end{table}

In Table \ref{tabspec} we list the wavelength ranges covered by the various
CCD, grating and grism choices and the FWHM resolution for a given slit.
Setups A, B and C are for LRIS on Keck, and D is for RCSP on the CTIO and KPNO
4-m telescopes.
We see significant differences between the resolution of
individual spectra even when we use the same gratings and slit, because of
differences in the focus, seeing and guiding.

\begin{table}
\caption{Spectra Characteristics. }
\label{tabspec}
\begin{tabular}{lcccllr}
\hline
  Setup &  Wavelength  & Pixel &  Slit & \multicolumn{2}{c}{FWHM}& $N$\footnotemark[1]  \\
&  Range (\AA )&  (\AA ) & ('') & (\AA )  &(\kms ) \\
\hline
 A1  & 3200$-$3880  & 0.24  & 0.7 & 0.98\footnotemark[2]  & 83\footnotemark[3] & 26 \\
 A1  & 3200$-$3880  & 0.24  & 1.0 & & 108\footnotemark[4] & 27\\
 A1  & 3200$-$3880  & 0.24  & 1.5 & & 162\footnotemark[4] & 4\\
 B1  & 3000$-$3770  & 0.41  & 0.7 & & $105 \pm 10 ^{5,6}$ & 13\\
 A2  & 3180$-$5800  & 1.07  & 0.7 & 4.21\footnotemark[2]  & 281\footnotemark[3] &7 \\
 B2  & 3000$-$5300  & 1.46  & 0.7 & & $260 \pm 30^{5,7}$ & 21\\
 C3  & 5500$-$8070  & 1.28  & 0.7 & 3.59\footnotemark[8]& 166\footnotemark[3] &14\\
 C4  & 4600$-$6330  & 0.85  & 0.7 & 2.39\footnotemark[8]& 131\footnotemark[3]&53\\
 C4  & 4600$-$6330  & 0.85  & 1.0 & 3.10\footnotemark[8] & 170\footnotemark[3]&29\\
 C4  & 4600$-$6330  & 0.85  & 1.5 &  & 255\footnotemark[9]&4\\
  D1$^{10}$  & 3180$-$6240  & 1.01 & -- & 3.2 & 199\footnotemark[3]&2 \\
  D2$^{11}$  & 1600$-$4720  & 0.52 & --  & 3.2 & 234\footnotemark[3]&2\\
  D3$^{12}$  & 3150$-$4720  & 0.76 & --  & 1.8 & $135 \pm 10$\footnotemark[5]&3\\
  D4$^{10}$  & 3600$-$6700  & 1.01 & --  & 3.2 & 163\footnotemark[3]&4 \\
 SDSS & 3800$-$9200  & 0.90 &      & & 165.5 &164\\
  LICK & 3175$-$5880 & 1.15 & 2.3 & & 250\footnotemark[5] &45 \\
\hline
\end{tabular}

$^1$ $N$ is the number of spectra with that setup.\\
$^2$ Values from Chuck Steidel 2002 July on LRIS web site: \\
\verb http://www.astro.caltech.edu/~ccs/lrisb/new_numbers.txt \\
$^3$ Converted from the FWHM in \AA\ at the central wavelength of the spectrum,
or the central wavelength to the red of \lya\ emission when only one or two pairs of QSOs,
or using 6500~\AA\ for C3.\\
$^4$ Guessed value. For the 1.0" slit we multiply the value for A1 by 1.3, the factor
increase reported by \citet[\S 2]{tonry98} for the  600 g/mm grating. For the 1.5"
slit we multiply the 1.0" value by 1.5.\\
$^5$ Measured by comparing the \lyaf\ in Keck HIRES spectra
of bright QSOs, following \cite{suzuki03b}. \\
$^6$ Constant from 3200 -- 3600~\AA . \\
$^7$ Decreasing from 300 \kms\ at 4350~\AA\ to 235~\kms\ at 5320~\AA . \\
$^8$ From \citet[\S 2]{tonry98} for the 600 g/mm grating and times 2/3 for the
 900 g/mm.\\
$^9$ Guessed value. We multiply the value for the 1.0" slit by 1.5.\\
$^{10}$  CTIO 4m RCSP KPGL-1 632 g/mm grating blazed at 4200~\AA .\\
$^{11}$ As D1 and including an HST FOS G270H spectrum.\\
$^{12}$ KPNO 4m RCSP BL 420 grating with 600 g/mm and blazed at 8000~\AA .\\
\end{table}

In Table \ref{tabobs} we give the observation date,
the width of the slit we used, the setup, exposure times, and the
SNR per pixel for each spectra. Unless otherwise noted, we measured the SNR
at 4200 \AA\ for spectra observed with the A2 or B2 setup, and at 5200 \AA\
for spectra observed with the A1 or B1 setup.
We give the SNR for only a sub-sample of all spectra.
The majority of the QSOs were observed using the 0.7 arcsecond slit but
we used a  1.0 arcsecond slit for twenty  nine QSOs.
Exposure times for the objects ran from 460 seconds to 8000 seconds. Spectra
were extracted using the standard IRAF extraction packages and our own software that
is designed to give accurate flux calibration with the optimal SNR.
In Table \ref{tabobs} we only list a pair if we obtained spectra of one
or both QSOs. We do not list the pairs for which we used SDSS and no
other spectra.

The spectra from Lick observatory used the Kast Double Spectrograph
on the 3-m Shane telescope. We typically used the 830 groves/mm grism blazed at
3460~\AA\ in the blue camera and a 1200 grove/mm grating blazed at 5000~\AA\
in the red camera. A dichroic with a 50\% transmission near
4600~\AA\ was used, and the wavelengths are setup to cover  3175 -- 5880~\AA\
with no gaps. The dispersion are 1.13~\AA\ per pixel
in the blue and 1.17~\AA\ in the red. The typical slit gives approximately
2.5 pixels per FWHM depending on the wavelength and focus setting.
We show similar spectra   in \citet{tytler04a}.

\begin{table}
\caption{Keck LRIS Observations}
\label{tabobs}
\begin{tabular}{llllccc}
\hline
 &  &  &  & \multicolumn{2}{c}{Exposure Time (s)} & \\
  Label  &  Date &  Slit  &  Setup & Blue & Red & S/N \\
\hline
  P1a &  11/2001 & 0.7  & B2,C4  & 1200 & 1200 & 35 \\
  P1b &  11/2001 & 0.7  & B2,C4  & 900 & 900 & 54 \\
  P2a &  11/2001 & 0.7  & B2,C4  & 2700 & 2700 & 46 \\
  P2b &  11/2001 & 0.7  & B2,C4  & 3600 & 3600 & 55 \\
  P3a &  10/2001 & 0.7  & B2,C4  & 2700 & 2700 & 17 \\
  P3a &  08/2002 & 0.7  & A1,C4  & 2400 & 2400 & \\
  P3a &  09/2004 & 1.0  & A1,C4  & 2200 & 2200 & \\
  P3b &  10/2001 & 0.7  & B2,C4  & 3600 & 3600 & 45 \\
  P3b &  08/2002 & 0.7  & A1,C4  & 462  & 462 & \\
  P3b &  09/2004 & 1.0  & A1,C4  & 3600 & 3600 & \\
  P4a &  10/2001 & 0.7  & B2,C4  & 900 & 900 & 6 \\
  P4a &  11/2001 & 0.7  & B2,C4  & 900 & 900 & \\
  P4a &  09/2004 & 1.0  & A1,C4  & 1800 & 1800 & \\
  P4a &  09/2004 & 1.0  & A1,C4  & 2400 & 2400 & \\
  P4b &  10/2001 & 0.7  & B2,C4  & 600 & 600 & 42 \\
  P4b &  08/2002 & 0.7  & A1,C4  & 300 & 300 & \\
  P5a &  10/2001 & 0.7  & B2,C4  & 424 & 423 & 32 \\
  P5a &  10/2001 & 0.7  & B2,C4  & 900 & 900 & 32 \\
  P5a &  08/2002 & 0.7  & A1,C4  & 2400 & 2400 & \\
  P5b &  10/2001 & 0.7  & B2,C4  & 1800 & 1800 & 51 \\
  P6a &  08/2001 & 0.7  & B2,C3  & 1200 & 1200 & 10 \\
  P6b &  08/2001 & 0.7  & B2,C3  & 1200 & 1200 & 16 \\
  P7a &  03/2003 & 0.7  & A1,C4  & 5400 & 5250 & 23 \\
  P7b &  03/2003 & 0.7  & A1,C4  & 5600 & 5400 & 26 \\
  {\bf P8a} &  03/2003 & 0.7  & A1,C4  & 3600 & 3450 & 33 \\
  {\bf P8a} &  03/2003 & 1.0  & A1,C4  & 2300 & 2300 &  \\
  {\bf P8b} &  04/2001 & 0.7  & B1,C4  & 1800 & 1800 & 17 \\
  {\bf P8b} &  03/2003 & 1.0  & A1,C4  & 3600 & 3600 &  \\
  {\bf P8c} &  04/2001 & 0.7  & B1,C4  & 1800 & 1800 & 12 \\
  P9a &  03/2003 & 1.0  & A1,C4  & 1000 & 1000 & 33 \\
  P9b &  03/2003 & 0.7  & A1,C4  & 3600 & 3420 & 21 \\
  P10a & 04/2001 & 0.7  & B1,C4  & 3000 & 3000 & 10 \\
  P10b & 04/2001 & 0.7  & B1,C4  & 3000 & 3000 & 7 \\
  P11a & 03/2003 & 0.7  & A1,C4  & 5400 & 5250 & 35 \\
  P11b & 03/2003 & 0.7  & A1,C4  & 5400 & 5250 & 16 \\
  P11b & 04/2004 & 0.7  & A1,C4  & 4600 & 4600 &  \\
  P12a & 03/2003 & 0.7  & A1,C4  & 3300 & 3150 & 30 \\
  P12b & 03/2003 & 0.7  & A1,C4  & 3300 & 3150 & 20 \\
  P13a & 03/2003 & 0.7  & A1,C4  & 2700 & 2550 & 41 \\
  P13b & 03/2003 & 0.7  & A1,C4  & 2700 & 2550 & 36 \\
  P14a & 03/2003 & 0.7  & A1,C4  & 2900 & 2750 & 13 \\
  P14b & 03/2003 & 0.7  & A1,C4  & 1500 & 1350 & 13 \\
  P15a & 07/2002 & 0.7  & A1,C4  & 7262 & 7262 & 22 \\
  P15b & 08/2002 & 0.7  & A1, $-$& 3200 & $-$  & 5\footnotemark[1] \\
  P16a & 07/2002 & 0.7  & A2,C3  & 1800 & 1800 & 47 \\
  P16b & 07/2002 & 0.7  & A2,C3  & 1800 & 1800 & 40 \\
  P17a & 04/2001 & 0.7  & B1,C4  & 2400 & 2400 & 24 \\
  P17b & 04/2001 & 0.7  & B1,C4  & 2700 & 2700 & 39 \\
  P18a & 07/2002 & 0.7  & A2,C3  & 2700 & 2700 & 21\footnotemark[2] \\
  P18b & 08/2001 & 0.7  & $-$ ,C3& $-$ & 3600 & 19\footnotemark[2] \\
  P19a & 07/2002 & 0.7  & A2,C3  & 2300 & 2300 & 34 \\
  P19b & 07/2002 & 0.7  & A2,C3  & 2100 & 2100 & 35 \\
  P20a & 07/2002 & 0.7  & A2,C3  & 900 & 900 & 8 \\
  P20a & 08/2002 & 0.7  & A1,C4  & 3421 & 3421 & \\
  P20b & 07/2002 & 0.7  & A2,C3  & 1500 & 1500 & 31 \\
  P21a & 07/2002 & 0.7  & A1,C4  & 5400 & 5400 & 33 \\
  P21b & 07/2002 & 0.7  & A1,C4  & 3600 & 3600 & 40 \\
  {\bf P22a} & 08/2001 & 0.7  & B2,C3  & 1200 & 1200 & 29 \\
  {\bf P22b} & 08/2001 & 0.7  & B2,C3  & 3600 & 3600 & 42 \\
  {\bf P22c} & 08/2001 & 0.7  & B2,C3  & 1901 & 1046 & 73 \\
  P23a & 08/2001 & 0.7  & B2,C3  & 900 & 900 & 25 \\
  P23b & 10/2001 & 0.7  & B2,C4  & 2700 & 2700 & 6 \\
  P24a & 11/2001 & 0.7  & B2,C4  & 3300 & 3300 & 23 \\
  P24b & 11/2001 & 0.7  & B2,C4  & 1500 & 1500 & 22 \\
\end{tabular}
\end{table}

\addtocounter{table}{-1}

\begin{table}
\caption{Continued.}
\begin{tabular}{llllccc}
\hline
 &  &  &  & \multicolumn{2}{c}{Exposure Time (s)} & \\
  Label  &  Date &  Slit  &  Setup & Blue & Red & S/N \\
\hline
  P25a & 04/2004 & 0.7  & B1,C4  & 3600 & 3600 &  \\
  P25b & 04/2004 & 0.7  & B1,C4  & 600 & 600 &  \\
  P26a & 04/2004 & 0.7  & B1,C4  & 2160 & 2160 &  \\
  P26b & 04/2004 & 0.7  & B1,C4  & 4100 & 4100 &  \\
  P27a & 04/2004 & 0.7  & B1,C4  & 3914 & 3914 &  \\
  P27b & 04/2004 & 0.7  & B1,C4  & 4900 & 4900 &  \\
  P28a & 08/2003 & 3.3  & LICK  & 3601 & 3601 &  \\
  P28b & 04/2004 & 0.7  & B1,C4  & 2800 & 2800 &  \\
  P29a & 07/2002 & 0.7  & A1,C4  & 4600 & 4600 &  \\
  P29b & 07/2002 & 0.7  & A1,C4  & 2880 & 2880 &  \\
  P30a & 09/2004 & 1.0  & A1,C4  & 2700 & 2700 &  \\
  P30b & 09/2004 & 1.0  & A1,C4  & 4500 & 4500 &  \\
  P31a & 09/2004 & 1.0  & A1,C4  & 1800 & 1800 &  \\
  P31b & 09/2004 & 1.0  & A1,C4  & 1800 & 1800 &  \\
  P32a & 09/2004 & 1.5  & A1,C4  & 2400 & 2400 &  \\
  P32b & 09/2004 & 1.5  & A1,C4  & 2200 & 2200 &  \\
  P34a & 09/2004 & 1.0  & A1,C4  & 2300 & 2300 &  \\
  P34b & 09/2004 & 1.0  & A1,C4  & 3200 & 3200 &  \\
  P35a & 09/2004 & 1.5  & A1,C4  & 2400 & 2400 &  \\
  P35b & 09/2004 & 1.5  & A1,C4  & 1200 & 1200 &  \\
  P36a & 09/2004 & 1.0  & A1,C4  & 1800 & 1800 &  \\
  P36b & 09/2004 & 1.0  & A1,C4  & 5400 & 5400 &  \\
  P37a & 09/2003 & 1.0  & A5,C4  & 2200 & 2200 &  \\
  P37b & 09/2003 & 1.0  & A5,C4  & 4000 & 4000 &  \\
  P38a & 09/2003 & 1.0  & A1,C4  & 4800 & 4800 &  \\
  P38b & 09/2003 & 1.0  & A1,C4  & 2500 & 2500 &  \\
  P39a & 09/2003 & 1.0  & A1,C4  & 2500 & 2500 &  \\
  P39b & 09/2003 & 1.0  & A1,C4  & 3050 & 3050 &  \\
  P40a & 09/2003 & 1.0  & A1,C4  & 5100 & 5100 &  \\
  P40b & 09/2003 & 1.0  & A1,C4  & 3700 & 3700 & \\
  P41a & 09/2003 & 2.3  & LICK  & 4805 & 4805 &  \\
  P41b & 09/2003 & 2.3  & LICK  & 4800 & 4800 &  \\
  P42a & 09/2003 & 1.0  & A1,C4  & 3000 & 3000 &  \\
  P42b & 09/2003 & 1.0  & A1,C4  & 5400 & 5400 &  \\
  P43a & 09/2003 & 1.0  & A1,C4  & 3600 & 3600 &  \\
  P43b & 09/2003 & 1.0  & A1,C4  & 2300 & 2300 &  \\
  P44a & 09/2003 & 1.0  & A1,C4  & 2000 & 2000 &  \\
  P44b & 09/2003 & 1.0  & A1,C4  & 1500 & 1500 &  \\
  P45a & 08/2003 & 3.3  & LICK  & 4802 & 4802 &  \\
  P45b & 08/2003 & 3.3  & LICK  & 6303 & 6303 &  \\
  P46a & 12/2004 & 2.3  & LICK  & 4802 & 4802 &  \\
  {\bf P49a} & 03/2005 & 2.3  & LICK   & 5421 & 5421 &  \\
  P50b & 03/2005 & 2.3  & LICK  & 5403 & 5403 &  \\
  P51a & 03/2005 & 2.3  & LICK   & 1451 & 1451 &  \\
  P51b & 03/2005 & 2.3  & LICK   & 4515 & 4515 &  \\
  P52a & 12/2004 & 2.3  & LICK   & 9005 & 9005 &  \\
  P52b & 03/2005 & 2.3  & LICK   & 5402 & 5402 &  \\
  P53a & 03/2005 & 2.3  & LICK   & 5407 & 5407 &  \\
  P54a & 03/2005 & 2.3  & LICK   & 3622 & 3622 &  \\
  {\bf P55b} & 03/2005 & 2.3  & LICK   & 5406 & 5406 &  \\
  {\bf P56c} & 03/2005 & 2.3  & LICK   & 5405 & 5405 &  \\
  P57a & 03/2005 & 2.3  & LICK   & 5404 & 5404 &  \\
  {\bf P58b} & 03/2005 & 2.3  & LICK   & 5413 & 5413 &  \\
  P60a & 03/2005 & 2.3  & LICK   & 4513 & 4513 &  \\
  P61a & 03/2005 & 2.3  & LICK   & 5403 & 5403 &  \\
  P61b & 03/2005 & 2.3  & LICK   & 5402 & 5402 &  \\
  P62a & 03/2005 & 2.3  & LICK   & 5400 & 5400 &  \\
  {\bf P63b} & 03/2005 & 2.3  & LICK  & 5410 & 5410 &  \\
  P64b & 03/2005 & 2.3  & LICK   & 5408 & 5408 &  \\
  P65b & 03/2005 & 2.3  & LICK  & 4524 & 4524 &  \\
  P66b & 08/2003 & 3.3  & LICK  & 4000 & 4000 &  \\
  P67a & 08/2003 & 3.3  & LICK  & 3602 & 3602 &  \\
  P69a & 09/1998 & 2.0  & LICK  & 5400 & 5400 &  \\
  P70b & 09/2003 & 2.0  & LICK  & 7200 & 7200 &  \\
\end{tabular}
\end{table}

\addtocounter{table}{-1}

\begin{table}
\caption{Continued.}
\begin{tabular}{llllccc}
\hline
 &  &  &  & \multicolumn{2}{c}{Exposure Time (s)} & \\
  Label  &  Date &  Slit  &  Setup & Blue & Red & S/N \\
\hline
  P72b & 03/2005 & 2.0  & LICK  & 5400 & 5400 &  \\
  P73b & 04/2002 &  & LICK  & 3600 & 3600 &  \\
  P74a & 06/2006 & 2.0  & LICK  & 5400 & 5400 &  \\
  P74b & 06/2006 & 2.0  & LICK  & 3600 & 3600 &  \\
  P75a & 06/2006 & 2.0  & LICK  & 3600 & 3600 &  \\
  P75b & 06/2006 & 2.0  & LICK  & 4500 & 4500 &  \\
  P76b & 06/2006 & 2.0  & LICK  & 2700 & 2700 &  \\
  P77a & 04/2004 & 0.7  & A1,C4 & 1000 & 1000 &  \\
  P77a & 06/2006 & 2.0  & LICK  & 6000 & 6000 &  \\
  P77b & 06/2006 & 2.0  & LICK  & 6000 & 6000 &  \\
  P78a & 06/2006 & 2.0  & LICK  & 3600 & 3600 &  \\
  P79a & 06/2006 & 2.0  & LICK  & 6000 & 6000 &  \\
  P79b & 06/2006 & 2.0  & LICK  & 6000 & 6000 &  \\
  P80a & 06/2006 & 2.0  & LICK  & 2800 & 2800 &  \\
  P80b & 04/2005 & 2.0  & LICK  & 2900 & 2900 &  \\
  P81a & 06/2006 & 2.0  & LICK  & 6000 & 6000 &  \\
  P81b & 06/2006 & 2.0  & LICK  & 4600 & 4600 &  \\
  P82a & 04/2004 & 0.7  & A1,C4  & 3602 & 3602 &  \\
  P116a & 10/1998 &  & D1 &  &  & 4 \\
  P116b & 10/1998 &  & D1 & & & 9 \\
  P117a & 12/1994$^3$ &  & D2 & & & 18 \\
  P117b & 12/1994$^3$ &  & D2 & & & 17 \\
  P118a & 06/1995 &  & D3 & & & 17 \\
  P118b & 06/1995 &  & D3 & & & 13 \\
  P118c & 06/1995 &  & D3 & & & 21 \\
  P119a & 10/1998$^4$ &  & D4 & & & 8 \\
  P119b & 10/1998$^4$ &  & D4 & & & 26 \\
  P120a & 10/1998$^4$ &  & D4 & & & 19 \\
  P120b & 10/1998$^4$ &  & D4 & & & 18 \\
\hline
\end{tabular}
$^1$ S/N measured at 3800 \AA . \\
$^2$ S/N measured at 6500 \AA . \\
$^3$ Also 1995 Jan 25. \\
$^4$ Also 1999 June 16 -- 17. \\
\end{table}

\section{Absorption Systems}
\label{abssys}

We take care to describe the types of absorption systems that we see
in the spectra, since the spectra have various resolutions, wavelength
coverage and SNR, all of which have a major effect on whether we
detect an absorption system at a given \zabs\ value. We will describe
the procedure, what we found and what we could have found.

We began by looking for absorption to the red of the \lya\ emission
where we expect only metal ions. We identify many systems from the
doublet lines and we then searched for other ions and \lya\ absorption
at the same redshift.  Absorption systems with a single doublet will
usually be reliable.  The least reliable are perhaps those with N~V
alone, since they tend to lie on the red side of the peak of the \lya\
emission where the flux changes rapidly and the continuum level is
least reliable. Most will be real, and we could but we did not conduct
a systematic search for Si~III 1206 to attempt to confirm them.

We fit the absorption lines with Voigt profiles convolved to the
spectral resolution.  At the resolution of these spectra some lines
are approximately a single unresolved Gaussian, but many others show
velocity structure.  If the velocity structure is clearly resolved, we
list separate \zabs\ values for each component. When components are
not well resolved, which in practice means velocity separations of $<$
FWHM of the spectrum, we select a single redshift for all the lines.
We do not average the velocities of the components, but rather we seek
to identify a single component, typically that with the largest column
density.  If all the lines are single components, we chose the \zabs\
of the line that is the best defined and most like a single Voigt
profile.  If we see multiple components we take the \zabs\ from the
strongest component, and when we see many ions each with components we
choose the ion with the least components and the location of the
highest optical depth.

In Table \ref{tababs}
we list the 691 metal line redshift systems that we found, including
34 that we will classify as BAL.  We give their \zabs\ values, the
ions we saw and the rest frame equivalent widths \wrest\ values for
the stronger lines in the doublets; $\lambda 1548.19$ for C~IV,
$\lambda 1238.82$ for N~V, and $\lambda 2796.35$ for Mg~II.  We list
the W$_{\rm rest}$ for C~IV if N~V is also seen.  We give \wrest\ for
587 non-BAL systems, less than the 657 total number of non-BAL
systems, since we did not measure any lines for some systems because
of low SNR or lack of C~IV, N~V and Mg~II. We also list under the
heading \wpm\ the approximate minimum \wrest\ value that we could have
seen in the partner spectrum at that redshift and for the same ion,
where a value of $-1$ means that that ion could not be seen and hence
there is a significant chance that the system would not be detected in
any ions.  The list is in order of $z$ for each pair, including the
\zem\ values, to make it easier to see coincident redshifts.

\begin{figure}
\includegraphics[width=84mm]{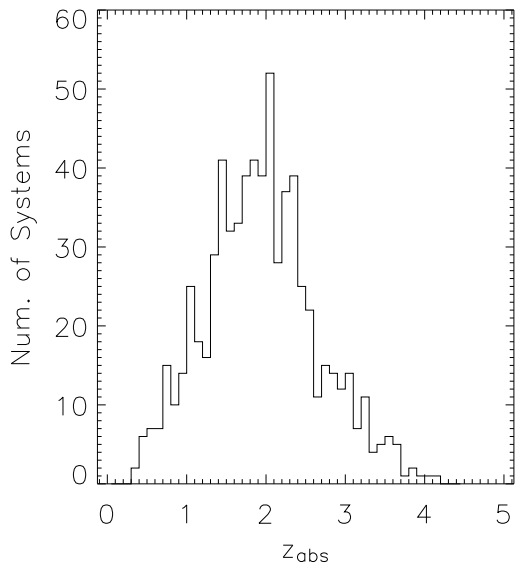}
   \caption{Distribution of all \zabs\ values in bins of size 0.01.}
   \label{figzabshist}
\end{figure}

In Fig. \ref{figzabshist}
we show the distribution of the \zabs\ values, which is very broad
from 0.2 to 4.0 with a mode of 2.0.

\begin{figure}
  \includegraphics[width=84mm]{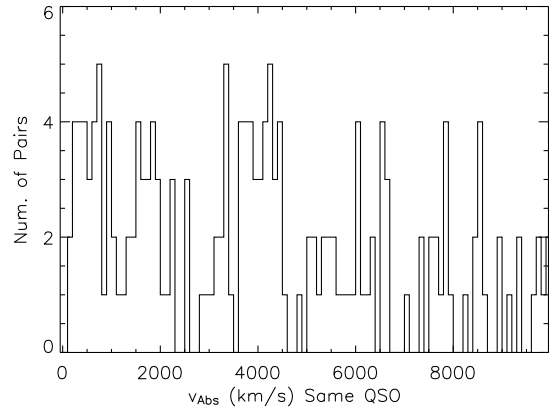}
  \caption{Histograms that show the distribution of redshift differences of
  absorption systems in individual QSOs. We include associated absorbers and use
  bins of 100~\kms .
}
  \label{figvaa}
\end{figure}

The effective mean FWHM of our sample is near 170~\kms , although the
spectra range from 83 -- 281~\kms . The mean FWHM value from Table
\ref{tabspec} weighted by the number of QSOs with spectra of each
spectral resolution of 173~\kms .  In Table \ref{tabspec} the $N$
parameter gives the number of spectra with each resolution. We
multiply these numbers by two for all setups where one rather than two
setups are quoted for that QSO (D1, D2, D3, D4, SDSS and Lick).  This
mean is larger than the effective FWHM value because we expect to see
more absorption systems in spectra with smaller FWHM values.  In
Fig. \ref{figvaa} we show the distribution of the separations of the
absorption systems in individual lines of sight. We see only two pairs
of systems with separations of $< 200$~\kms , consistent with an
effective resolution of near 170~\kms .

In Table \ref{tababs} we mark the 34 systems that we consider to be
broad absorption line BAL, because they show strong wide C~IV lines.
We arbitrarily choose to call BAL all systems with a total rest frame
equivalent width for 1548 and 1550 of C~IV $W_{rest} (C~IV) > 5$~\AA .
We find that 32 of the QSOs contain one or more BAL systems.  We will
present separate analysis for the BAL and other systems because they
are located near to their QSOs and not at the distance suggested by
their \zabs .  We exclude the BAL systems from our main analysis.

\begin{table}
\caption{The 619 Metal Lines Absorption Systems in the spectra of the 310 QSOs.
Definitions for the symbols given under  Notes are in \S 5.2 for EA and EAV,
\S 5.3 for AA and \S 9 for AAV and AAA.
An ``EA'' for two redshifts means a coincidence between the emission
redshift of one QSO and an absorber redshift in the partner QSO.
An ``AA''  for two redshifts means a coincidence
between two absorber redshifts, one in each QSO of a pair. We give 5 decimal
places for the mean redshifts of absorption systems in the AA AAV and AAA
coincidences, and 4 decimals for most other absorbers.
Errors on redshifts are given in Table \ref{velerr}
discussed in \S 9 below. We comment on the AA coincident systems in the appendix.
\wrest\ is the rest frame equivalent width for one of the stronger doublet lines
(1548, 1238 or 2796) and \wpm\ is the minimum \wrest\ that we could have seen for
the same line in the spectrum of the partner QSO at the same $z$. Here we give
only the first few lines of the table which will be distributed electronically.
}
\label{tababs}
\begin{tabular}{llllcc}
\hline
 QSO & $z$ & Ions or \zem\ & Notes & \wrest\ & \wpm\ \\
\hline
 P1 a & 0.9397 & Mg~II,Fe~II & & 2.29 & 0.08 \\
 P1 b & 1.0091 & Mg~II,Fe~II  & & 0.68 & 0.1 \\
 P1 a & 1.1355 & Mg~II,Fe~II   & & 1.59 & 0.10 \\
 P1 b & 1.4267 & C~IV,Mg~II    & & 0.68 & 0.15 \\
 P1 a & 1.541  & Mg~II         & & 0.31 & 0.20 \\
 P1 b & 1.7244 & H~I,C~II,C~IV, & & 0.86 & 0.15 \\
      &        & Si~II,Si~III,  & &  \\
      &        & Si~IV,Al~II,  & &  \\
      &        & Al~III,Mg~II,Fe~II & &  \\
 P1 b & 1.7584 & C~IV & & 0.39 & 0.15 \\
 P1 b & 1.85837 & H~I,C~IV,Si~IV & AAA21 & 0.62 & 0.18 \\
 P1 b & 1.866 & \zem\ & EA1 &  \\
 P1 a & 1.86730 & H~I,N~V\footnotemark[2],C~IV & EA1 & 0.38 & 0.10 \\
      &&& AAA21 &\\
 P1 a & 1.878 & \zem\ & &  \\
       &      &    &  &  \\
 P2 a & 2.1142 & H~I,C~IV,Si~IV & & 0.63 & 0.10 \\
 P2 b & 2.125 & \zem\ & &  \\
 P2 a & 2.2193 & H~I,C~IV,Si~IV & & 1.05 & 0.10 \\
 P2 a & 2.250 & \zem\ & &  \\
       &      &    &  &  \\
 P3 a & 0.96767 & Mg~II,Fe~II & AA1 & 1.29 & 0.10  \\
 P3 b & 0.96770 & Mg~I,Mg~II,Fe~II & AA1 & 1.33 & 0.10 \\
 P3 b & 1.0322 & Mg~II & & 0.26 & 0.10 \\
 P3 b & 1.5483 & C~IV\footnotemark[2] & & 0.34 & 0.30 \\
 P3 b & 1.7854 & H~I,C~IV,Si~III,  & & 0.50 & 0.25 \\
      &        & Si~IV & &  \\
 P3 a & 1.954 & N~V,C~IV,Si~IV & BAL & 11.3 & 0.15 \\
 P3 a & 1.970 & \zem\ & & \\
 P3 b & 1.976 & \zem\ & &  \\
\hline
\end{tabular}
 \\
\footnotemark[1] \wrest\ value for C~IV 1550 or N~V 1242.\\
\footnotemark[2] A significantly blended absorption line.\\
 Full version of this 11 page table is in the electronic version of the paper.
\end{table}

In Table \ref{tabocc}
we list the number of times that we see each ion, $N_{tot}$,
and the fraction of the absorption systems $f_{tot}$ that show each ion.
These distributions resemble those seen in other samples, such as
Table VI of \citet{barthel90}. However, we see fewer ions per system, and
hence fewer instances of most of the ions than \citet{milutinovic07a}
saw in high resolution HST spectra of systems at \zabs\ $ \simeq 1$.

In addition to the BAL systems, we expect that other systems, especially
those with \zabs\ $\sim $ \zem\ will be intrinsic to the QSOs and not
at the distances from the QSOs implied by their \zabs\ values
\citep{misawa07b,ganguly07a}. We make the following definition to help isolate
such systems.
\begin{itemize}
\item {\bf Associated absorbers} are at velocities  $< 3000$~km/s in the frame
of the \zem\ value of their QSOs.
\end{itemize}
We will explicitly state when we include or
exclude associated systems from the samples that we analyse.

\begin{table}
\caption{Ions seen in absorption systems. We have not conducted a thorough
search, and we only occasionally looked for ions such as Si~III, C~III
with lines at rest wavelengths $< 1216$~\AA .
}
\label{tabocc}
\begin{tabular}{lrcrrcrc}
\hline
 Ion  & N$_{tot}$ & f$_{tot}$ & N$_{\rm BAL}$ & N$_{AA}$ & f$_{AA}$
 & N$_{EA}$ &f$_{EA}$ \\
\hline
 C~IV &   416 & 0.602 &  32 & 21 & 0.656 & 16 & 0.89\\
 H~I  &   256 & 0.370 & 27 & 5 & 0.156   & 14 & 0.78\\
 Mg~II &  192 & 0.278 & 0 & 12 & 0.375   & 0 & 0.00\\
 Si~IV &   117 & 0.169 &  11 & 3 & 0.094 & 4 & 0.22\\
 Si~III &  114 & 0.165 &  0 & 2 & 0.063  & 2 & 0.11\\
 N~V  &   106 & 0.153 &  26 & 1 & 0.031  & 4 & 0.22\\
 Fe~II &  106 & 0.153 &  0 & 7 & 0.219   & 0 & 0.00\\
 Si~II &  56 & 0.081 &  0 & 3 & 0.094    & 1 & 0.06\\
 Al~II &  45 & 0.065 &  0 & 2 & 0.063    & 2 & 0.11\\
 C~II &   44 & 0.064 &  0 & 2 & 0.063    & 1 & 0.06\\
 O~VI &   31 & 0.045 &  16 & 0 & 0.000   & 0 & 0.00\\
 Al~III &  17 & 0.025 &  0 & 1 & 0.031   & 2 & 0.11\\
 O~I &    10 & 0.014 &  0  & 0 & 0.000   & 0 & 0.00\\
 Mg~I &   6 & 0.009 &  0 & 1 & 0.031     & 0 & 0.00\\
 C~III &  2 & 0.003 &  0 & 0 & 0.000     & 0 & 0.00\\
 Ca~II &  1 & 0.001 &  0 & 0 & 0.000     & 0 & 0.00\\
\hline
\end{tabular}
\end{table}

\section{Analysis of non-BAL Metal Line Absorption Systems}
\label{sanal}

In this section we examine the distribution of the absorbers relative to
each other and relative to the emission redshifts. We look for correlations
along the individual sight lines and especially between the sight lines.
We work in both redshift and velocity along the line of sight.
We will see that many of the systems have \zabs $\sim $ \zem , and we will
establish that we see significant correlation between absorbers in the paired
sight lines and between absorbers in one sight line and the
\zem\ in the other sight line.

\begin{figure*}
  \includegraphics[width=180mm]{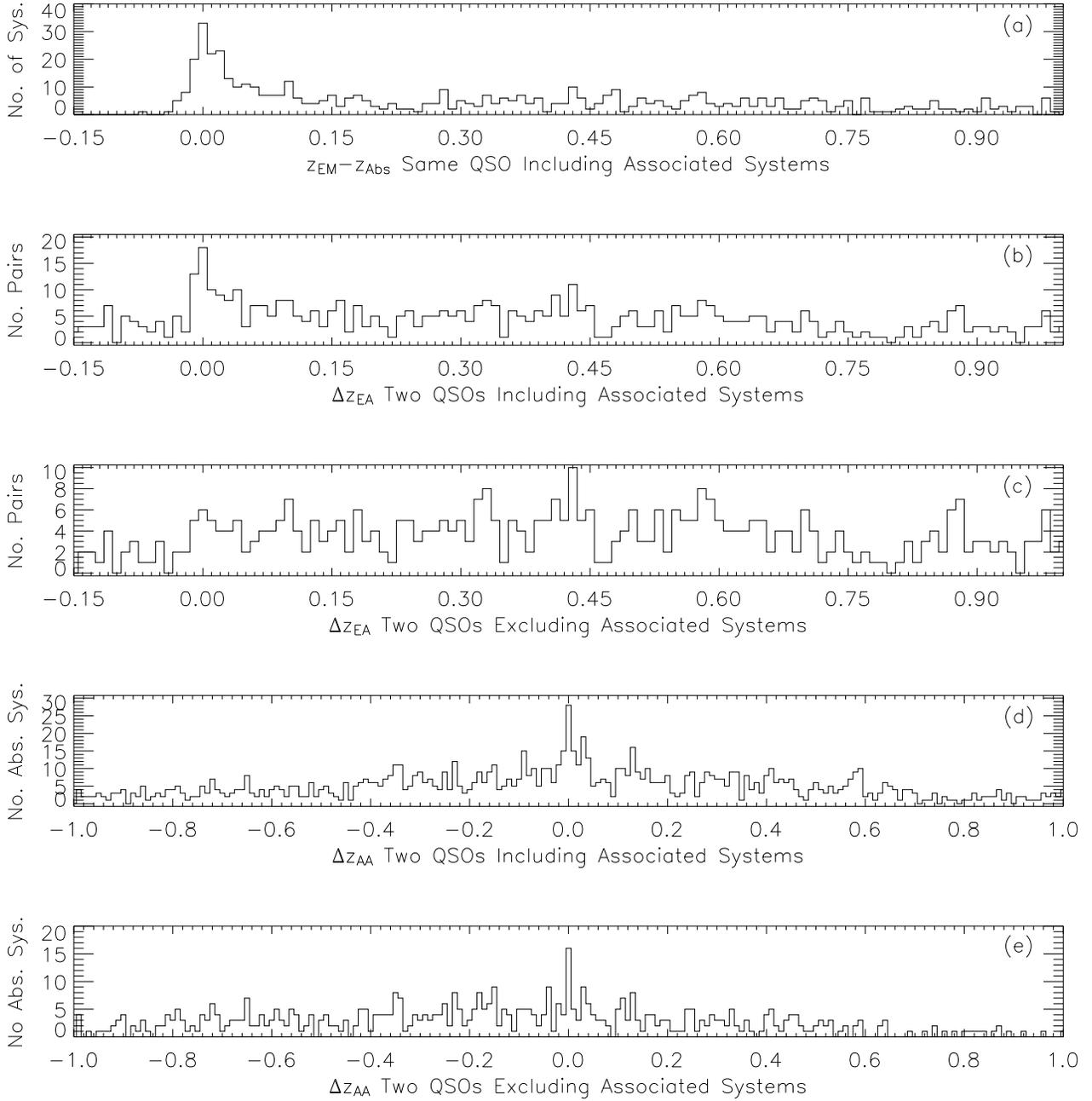}
  \caption{Histograms that show the distribution of redshift
    differences.  The top panel (a) shows the distribution of the
    \zabs\ values from the \zem\ of the same QSO.  The distributions
    in the remaining four panels all use to one redshift from each of
    the pair of QSOs.  The 2nd and 3rd panels (b) and (c) take a \zem\
    value from one QSO in a pair, and a \zabs\ value from the other
    QSO.  The 4th and 5th panels (d) and (e) take one absorber from
    each QSO.  In the 2nd and 4th panels use all \zabs\ values, while
    the 3rd and 5th panels use subsets that exclude all absorbers
    within 3000 \kms\ of their QSOs \zem\ value.  All bins have a
    width of \dz $=0.01$, approximately 1000~\kms\ for QSOs at $z=2$.
}
  \label{figangdistsep}
\end{figure*}

\begin{figure*}
   \includegraphics[width=180mm]{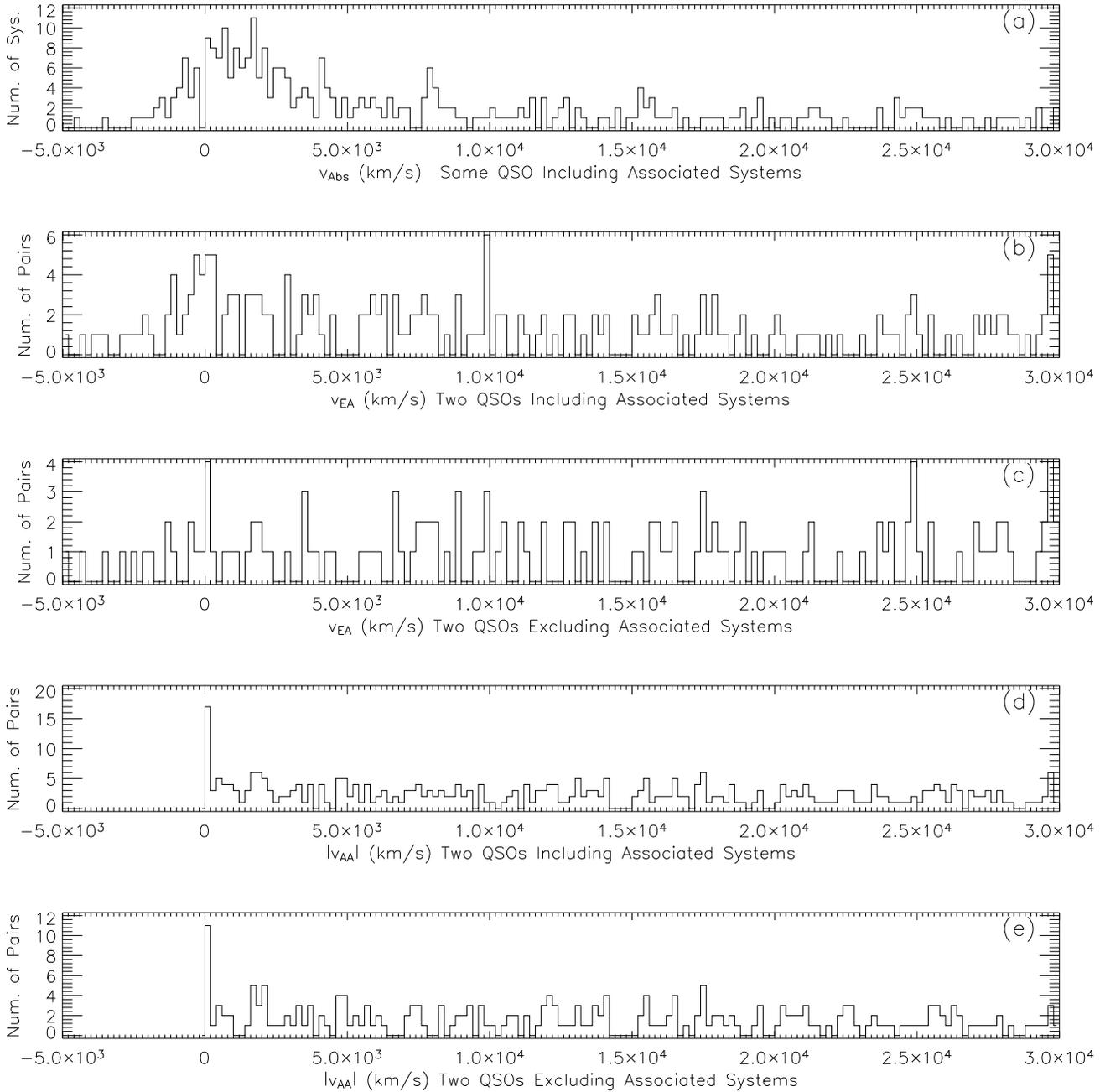}
   \caption{As Fig. \ref{figangdistsep} but using bins 200~\kms\ wide.
}
\label{figcloseup}
\end{figure*}

Many of our results on the distribution of the absorption and emission
redshifts
are shown in the five panels of Fig. \ref{figangdistsep}
and the related panels of Fig. \ref{figcloseup}
that we will discuss one by one, and in comparison.
The panels of Fig. \ref{figcloseup} are closeups of the panels
from \ref{figangdistsep}, using the relative velocity instead of
redshift differences. The bin size is 5 times smaller in Fig. \ref{figcloseup}
 for $z=2$.
We include QSOs with BAL absorption,
because a large fraction of the QSO groupings, 29/140,
show one or more BAL systems; but we exclude the BAL systems themselves.

\subsection{Absorbers Distributed Along Individual Lines of Sight}

In Fig. \ref{figangdistsep}(a) we show how the absorbers are
distributed relative to the \zem\ of their QSO. For each absorption system
we plot \zem\ $-$ \zabs , where both redshifts are for the spectra of the same QSO.
We see approximately 7 times more systems with \zabs\ $\simeq $ \zem\
compared to $0.15 <$ (\zem $-$ \zabs ) $<0.3$.
The excess is conspicuous at \zem\ $-$ \zabs\ $ < 0.05$
(5,000 km/s for $z=2$) and continues with lower amplitude to
\zem\ $-$ \zabs\ $ > 0.1$ (10,000~\kms\ at $z=2$).

Fig. \ref{figcloseup}(a) shows the same data as Fig. \ref{figangdistsep}(a)
but
in terms of velocity \vabs\ of the absorption system relative to the \zem\
of its QSO,
\begin{equation}
\vabs\ = c \times { (1+ \zem )^{2}-(1+\zabs )^{2}
\over (1+\zem )^{2}+(1+\zabs )^{2} },
\end{equation}
where absorbers that appear to be falling into the QSOs have negative velocities.
The \vabs\ values are
approximately uniformly distributed from 0 -- 2000~\kms , and the
distribution is  centred at approximately \vabs $\sim 1300$~\kms , and not
at \vabs $=0$.

The excess absorbers with \zabs $\sim $ \zem\ is similar to that
reported by \citet{weymann81} for a sample like our that does not
employ an equivalent width cutoff. The velocity range $-4000 < $\vabs\
$ < 4000$~\kms\ includes 42\% of the \citet{weymann81} sample of C~IV
systems and 50\% of our mixed absorption systems.  Samples that
contain only lines with \wrest\ exceeding some fixed minimum show much
smaller excesses (Fig. 3 of \citet{young82a} and Fig. 2a of
\citet{sargent88a}) because they exclude the additional absorption
lines that are easiest to see in the regions with the highest SNR,
especially in and near to the C~IV emission line.

\subsection{Emission redshift errors and ``infalling'' Absorbers}
 \label{secnegzem}

Some QSOs show absorption with \zabs\ larger than their \zem\ with \vabs $>1000$ \kms .
Peculiar velocities will account for many of the
 smaller ``infall'' velocities, but not the largest ones \citep{sargent82a}.
We believe that when we see a large negative \vabs\ the \zem\
 value is too small by about $-$\vabs . This idea due to
 \citet{gaskell83} is credible because we know
 that  large negative blueshifts of the C~IV emission lines are common
 \citep{gaskell82,tytler92,richards02a}. With this interpretation
 many of the absorbers at positive velocities relative to the
QSO systemic redshift and  the negatives \vabs\ values do not
need special treatment, other than noting that the \zem\ values are too small.

The distribution of the velocities of the absorption systems in the
frame of the QSOs give information on the \zem\ errors.  In
Fig. \ref{figcloseup}(a) we see only 15 systems from 310 QSOs at
velocities $< 1000$~\kms . This suggests that \zem\ errors are
typically $<1000$~\kms . We say more about \zem\ errors below.

\subsection{Redshift ordering of absorption and emission redshifts}

The remaining panels in Fig. \ref{figangdistsep} all compare one redshift
from each QSO of a pair. We will discuss pairings of absorbers with
emission redshifts first, then separately, absorbers with absorbers.

\begin{figure}
  \includegraphics[width=84mm]{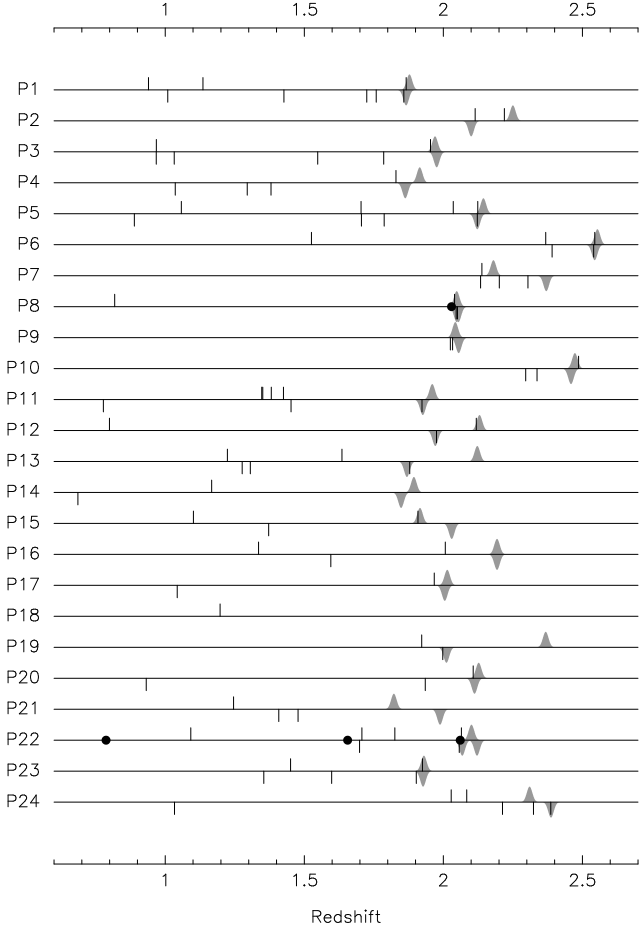}
  \caption{The absorption systems and \zem\ from Table \ref{tababs}
    for the first 24 QSO pairs. The \zem\ is shown as a gaussian grey
    region with $\sigma_{\rm z} = 0.01$ (1000~\kms\ at $z=2$).  The
    upwards tick marks represent the redshifts in the `a' QSO and the
    downward tick marks represent the redshifts in the `b' QSO.  For
    the triples P8 and P22 the filled black circles represent the
    absorption redshifts in the `c' QSO, `c' \zem\ are shown below the
    line. P18a and P18b have \zem\ $> 2.7$ and one system has \zabs\
    $< 0.6$. The two highest \zabs\ in P1 are coincident and named
    AAA21 in Table \ref{tababs}.  The highest \zabs\ in P1a (above the
    line) is coincident with the \zem\ of P1b below the line (EA1).
    The lowest two \zabs\ in P3 are coincident (AA1). P5 has
    coincident absorbers at 1.7 (AA2) and 2.2 (AAV17). In P6 absorbers
    at 2.54 are coincident with each other (AAA23) and with the \zem\
    of the partner QSO (EA3, EAV19).  P7 has a coincident absorbers at
    2.14 (AA4). The triplet P8 has multiple coincidences near the
    \zem\ values. The triple P22 has a wide coincidence at 1.7 (AAV18)
    and multiple coincidences near the 3 \zem\ values. The absorber in
    P23a at 1.9 is coincident with the \zem\ of P23b.
}
  \label{f1b}
\end{figure}

\begin{figure}
 \includegraphics[width=84mm]{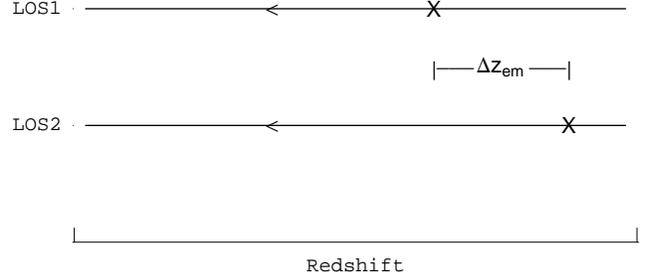}
 \caption{Visualisation of the range of possible and most common
   redshift differences.  The two horizontal lines represent the lines
   of sight to a pair of QSOs, with light travelling to the left and
   redshift increasing to the right.  The X's represent the emission
   redshifts.  In Fig.  \ref{figsepzem} we saw that the \zem\ values
   of the pair are often similar.  In Fig.  \ref{figangdistsep}(a) we
   saw the distribution of \zabs\ values relative \zem\ values. It is
   also common for \zabs\ values to the similar to the \zem\ values.
   If we take \zem\ from LOS1 and compare to a \zabs\ value from LOS2
   which is at $z_{em2}+0.05$, then the \dzea\ value is negative:
   $z_{em1} - z_{em2}-0.05$.  Similar arrangements may explain why the
   \dzea\ distribution in Fig.  \ref{figangdistsep}(b) is asymmetric
   around zero.  }
 \label{figasym}
\end{figure}

We provide two figures to help visualise the arrangement of the redshifts.
In Fig.  \ref{f1b}
we show the emission and absorption redshifts in the first 24 QSO pairs.
In Fig. \ref{figasym}
we give a sketch to help clarify the possible
arrangements of the \zem\ and \zabs\ values.
When we add an absorber to Fig. \ref{figasym} we can distinguish 4 binary
choices:
\begin{itemize}
\item Does the pair of QSOs have similar \zem ?
\item Is the absorber in the QSO with the higher \zem ?
\item Is the \zabs\ less than the partner QSO's \zem ?
\item Is the absorber associated with its QSO?
\end{itemize}
Not all of these 16 combinations are possible, but those that are can
populate and can explain much of the shape of the distributions that we will
now discuss.

\subsection{Absorption in one QSO near the Emission Redshift of the Partner: EA}
\label{secea}

Fig. \ref{figangdistsep}(b)
shows the distribution of \dzea\ values that we define as
\begin{equation}
\dzea\ =z_{\rm em1}-z_{\rm abs2}.
\end{equation}
This panel shows the tendency of absorbers in the spectrum of one QSO to
lie near the emission redshift of the other QSO.

We do not place any constraints on the relative values of the \zem\ of
the paired QSOs, hence we see a large number of highly negative \dzea\
values, most of which are of no interest to us.  If there were no
peculiar velocities, and the QSO emission lines gave the systemic
redshifts of the QSOs (that of the host galaxies), then all cases with
$\dzea\ < 0$ would be of no interest because they would come from
absorption systems in the background QSOs which are at higher redshift
than the foreground QSO.  However, we expect peculiar velocities are
frequently a few hundred \kms .  If the \zem\ values for a pair of
QSOs are very similar, and we allow for measurement errors, the light
from the QSO showing the \zabs\ is passing through at least part of
the volume around the other QSO.

Fig. \ref{figangdistsep}(b) shows how the absorbers in one QSO are
distributed relative to the \zem\ of the partner QSO.  We see 18
absorbers in the bin centred at zero, at $-0.05 <$ \dzea\ $< 0.05$,
which is $\pm 500$~\kms\ at $z=2$. Since the errors on the \zem\
values are likely in the range 400 -- 800 or more \kms\ we expect any
clustering of absorbers next to a QSO to be spread over the three bins
centres on zero. There at $-0.015 < $\dzea\ $< 0.015$ we see about 23
pairings in excess of the background level at larger \dzea\
values. This excess is highly significant.

In Fig.\ref{figcloseup}(b) we use velocity $v_{EA}$ instead of redshift, where
\begin{equation}
v_{\rm EA} = c \times { (1+ \zem )^{2}-(1+\zabs )^{2}
\over (1+\zem )^{2}+(1+\zabs )^{2} },
\end{equation}
and the emission and absorption redshifts are from the different QSOs
in the pair.  Fig. \ref{figcloseup}(b) shows that the excess extends
over approximately $\pm 400$~\kms\ and perhaps farther; the extend is
not well determined in this small sample when we do not have definite
model to test. We fit the distribution with the sum of a straight line
plus a Gaussian. The Gaussian that represents the excess has a mean
near $-150$~\kms\ (consistent with zero) and a dispersion of
approximately 525~\kms .

Both the dispersion and the mean are of physical interest.  The
dispersion gives an upper limit on the random errors in the \zem\
values of the QSOs that show systems with \zabs $\sim $\zem . In
addition to the \zem\ errors, the dispersion includes the pair-wise
random velocity difference of the absorbing galaxies relative to the
QSOs galaxy. From references that we discuss in \S 7.7, this pair-wise
velocity is likely in the range 200 -- 400~\kms\ if the QSOs are in
blue galaxies and 500 -- 800~\kms\ if they are in red galaxies. Hence
much of the dispersion we measure can be from pair-wise velocities
leaving little for the \zem\ errors. If we know the \zem\ errors we
should be able to decide if the QSOs are in red or blue galaxies. We
will see below that the absorbers far from QSOs
are typically in a population with
unusually small pair-wise velocity dispersions.

The mean velocity of the excess of absorbers around the partner QSOs
is an interesting new way to measure the difference of the \zem\ values from
the systemic redshifts. At least, this is true if
the absorbers are symmetrically distributed in velocity relative to
the QSO systemic velocity. We will discuss below why absorbers would be
preferentially behind the QSOs if QSOs are luminous for $\sim 3$~Myr and their UV
radiation destroys absorbers.

Fig. \ref{figangdistsep}(c) repeats the panel above, but
now excluding all associated systems, those within 3000~\kms\ of the \zem .
The excess near \dzea $=0$ has gone leaving an approximately constant
number of pairings at $ -0.015 <$ \dzea\ $< 0.6$.

Fig. \ref{figcloseup}(c) shows \dvea\ excluding associated absorbers.
While there are more pairs in the $v=0$ -- 200~\kms\ bin than in most other bins,
two other bin also show 4 pairs, hence from an {\it a posteriori}
perspective, where we do not know which velocity to consider,
this is not a significant excesses.
However, we have {\it a priori} reasons to look for the excess in the two
bins covering $\pm 200$~\kms . Here  we see 5 coincidences where we expect 1.5, 
from the 18
seen at 200 - 5000~\kms . The probability of 5 or more is 1.9\%.

We distinguish two possible explanations for the excess pairings
of absorbers with \zabs\ similar to the \zem\ of the partner QSO.

1) {\bf Normal  Line of Sight (los) associated absorbers.}
These are the excess C~IV absorbers at \dvea $< 3000$~\kms\ seen in
individual lines-of-sight, and seen in Figs. \ref{figangdistsep}(a) and
\ref{figcloseup}(a).
The excess pairings that we see in
Fig. \ref{figangdistsep}(b)  may be normal los associated absorbers that
are selected because the QSO pairs often have very similar \zem\ values.

2){\bf  Transverse associated absorbers.}
These are a new  type of absorption connected to and near to the QSOs.
Their existence has already been established
by the 4 Mg~II pairings reported by \citet{bowen06a}
and probably also for LLS found by \citet{hennawi07}.
They are different because most of them are not seen along the line of sight
to individual QSOs. The transverse absorbers
know with more precision the systemic velocity of the other QSO, 
that giving the \zem , than do its own los associated absorbers.
The transverse absorbers might arise in the host galaxy, but considering the sky
separations, they are more far likely from galaxies clustered around the QSO.
\citet[Fig. 8]{chelouche07a} 
shows a possible arrangement of the transverse
absorbers.

We suspect that many of the excess EA pairings at \dz $\simeq 0$ are
 transverse associated absorbers.
 The excess of EA absorbers Fig. \ref{figcloseup}(b) appear to be
 more concentrated around zero than are the normal los associated absorbers in
 Figs. \ref{figcloseup}(a). The four absorbers that remain at v=0 in
Figs. \ref{figcloseup}(c) when we have removed the associated absorbers hint
the same. 

Several pieces of weak evidence suggest that the transverse associated absorbers are
not just a subset of the los associated absorbers.
For Mg~II systems and LLS the arguments given by \citet{bowen06a} and
\citet{hennawi07} are that the transverse absorbers are more common than the
los associated. We can not claim this for the Mg~II and C~IV systems we study here
because we do not have a complete sample with a defined \wrest\ limit. Indeed
Figs. \ref{figcloseup}(a) and (b) show that there
are more los associated systems in our sample per \kms\ than there are transverse
QSO-absorber coincidences. However we do expect that many of the transverse
Mg~II and C~IV  systems  that we study
are similar to the Mg~II and LLS studied by \citet{bowen06a} and
\citet{hennawi07}, and hence their arguments should apply our sample.

We suspect that we would also see a difference in the \dvea\ of the
los and transverse associated systems if we had a complete sample of absorbers.
This complete sample would need to come from a much larger total sample
than we have here,  since complete
samples tend to contain only a fraction of all systems seen.
\citet[Fig. 2a]{sargent88a} shows that the excess of los associated
systems extends over
about $\pm 2000$~\kms , a larger range than the $\pm 400$~\kms\
indicated for the coincidences in Figs. \ref{figcloseup} (b), however
this comparison is insecure because of the small samples and differences
in the ways in which \zem\ values are measured. \citet{sargent88a} measured
\zem\ values using laboratory rest frame wavelengths for emission lines of
H~I, N~V,  Si~IV+OIV], C~IV, and their QSOs are significantly more luminous
leading to less distinct emission lines, factors that might lead to larger
errors in the \zem\ values and hence a wider range of \dvea\ for the
excess C~IV absorbers in their complete sample.

We now introduce a definition to simplify our discussion.
\begin{itemize}
\item {\bf EA } (emission-absorption) coincident absorbers are close to the
emission redshift of the partner QSO, with $-0.005 < $ \dzea\ $ < 0.005$. We
include associated systems. We do not require that the absorber is in the QSO
with the higher \zem\ value.
\end{itemize}

In Table \ref{tababs} we mark with EA1, EA2,.. the 18 pairings of \zem\ and
\zabs\ values and we give notes on each in the appendix.
In Table \ref{tabeavel}
 we list some properties of these 18 EA coincident absorbers, including the
 separation in  Mpc in the plane of the sky and in velocity. The triple P8
 includes 3 EA pairings, one an absorber in P8a with QSO P8c, and two from
 absorbers in P8b paired with QSO P8a. The other EA parings are each from
 a different pair of QSOs.

 For 2 of the 18 EA pairings (P116 and P119) the absorber is in the QSO with the
 lower \zem\ value. For P116
 the absorber is at \vabs $=-3,435$~\kms\ and \dvea $=13$~\kms . If both
 \zem\ values have negligible errors, this can not be a physical coincidence.
 However, if the \zem\ value of the QSO showing the absorber is too low by
 $\sim 3,400$~\kms , depending on peculiar velocities,
 we may still have a physical coincidence. P119 is similar
 but less extreme with \vabs $=-444$~\kms\ and \dvea $=266$~\kms .
 In a third case, EA18, the absorber is in the QSO with the higher \zem\ value,
 and the \zabs\ is larger than the \zem\ of its QSO by \vabs $= -239$~\kms .
 These coincidences between absorbers apparently ``infalling'' into QSOs
 with the \zem\ values in partner QSOs strengthen our belief
 that the systemic redshifts for these two QSOs, and by implication most QSOs
 with absorbers at large negative velocities,
 are significantly higher than their \zem\ values.

We define a second type of coincidences that explore a larger range of
separations.
\begin{itemize}
\item {\bf EAV coincident absorbers} (V for velocity) are like EA coincidences
but they have \dvea $< 1000$~\kms\ instead of $-0.005 < $ \dzea\ $ < 0.005$.
\end{itemize}
In Table \ref{tabeavel} we list 12 EAV coincidences.

Examining Table \ref{tabeavel}, we see
contradictory evidence as to whether we see the transverse absorbers
along a single line-of-sight. Arguing in the negative,
both \citet{bowen06a} and \citet{hennawi07} claim that
they see far more transverse absorbers than los associated ones.
Our spectra also show no excess of los associated
absorbers confined to \vabs $< \pm 400$~\kms\ in Fig. \ref{figcloseup}(a).
For example, if 20\% of QSOs showed absorption within $\pm 400$~\kms\ of
their \zem\ value, from the population of transverse absorbers that happen to
be in the los, then we would see 62 excess absorbers at these \vabs\ values
where we see only 23, which is no excess compared to $400 < \vabs < 2000$~\kms.

However, examining Table \ref{tabeavel} we see that many of the absorbers in
the EA and EAV coincidences have small \vabs\ values that presumably place
them very close to their QSOs, where we have just argued we do not typically see
excess absorption. The distribution of the \vabs\ values of the EA and EAV
coincidences seems similar to that for absorbers as a whole.
We are surprised to see absorption in a QSO coincident with the \zem\ of
the partner QSO, when the absorber and the two QSOs
are all have similar redshifts. In Fig. \ref{figasym} imagine that both QSOs
are embedded in a spherical halo of (transverse) absorbers that also
overlaps the other line of sight. We see excess absorption in LOS2 at the \zem\ of
QSO1, but no excess in LOS1 due to its own halo of absorbers.
How can we see transverse absorption associated
with the partner QSO1 at small \vabs\ values, given that we do not often see
such absorption, as similar small \vabs\ values, in either individual line-of-sight?
 We expect that whatever prevents us from seeing
the transverse absorbers along the line-of-sight to most individual QSOs
would also prevent us from seeing them around the partner QSO when the similarity of
the \zem\ values places them at small \vabs\ values.
This is a mystery.

\begin{table}
\caption{EA Emission-absorber coincidences. The first column gives the QSO
that contributes the \zem\ value (listed in Table \ref{tababs}). The letter
in front of the \zabs\ in the second column identifies the QSO that
gives the \zabs\ value. We designate a pairing EA if
$-0.005 <$ \dzea $< 0.005$, otherwise it is EAV which means
that \dzea $> 0.005$ and \dvea $< 1000$~\kms . Velocities are in
(\kms ) and $v_{\rm abs}$ is the velocity of the absorber in the frame of the
\zem\ value of its QSO. The separation of the two sight lines in the plane
of the sky is the $b$ value in proper Mpc.
}
\label{tabeavel}
\begin{tabular}{lllrrcc}
\hline
 QSO & QSO    & Pair & $v_{\rm abs}$ & \dvea\ & $b$ & W$_{\rm rest}$ \\
  em & \zabs\ &      &       &  & (Mpc) & (\AA ) \\
\hline
 P26b & a 1.8896  & EAV22 & -815 & -919 & 0.2114 & 0.80 \\
 P26b & a 1.89084 & EAV23  & -686 & -790 & 0.2114 & 0.80 \\
 P64a & b 2.00451 & EAV24  & 26886 & -651 & 2.1137 & 0.54 \\
 P8c  & b 2.04602 & EAV20 & 785 & -593 & 0.9202 & 0.77 \\ 
 P147a & b 2.3708 & EAV28 & -2484 & -517 & 0.9911 & 0.28 \\
 P147b & a 2.34871 & EAV29 & 1456 & -512 & 0.9927 & \\
 P94a  & b 3.2081 & EAV25 & 42063 & -363 & 1.9958 & \\
 P28a  & b 2.1686 & EA9 &  1831 & -341 & 1.5089 & 2.72 \\
 P153a & b 2.5178 & EA18 & -239 & -324 & 1.1315 & 0.75 \\
 P8a  & b 2.05115 & EA6 & 280 & -310 & 1.1069 & 0.52 \\
 P6b  & a 2.54697 & EA3 & 594 & -251 & 0.5880 & 0.17 \\
 P43b  & a 2.0665 & EA12 & 13151 & -147 & 0.4292 & 0.32 \\
 P1b  & a 1.8673 & EA1 & 1117 & -136 & 0.7556 & 0.38 \\
 P123a & b 2.25927 & EA17 & 12762 & -117 & 2.1715 & 0.40 \\
 P5b  & a 2.12306 & EA2 & 2005 & -102 & 0.5075 & 0.28 \\
 P116b & a 1.79988 & EA15 & -3435 & 13 & 0.6928 & 2.00 \\
 P38b & a 1.91658 & EA11 & 15366 & 43 & 0.1869 & \\
 P8c  & a 2.03912 & EA4 & 875 & 87 & 1.8363 & 0.16\\
 P36a & b 1.96688 & EA10 &  32073 & 113 & 0.3425 & \\
 P8a  & b 2.04602 & EA5 & 785 & 195 & 1.1072 & 0.77 \\
 P23b & a 1.9249 & EA8 & 523 & 215 & 0.4211 & 0.14 \\
 P95c  & b 3.14571 & EA14 & 1825 & 238 & 2.2246 & 0.69 \\
 P119a & b 3.22625 & EA16 & -444 & 266 & 0.4713 & 0.57 \\
 P22b & a 2.06449 & EA7 & 3553 & 343 & 1.3093 & 1.25 \\
 P82b  & a 1.98055 & EA13 & 2151 & 347 & 1.4410 & 0.58 \\
 P153b & a 2.50758 & EAV30 & 549 & 634 & 1.1325 & \\
 P22b  & c 2.05997 & EAV21 & 5924 & 786 & 1.2880 & 0.16 \\ 
 P141a & b 4.1160 & EAV27 & 3001 & 845 & 2.0000 & 2.85 \\
 P6a  & b 2.54327 & EAV19 & 62 & 907 & 0.5883 & 0.26 \\ 
 P121b & a 2.2882 & EAV26 & 347 & 984 & 2.1815 & 1.32 \\
\hline
\end{tabular}
\end{table}

\subsection{Absorption in one QSO near Absorption in the Partner: AA}

We now discuss absorber-absorber coincidences, where one absorber comes from each
of a pair of QSOs. Fig. \ref{figangdistsep}(d) shows the distribution
of \dzaa\ which we define as
\begin{equation}
\dzaa\ =z_{\rm abs1}-z_{\rm abs2},
\end{equation}
where one absorber is from each QSO in a pair. We arbitrarily chose which QSO
is 1 for the subtraction of the absorption
redshifts and hence the signs have no physical meaning.
However, the QSO ordering in the main table is not random, and hence
the \dzaa\ distribution has a clear, unintended asymmetry. Some of the peak
near the \dzaa $= 0$ value again comes from the tendency of the two
QSOs in a pair to have similar \zem\ values and the
excess of absorbers with \zabs\ $\simeq $ \zem . The counts are low but
hint that the excess extends over several bins, perhaps covering a range of
\dz\ similar to that of the excess of absorbers in the individual sight lines
in the top panel.

Fig. \ref{figangdistsep}(e) is like panel \ref{figangdistsep}(d), but now
excluding the associated systems from both QSOs.
The excess remains and is now restricted to the 16 systems in the
single bin at
$-0.005 < $ \dzaa\ $< 0.005$ ($\pm 500$~\kms\ at $z=2$). We now make a
definition to simplify our discussion.
\begin{itemize}
\item {\bf AA} (absorption-absorption) coincident systems have a partner in the paired QSO with
$-0.005 < $ \dzaa\ $< 0.005$. They exclude  associated systems, those with
\vabs $< 3000$~\kms\ relative to their QSOs \zem\ value.
\end{itemize}

The 16 AA coincident systems constitute a highly significant detection of the
correlation of metals between the paired sight lines.
A straight line fit to the distribution of the
absolute \dzaa\ values in Fig. \ref{figangdistsep}(e),
excluding absolute values $< 0.005$,
gives $y=-0.037x+4.645$ absorption systems per $\Delta z = 0.01$. Hence we
expect 4.645 system pairs
with $-0.005 < $ \dzaa\ $< 0.005$. The Poisson probability of observing
$\geq 16$ given an expectation of 4.645 is $3 \times 10^{-5}$.
The 16 pairings are a
factor of $16/4.645 = 3.44$ above the expected number of chance pairings.
We give notes on each of these 16 systems in the appendix, and we label and
number them with AAxx in Table \ref{tababs}.

Fig. \ref{figcloseup}(e) shows the absolute values of \dvaa\
values, excluding all
associated systems. The excess that we saw in terms of redshift is still
present, and now gives 12 pairs in the first bin covering
$0 < v < 200$~\kms .
The peak at $v=0$ is largely confined to this first bin and is clearly highly
significant.

\section{Rest Frame Equivalent Widths and Ions Seen in the Coincident Systems}

In this section we see that both the AA and the EA coincident systems are
unremarkable spectroscopically. Rather they seem typical of all systems that
we see.

\begin{figure}
  \includegraphics[width=84mm]{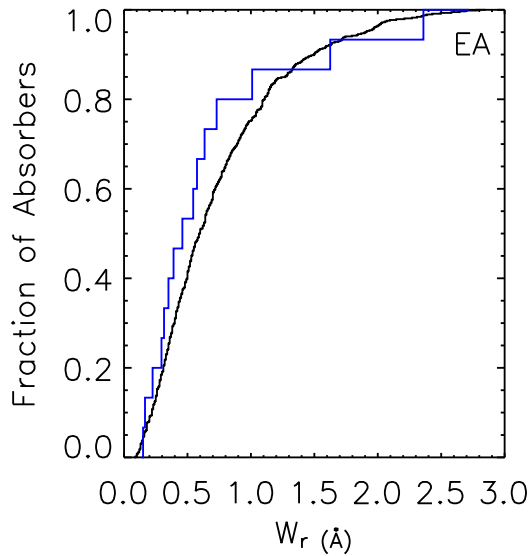}
  \caption{Cumulative distribution of the rest frame equivalent width of
572 of the non-BAL absorption lines from Table \ref{tababs},
excluding the 15 EA coincidences but including
associated systems (smoother black line).
We show separately the \wrest\ values of 15 of the EA coincidences
(stepped blue line).
}
  \label{figwrcontea}
\end{figure}

In Fig. \ref{figwrcontea}
we show the cumulative distribution of the
\wrest\ values of the main ions of the
non-BAL absorption systems listed in Table \ref{tababs}.
There is a surprisingly wide range of \wrest\ values
from 0.1 -- 2.8~\AA , with a mean near 0.74~\AA .
We show separately the \wrest\ distribution for 15 of the 15 EA coincident
absorbers. We did not measure \wrest\ values for the other 3 EA systems
for various reasons.
The EA systems tend to have smaller \wrest\
values than the sample as a whole,
 but the maximum difference of 0.208 is not significant according to the KS test.

\begin{figure}
  \includegraphics[width=84mm]{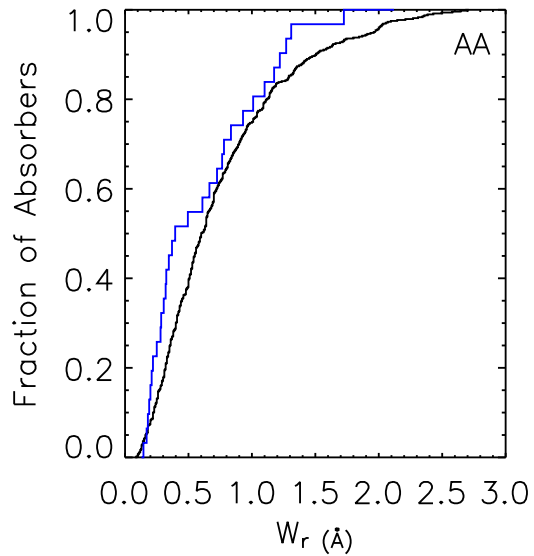}
  \caption{Cumulative distribution of the rest frame equivalent widths
  of the AA systems (stepped blue line)   and the
  sample from which they were drawn (smoother black line).
We use the \wrest\ values from Table \ref{tababs} of the main ions in the 452
non-BAL absorption systems, excluding
associated systems and excluding
the 32 absorption systems from the 16 paired AA coincidences.
This figure differs from Fig. \ref{figwrcontea} in that associated absorbers
are now removed and the samples are divided by whether an absorber is an
AA rather than EA coincidence.
}
  \label{figwrcont}
\end{figure}

In Fig. \ref{figwrcont}
we compare the AA systems with the sample from which they were selected.
Like the EA absorbers, the AA coincident absorbers also
include a larger fraction of
smaller than average \wrest\ values. The maximum difference between the two
distributions is 0.243 near \wrest\ = 0.41~\AA . This difference is barely
significant at the 95\% level.

We are not surprised that the coincident absorbers have smaller \wrest\
values,
since higher quality spectra that can reveal smaller \wrest\ values will tend
to have more systems per spectrum and hence a larger chance of showing
coincidences. They will not all have the smallest \wrest\ values, because too
few spectra could show such values.

\subsection{Ionisation of the Coincident Systems}

In Table \ref{tabocc}
we list the frequencies of ions in the EA and AA
systems. We see nothing unusual about the AA coincidences.
Most AA coincidences show the same ions in both systems. In only two cases (AA5, AA15)
 do coincident systems have no ions in common. The AA coincidences show C~IV 13
times and Mg~II 8 times, and 11 of the AA systems show other ions, most
often Fe~II, but also  Mg~I, Si~III, Si~IV, N~V, Al~II and Al~III.

The 16 of the 18 EA systems show C~IV while the other two show Al~II and Al~III.
They do not show Mg~II or other lines with large rest frame wavelengths because
the \zem\ values are too high for these lines to be in our spectra.
We do not see any excess of N~V, unlike
\citet{hennawi06a} who note that several of the LLS that they find near to QSOs
exhibit N~V, an ion rarely seen far from QSOs.

\section{Spatial Distribution of Absorbers around Absorbers}

We now give a more detailed discussion of the absorber-absorber
correlation, in both redshift and impact parameter in the plane of the sky.
We see that the absorbers are strongly correlated on scales $< 0.6$~Mpc. We detect the
 redshift-space distortion, in which peculiar velocities make the correlation
 elongated along the line of sight.
We show that the peculiar velocities are low, much less than the velocities
that \citet{adelberger05} see in the C~IV and H~I absorption from star forming
Lyman break galaxies (LBGs).

In Table \ref{tabaavel}
we list the velocity separations for the 16 AA
system pairs that we defined as all non-BAL systems with
$-0.005 < $ \dzaa\ $< 0.005$, which is 500~\kms\
or a distance of 2.49 Mpc at $z=2$.
Of the 16 AA such systems, only AA12 has a velocity separation
\begin{equation}
\dvaa\ = c \times { (1+ z_{\rm abs1} )^{2}-(1+z_{\rm abs2} )^{2}
\over (1+ z_{\rm abs1} )^{2}+(1+z_{\rm abs2} )^{2} },
\end{equation}
larger than 500~\kms , which somewhat less than the
correlation length seen in
individual sightlines for strong C~IV and Mg~II absorption systems
\citep{young82a, petitjean90a,scannapieco06a}.
We make two further definitions that expand upon the AA definition.
\begin{itemize}
\item {\bf AAV coincident systems} (V for velocity) have \dvaa\ $<1000$~\kms\
but
$|$\dzaa $| > 0.005$, so that they were not listed as AA. As with the AA
systems, we require that both absorbers be farther than
\vabs\ $ =3000$~\kms\ from their QSO.
\item {\bf AAA coincident systems} (Associated Absorber-Absorber)
have \dvaa\ $<1000$~\kms\ and one or both absorbers
at \vabs\ $ < 3000$~\kms\ from their QSO.
\end{itemize}
We find 4 AAV coincidences (AAV17 -- 20)
and 14 AAA coincidences (AAA21 -- 34). We list them
 in Tables \ref{tababs} and \ref{tabaavel}
and we describe some of them in the appendix.
Seven pairs (P5, P6, P25, P31, P38, P42, P155) and two tripples (P8, P22)
contribute two or more coincidences each.

\begin{table}
\caption{The 34 absorber-absorber coincidence separations
in order of velocity difference \dvaa .
In all cases one absorber is from each of a pair of QSOs.
We label pairs of absorbers with
\dzaa $<0.005$ as AA (16 cases), and those with \dvaa $< 1000$~\kms\ as AAV
(4 cases).
When one or both system is within 3000~\kms\
of their QSO ($v_{\rm E1}$ or $v_{\rm E2} < 3000$) we call them AAA (14 cases).
The two letters after the pair number in the
first column identify the two QSOs containing the absorbers.
The first letter (e.g. P3b) refers to the $v_{\rm E1}$ value in the
3rd column, while the second letter refers to the $v_{\rm E2}$ value.
Velocities are in \kms , and the redshift difference is in units of $10^{-5}$.
At $z=2$, \dzaa $=10^{-5}$ corresponds to \dvaa $\simeq c$\dzaa $/(1+z) = 1$~\kms .
The last column shows the separation between the sight lines at the redshift of the
absorbers in proper Mpc.
}
\label{tabaavel}
\begin{tabular}{llrrrccl}
\hline
QSO\footnotemark[1] & Label & $v_{\rm E1}$ & $v_{\rm E2}$ &  \dzaa\ &\dvaa\ & Sep. \\
 pair           & &  & & ($10^{-5}$)  & & (Mpc) \\
\hline
 P3ba   &   AA1 &    117487 &       116978 &           3.0 &5 & 0.096 \\
 P7ba   &   AA4 &     21463 &         4098 &           9.0 &9 & 0.544 \\
 P38ab  &  AA10 &     69880 &        55187 &         20 & 25 & 0.187 \\
 P31ba  &   AA9 &     70389 &        60783 &          26 &32 & 0.858 \\
 P22cb  & AAA27 &      5924 &          745 &          40 &41 & 1.310 \\
 P31ba  &   AA8 &    119710 &       111093 &          36 &53 & 0.815 \\
 P36ba  & AAA30 &     32073 &          180 &          66 &67 & 0.343 \\
 P155ba & AAA34 &      2097 &         2708 &          90 &68 & 0.108 \\
 P25ba  &   AA6 &     64626 &        66842 &          54 &70 & 0.126 \\
 P70ab  &  AA15 &     18124 &        31855 &          72 &76 & 1.560 \\
 P155ba &  AA16 &     24237 &        24829 &          102 &84 & 0.111 \\
 P38ab  &  AA11 &     68229 &        53422 &         88 &108 & 0.187 \\
 P5ab   & AAA22 &      2005 &           26 &         133 &128 & 0.507 \\
 P5ab   &   AA2 &     44888 &        42697 &         119 &132 & 0.516 \\
 P44ab  &  AA14 &     76087 &        79820 &         100 & 140 & 0.382 \\
 P42ab  &  AA13 &     39010 &        66642 &         147 &179 & 0.228 \\
 P83ba  & AAA31 &      8135 &         1731 &         90 &198 & 1.148 \\
 P25ba  &   AA7 &     49553 &        52112 &         192 &238 & 0.128 \\
 P147ba & AAA32 &      -795 &         1456 &         316 &283 & 0.993 \\
 P6ab   & AAA23 &       594 &           62 &        370 &313 &  0.588 \\
 P22ab  & AAA29 &      3553 &          745 &        410 &402 &  1.310 \\
 P6ab   &   AA3 &    202347 &       201643 &        231 & 442 & 0.468 \\
 P22ab  &   AA5 &      3553 &         5923 &         452 &443 & 1.310 \\
 P22cb  & AAA26 &      5924 &         1253 &         476 &467 & 1.310 \\
 P125ab & AAV20 &     38663 &        39804 &        568 &558 &  0.831 \\
 P42ab  &  AA12 &     78387 &       103869 &         445 &621 & 0.222 \\
 P8ab   & AAA25 &       875 &          785 &        690 &680 &  1.836 \\
 P5ab   & AAV17 &     10597 &         7701 &         802 &792 & 0.510 \\
 P22ba  & AAV18 &     38224 &        40523 &         784 &870 & 1.326 \\
 P153ba & AAA33 &      -239 &          549 &         1022 &873 & 1.132 \\
 P22ab  & AAA28 &      3553 &         1253 &         928 &910 & 1.310 \\
 P1ab   & AAA21 &      1117 &          800 &         893 &936 & 0.756 \\
 P31ba  & AAV19 &     69457 &        60783 &        775 &953 &  0.858 \\
 P8ac   & AAA24 &       875 &         1048 &         959 &961 & 1.836 \\
\hline
\end{tabular}
\\
$^1$ In order of velocity separation. \\
\end{table}

\subsection{Absorption Redshift Errors}

We are especially interested in the errors on the absorption system \zabs\
values that happen to be close to another \zabs\ or \zem\ value.
For each system listed in Table \ref{tabaavel} we made optimised fits
to the main absorption lines. We list the results in Table \ref{velerr}.
For about 40\% of these systems we have
two or more ions per system. In those cases we calculate $\sigma (z)$,
the standard
deviation of the $z$ that we measured for each ion. The mean $\sigma (z)$
is $23 \pm 4$~\kms . We believe that this is representative of the internal
errors on the \zabs\ values for systems showing one ion. For systems with two
ions, we might attain errors smaller by a factor of up to $1/\sqrt{ 2}$.
For approximately 10\%
of the systems the lines are blended in ways that make
it hard to find a unique centre, and errors will be several times larger.
When we compare to a \zabs\ value measured in the partner QSO we need to
also account for systematic errors in the wavelength scales. We typically see the
same ions in both spectra, and hence we are measuring the \zabs\ values at the
same observed wavelengths. When these measurements are done with the
same instrument and grating these external errors should be similar
and hence they will have little effect on
relative velocities. Otherwise the external errors may be some
fraction of a pixel. We shall assume that our errors for a typical
\zabs\ value are 23~\kms ,
noting that they will be several times larger in some cases,
as seen from the large standard deviation of 19~\kms\ of the $\sigma (z)$ values in
Table \ref{velerr}.

\begin{table}
\caption{Redshift and velocity errors for 28 AA, AAV or AAA systems
in order of pair number.
We list only the
ions we use to obtain the mean redshift in column 2 and
the standard deviation of the two
\zabs\ values in column 3. In column 4 we have converted the
standard deviation of the \zabs\ values into velocity.
The $\sigma (z)$ is in units of $10^{-5}$.
Velocities are in \kms .
}
\label{velerr}
\begin{tabular}{lcccl}
\hline
 QSO & \zabs\ &  $\sigma (z)$ & $\sigma (v_{AA})$ & Ions \\
     &  mean  &  ($10^{-5}$)      & (\kms ) & \\
\hline
 P1a & 1.86730 & 13 & 14 & C~IV N~V\\
 P1b & 1.85837 & 26 & 27 & C~IV Si~IV\\
 P3a & 0.96770 & 1.4  & 2.1 & Mg~II Fe~II\\
 P3b & 0.96767 & 1.4 & 2.1 & Mg~II Fe~II\\
 P5a & 2.03513 & 20 & 20 & C~IV C~II\\
 P5a & 2.12306 & 7.2 & 6.9 & C~IV N~V\\
 P5b & 2.04286 & 75 & 74 & C~IV Si~III\\
 P5b & 2.12173 & 18 & 18 & C~IV Si~IV\\
 P6a & 2.54697 & 17 & 14 & C~IV N~V\\
 P7a & 2.13685 & 52 & 49 & C~IV Al~II\\
 P22a & 2.06449 & 23 & 22 & C~IV Si~IV\\
 P22a & 1.70703 & 6.0 & 6.6 & C~IV Si~IV\\
 P22a & 1.69908 & 26 & 29 & C~IV Si~IV\\
 P22c & 2.05997 & 44 & 42 & N~V Si~III\\
 P25a & 1.30321 & 27 & 35 & C~IV Si~IV\\
 P25b & 1.42593 & 52 & 65 & C~IV Mg~II\\
 P31a & 1.02822 & 4.7 & 6.9 & Mg~II Fe~II\\
 P36a & 1.96622 & 16 & 16 & N~V Si~III\\
 P36b & 1.96688 & 23 & 24 & Al~II Si~III\\
 P42a & 1.14819 & 6.2 & 8.6 & Mg~II Fe~II\\
 P44a & 1.14351 & 10  & 14 & Mg~II Fe~II\\
 P70a & 1.83710 & 41 & 43  & Mg~II Fe~II\\
 P147b & 2.35187 & 6.0 & 5.3 & C~IV Si~IV\\
 P153b & 2.51780 & 23 & 19 & C~IV Si~IV\\
 P155a & 2.66316 & 59  & 48 & C~IV Si~IV\\
 P155a & 2.94423 & 7.0 & 5.3 & C~IV Si~IV\\
 P155b & 2.66216 & 23 & 19 & C~IV Si~IV\\
 P155b & 2.94333 & 4.7 & 3.6 & C~IV Si~IV\\
\hline
\end{tabular}
\\
\end{table}

\subsection{Impact parameters of AA coincidences}
\label{secaaimpact}

\begin{figure}
  \includegraphics[width=84mm]{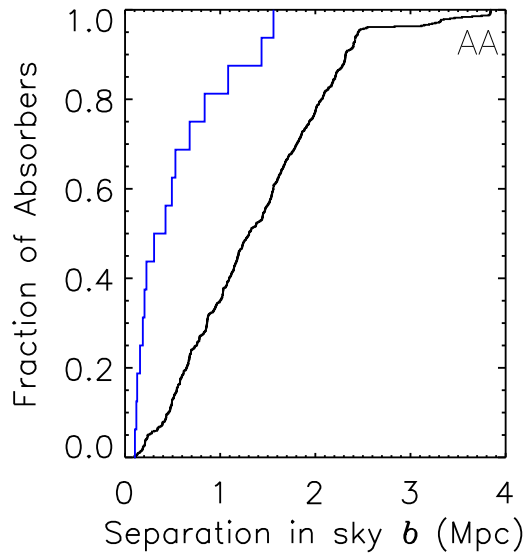}
  \caption{
The cumulative
distribution of the distances between lines of sight measured in the plane of the sky
$b$ proper Mpc. We show separately the separations between 15 of the
AA coincidences (blue, stepped line) and the partner sight line, and the
sample of 625 systems (black, smooth line) from which the AA sub-sample were drawn.
}
  \label{figcontmpcsepaa}
\end{figure}

\begin{figure}
  \includegraphics[width=84mm]{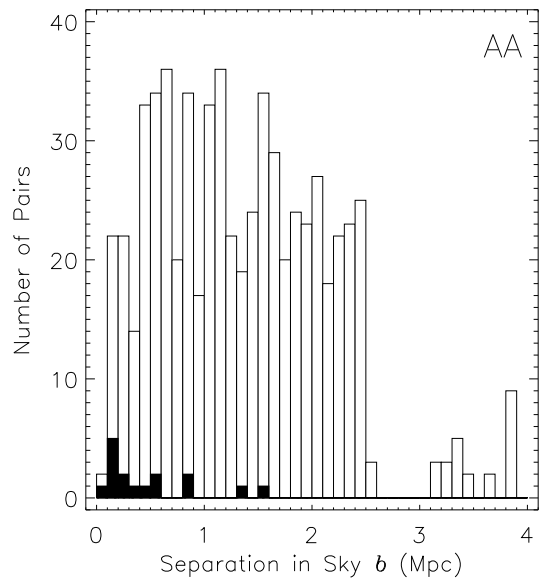}
  \caption{The distribution of distances between the 16 AA coincidences
  each with \dzaa $< 0.005$ (solid)
  and the sample from which they were selected (clear). The horizontal axis is
  the impact parameter, the distance from each absorber to its partner
  QSO in the plane of the sky, $b$ in proper Mpc.
  The clear histogram includes all absorption systems,
  including coincidences, but excluding all
  cases where \zabs\ $>$ 0.005 + \zem\ of partner QSO.
For both histograms we have removed all absorption systems within 3000 \kms\
of the host QSO, and we count each coincident pair of systems as one system.
}
  \label{figMpc_separation}
\end{figure}

In Fig. \ref{figcontmpcsepaa}
we see a highly significant tendency for the AA coincidences to be at unusually
small separations. Of the 16 pairs,  6 are  at $< 0.2$, and 12 at $<0.6$~Mpc,
where as the mean separations of all systems is 1.3~Mpc.
This is also seen in the histograms in Fig. \ref{figMpc_separation}.
When we see absorption in one sight line, the probability
of also seeing absorption with \dzaa $<0.005$ (approximately \dvaa $<500$~\kms )
in the partner sight line is high for small impact parameters and declines
rapidly. We see
1 coincidences from 2 systems (50\%) in sight lines separated by $<100$~kpc,
5 from 22 sight lines (23\%) at 100 -- 200~kpc,
6 coincidences in 103 systems (6\%) at 200 -- 600~kpc,
2 in 107 (1.9\%) at 0.6 -- 1~Mpc, and 2 in 264 (0.7\%) at 1 -- 2~Mpc.
We obtain these ratios by dividing the
two histograms in Fig. \ref{figMpc_separation}.
These ratios are very much lower limits since there
are many cases in which we have limited or no ability to see systems in
the partner QSO. High quality spectra with complete wavelength coverage
would show many more systems include those with small \wrest\ values
that would give
much higher probabilities. We could obtain higher probabilities if we include the
\wpm\ values in the calculation, but we have not done this.

\subsection{Redshift-space distribution of AA coincidences}

\begin{figure*}
  \includegraphics[width=180mm]{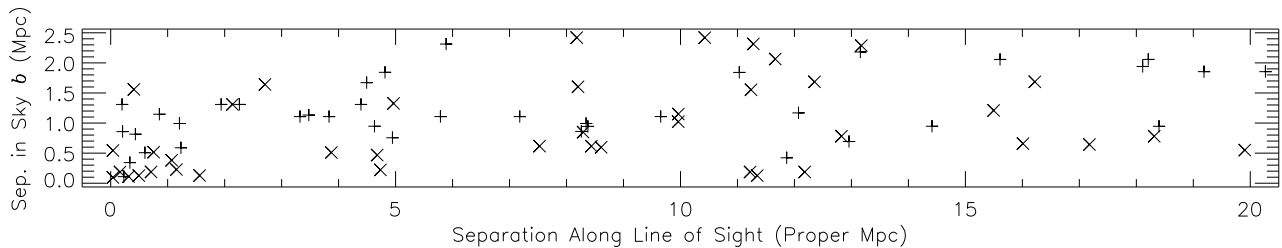}
  \caption{The separation in proper Mpc of two absorption systems, one
in each of a pair of QSOs. We plot each pairing once, with one system
 at the origin and the other is marked by
the $\times $, or a $+$ if either absorber is within 3000~\kms\ of its QSO.
The $x$-axis is distance from one absorption system to
that in the partner, obtained directly from the difference between
the two absorption redshifts. This is the proper equivalent of the comoving
$\pi $ used in galaxy literature. The $y$-axis is the distance between the
sight lines in the plane of the sky, at the redshift of the absorbers. This
is the impact parameter $b$ from Eqn. (1) and (2), and it is the proper equivalent
of the $r_p$ in galaxy literature.
We limit the vertical axis to 2.5~Mpc because we saw in
Fig. \ref{figMpc_separation}
that we have approximately constant number of sight lines per unit $b$
out to this distance.
}
  \label{figmpcsepscat}
\end{figure*}

In Fig.  \ref{figmpcsepscat}
we see the 2D distribution of the absorber-absorber
separations, in redshift space along the $x$-axis
and in the plane of the sky along the $y$-axis. We place one of the two
 absorbers in each pairing at the origin. The $y$-axis is the impact
parameter, and the $x$-axis is distance derived from \dzaa\ assuming Hubble flow
and ignoring peculiar velocities. Light rays travelling to us are
horizontal lines going to the right. We show separately pairings
that do not involve associated absorbers (including, but not
limited to all the AA and AAV pairings that extend to approximately 5~Mpc)
and those that do include one or two associated systems (including
and not limited to all AAA coincidences). We have  already seen projections
of the absorber-absorber correlations into the two axes of this Figure.
Fig. \ref{figcloseup} showed the projection along the $x$-axis ignoring the
sky separation, while Fig. \ref{figMpc_separation} is the projection in the
plane of the sky for coincidences with \dzaa $< 0.005$, the AA coincidences.

We see the over-density of  absorbers near the origin. We have already
established the statistical reality of this excess,
which seems to extend out to about 1~Mpc or perhaps $\sim 2.5$~Mpc.
There are 23 absorbers within 2.5~Mpc
(500~\kms\ at $z=2$) of the origin, including all 16 AA coincidences
that we list in Table \ref{tabaavel}.
We see the background of absorbers separated by 5 -- 20~Mpc that is
approximately uniform in density, although sparsely sampled.
The clumps of points far from the origin are accidental
because they are nearly all from different QSO pairs. Clustering of two of more
system pairings along the line
of sight to a given QSO pair would appear as a horizontal grouping.

The distribution of absorbers around absorbers can be determined from
their 3D spatial correlation function. If the absorbers arise in galaxies,
then the correlation function for those galaxies, modified by
peculiar velocities, our sampling
function, and our measurement errors
will describe the distribution in Fig. \ref{figmpcsepscat}.
The level of concentration of points near the origin of the plot then
depends on
the correlation length, the peculiar velocities and the measurement errors.

Errors in the measurement of absorption redshifts tend to
smear the $x$-coordinates of the plotted points. Measurement
errors of 23~\kms\
per absorption system will contribute a $1 \sigma $ dispersion of only
 $\sim 0.16$~Mpc along the $x$ direction, too small to have a major effect.

\begin{figure}
  \includegraphics[width=84mm]{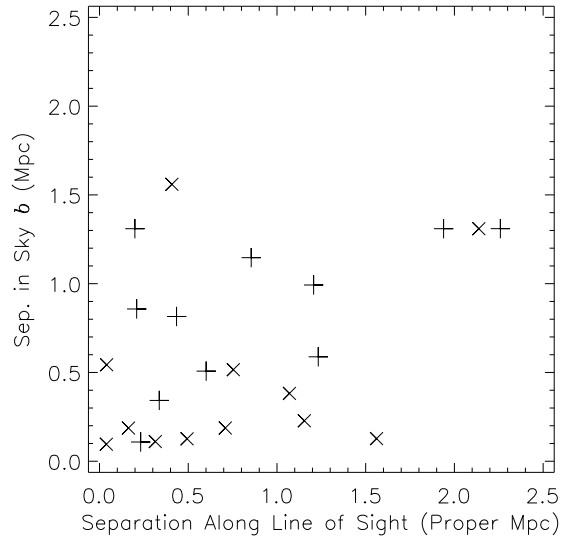}
  \caption{
  Enlargement of the inner 2.5~Mpc of  Fig. \ref{figmpcsepscat}.
}
  \label{figmpcsepscat25}
\end{figure}

We see  non-uniformity and asymmetry in the
distribution of absorbers in the inner 2.5~Mpc. We choose 2.5~Mpc because
we know from Fig. \ref{figmpcarcdist}
that we have approximately constant number of sight lines as a function of
separation in the sky out to 2.5~Mpc. In Fig. \ref{figMpc_separation}
we saw that the absorbers that we found also have an approximately
uniform distribution in impact parameter out to 2.5~Mpc.
This is because we deliberately observed
all known close pairs of QSOs but only a fraction of those at larger separations.
In this way we accidentally cancelled the increase with $b^2$
in the area  of annuli on the sky of radius $b$.
In the absence of clustering we expect the absorbers
to be approximately uniformly distributed along
the $y$-axis of Fig. \ref{figmpcsepscat}.
Fig. \ref{figmpcsepscat25} is an enlargement of the inner 2.5~Mpc
of Fig. \ref{figmpcsepscat}.

We now comment on the distribution of
the points. Although the excess of points near the origin is clearly
real, the asymmetry in the distribution of this excess could be
entirely accidental because the sample is small.
Since we noticed the features {\it a posteriori} we decline to give any
statistical assessment, and we leave the reader to decide if the
evidence combined with the physical implications are sufficient to motivate
 further comment or investigation.

We see a relative lack of absorbers in the upper left
at $x < 5$~Mpc and $y > 1.5$~Mpc. We have no explanation for this and
expect that it is an accident.  Within approximately 1~Mpc of the
origin we see that the excess is mostly at angles below $45 \deg$ from the
$x$-axis. We do not have any explanation for such angles.
However, we can instead think of the plot as showing a
tendency of the clustering around an absorber
to be more widely distributed along the line of
sight ($x$-axis) than in the plane of the sky ($y$-axis), and this has a well known
explanation.

\citet{crotts97a} found the same effect from a study of metal lines in HIRES
spectra of the triplet of QSOs, P118abc. They found that the correlation
between the lines of sight was much weaker than expected from the
two-point correlation seen along many individual lines of sight.
\citet{crotts97a} proposed that the effect
was caused by peculiar velocities that are large compared to the correlation length.
The peculiar velocities make the line of sight correlation in velocity space
appear to be more extended  than it is in proper Mpc multiplied by the Hubble
constant. This is the usual redshift-space distortion or anisotropy that makes
the ``fingers of God'' in maps of galaxy position in redshift versus sky position.
Clusters of
galaxies are elongated in the redshift coordinate \citep[Fig. 4]{davis83}
because of the large peculiar velocities.
Peculiar velocities move points
along the $x$-axis, decreasing the density near the $y$-axis. Peculiar
velocities along the $y$-axis have no effect since they do not change redshifts
or positions on the sky.
For our absorbers the peculiar velocities could include random motions, rotations and
winds flowing out from the absorbing galaxies.
To account for the asymmetry we see in Fig. \ref{figmpcsepscat25} we need
velocities along the line of sight of $\sim 100$~\kms . \citet{adelberger05}
also claim to see this redshift distortion in their Fig. 12 that we will discuss
below.

\subsection{Qualitative assessment of the redshift-space distortion}

\begin{figure}
  \includegraphics[height=65mm]{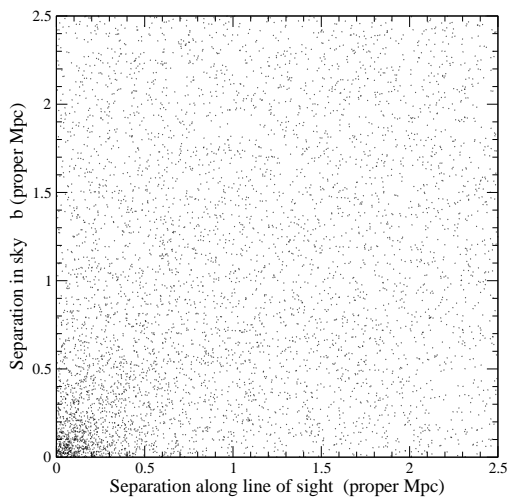}
  \caption{The expected distributions of absorbers about an absorber
  at the origin. We begin with a random distribution of points that sample the
  galaxy-galaxy two point correlation function. We ignore
  peculiar velocities but we add a Gaussian random deviate to the horizontal
  position of each point with $\sigma (v) = 23\sqrt{2}$~\kms\ or $\sim 0.16$~Mpc.
}
  \label{simaa}
\end{figure}

In Fig. \ref{simaa}
we show the expected distribution of AA separations derived from the 3D
2-point correlation function of LBGs from \citet{adelberger05c}. Their
correlation length measurements are larger but consistent with
those from \citet{cooke06a} who use redshifts for LBGs with
similar redshifts and magnitudes, but with
less sky coverage. We assume $ \xi (r) = [(r + r_m)/r_0]^{-\gamma }$
with $\gamma = 1.6$ from \citet{adelberger05c}. We set $r_m = 0.05$~proper Mpc
to account for our difficulty in distinguishing multiple absorbers inside
one halo. We convert the \citet{adelberger05c}
comoving correlation length measurements at $z=1.69$ and 2.24 to proper Mpc
and we linearly interpolate to $z=2$, giving $r_0 = 2.06$~Mpc for our Hubble
constant. We further reduce this by a factor of 0.6 to 1.24~Mpc at $z=2$ to convert
from the correlation length of their
LBG galaxies to the correlation of all galaxies in the DEEP sample. We
obtain the factor 0.6=3.2/5.4 from the comparison at $z \sim 1$ given in
Fig. 12 of \citet{adelberger05c}.  To mimic our
Fig. \ref{figmpcsepscat}, we assume a constant number of sight lines
per unit of impact parameter along the $y$-axis. Note that we have no paired QSOs
separated by $y < 0.09$~Mpc, and hence we can not see the full amplitude of
the peak near the origin. By design the
distribution is symmetric about the origin, and in this sense, and in the extent of
the concentration about the origin it looks different from our data.

\subsection{Adding gravitational infall}
\label{secinfall}

\citet[Fig. 2]{adelberger05b} discusses the distortion produced by different
types of peculiar velocity flows.
Gravitational infall of one absorbing galaxy towards the other is a systematic
effect that is correlated across the area of the plot. If the density field
were spherically symmetric about the absorber at the origin we would expect
radial towards the origin that moves most points closer to the $y$-axis.
The amplitude of the infall will depend on the masses of the halos of the absorbers.

\begin{figure}
  \includegraphics[width=65mm]{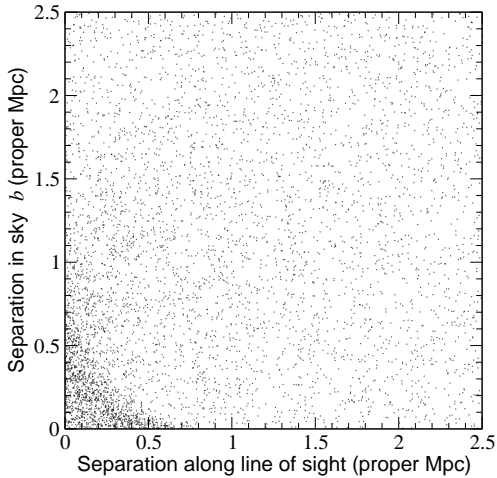}
  \caption{As Fig. \ref{simaa} with observational errors
 and now including radial infall due to a moderate distribution of halo masses,
 MidM from Table \ref{tabhalo}.
}
  \label{simaainfall}
\end{figure}

In Fig. \ref{simaainfall}
 we show the effect of adding systematic infall velocities. We use  infall
velocities for halo masses given in Table \ref{tabhalo}
from the lower portion of Fig. 2 of \citet{kim07a}.
The halos give infall velocities of approximately
70~\kms\ out to 2.8~Mpc, declining to 20~\kms\ by 10~Mpc.
We assume that the number of
halos is distributed as $M^{-2}$ \citep{vale04} and that the absorbing area of an
halo scales with the halo mass. \citet{chen01b} find that the area out to which
C~IV absorption is readily seen at low $z$
scales with the galaxy luminosity, and we assume constant M/L. Hence we assume that
the absorption we see samples a halo mass distribution $\propto M^{-1}$ that
we call the MidM distribution in Table \ref{tabhalo}.
We expect slightly larger infall velocities because we observe two galaxies that
are absorbing, implying a local mass density above that
around a typical galaxy that need not have an neighbour.

\begin{table}
\caption{Halo masses used for infall velocities in different models.
We assume the probability of absorption in a halo of mass
M is $\propto M^{-\beta }$. We list the percentage of halos of each mass listed
on the top row in solar units.
}
\label{tabhalo}
\begin{tabular}{lcccc}
\hline
Model & $\beta $ & $ 5.1 \times 10^9$ &  $1.6 \times 10^{11}$ & $9.1 \times 10^{11}$\\
\hline
LowM & 2    & 96.9 & 3 & 0.1\\
MidM & 1    & 75   & 23 & 2\\
HiM &  ...  & 0.1 & 3 & 96.9\\
\hline
\end{tabular}
\end{table}

 We add radial infall velocities directed towards the origin. These 3D velocities
are a function of the radial distance (before we add simulated
observational errors!) from the origin alone, with no random component.
The component of the velocity along our line of
sight decreases to zero as we rotate from the $x$-axis up to the $y$-axis, making a
caustic like density peak, as shown in
the upper right quadrant of Fig.~5 of \citet{kaiser87a}.
Since we use more than one halo mass and we
have smoothed in the $x$ direction by adding random errors
in the redshifts, the caustic is less distinct in Fig. \ref{simaainfall}.
The infall increases the density of
points along both the $x$ and $y$ axes. For $y < 0.1$~Mpc
the infall moves points away from the origin and along the $x$ axis,
but we have almost no sight lines that sample these small $y$ values.
For other $y$ values the infall moves points
towards the $y$-axis, giving a lower density of points at $x>0.4$~Mpc, $y < 0.5$~Mpc,
and a higher density at $x < 0.2$, $0.2 <y< 0.5$~Mpc. This elongation of the
density along the $y$-axis is the opposite of the asymmetry that we see in the
absorber-absorber correlation.

\subsection{Adding pair-wise random velocities}

\begin{figure}
  \includegraphics[width=65mm]{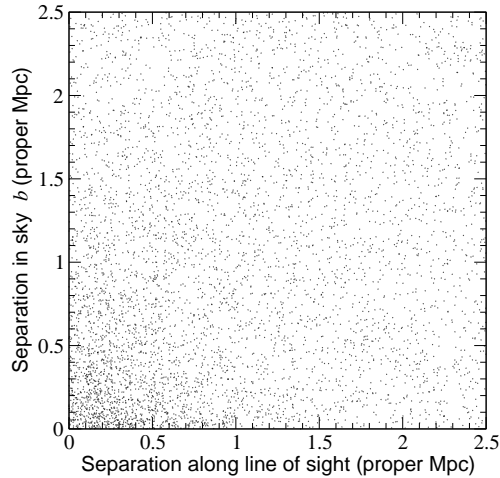}
  \caption{As Fig. \ref{simaa} including observational errors and
  radial infall (for the MidM halo mass distribution) but now adding
 random pair-wise velocities from the exponential distribution
  with a $\sigma_{12} = 240$~\kms\
  applied to the absorber that is not at the origin.
}
  \label{simaaran}
\end{figure}

In Fig. \ref{simaaran}
we add random velocities to represent the pair-wise velocity differences
of galaxies.
We select the velocities from the exponential distribution
\begin{equation}
\label{eqncoil}
f(v_{12}) = {1 \over \sqrt{2} \sigma _{12}}exp( -{\sqrt{2} \over \sigma _{12}}
| v_{12} - {\bar v_{12}} | )
\end{equation}
from \citet[Eqn 17]{coil07a}, where ${\bar v_{12}}$ is the mean infall velocity.
Since we  apply these random velocities to the undisturbed $x,y$ coordinates,
before we apply the infall velocities or simulated measurement errors,
we set the term ${\bar v_{12}}=0$.
This term is used if we fit the function to data comprising velocities that
will necessarily include infall.
We choose the dispersion $\sigma _{12} = 240$~\kms\
to represent the random pair-wise velocity differences.
The points are smeared along the $x$-axis giving the
usual redshift-distortion that makes the
``finger of God'' effect that is most readily seen in dense
groups and clusters where the peculiar velocities are large.
Now we see a tendency of the points to be more extended along the $x$-axis,
but it is unclear if this provides a better match to our data.

\subsection{Estimation of the pair-wise velocity dispersion}
\label{secest}

Is the distribution of absorber-absorber separation in
Fig.\ref{figmpcsepscat25} consistent with absorption arising in
ordinary galaxies with the expected galaxy-galaxy autocorrelation,
infall and the pair-wise peculiar velocity distributions? Given the
small size of our sample, we can only detect the absorber-absorber
correlation on very small scales where the peculiar velocities are
larger or comparable to the Hubble flow. Let us then assume that we
know the absorber-absorber correlation length, the infall velocities
and the measurement errors, and ask what value for the pair-wise
dispersion of the velocities is most consistent with the data. We
understand that the pair-wise velocity distribution, the infall and
the correlation length are all related to each other and to the history
of the density distribution \citep{scoccimarro04,slosar06}.

We calculate the likelihood of our absorber-absorber distribution, as
a function of the dispersion of the pair-wise velocities $\sigma
_{12}$.  We estimate the probability of each coincidence shown on
Fig. \ref{figmpcsepscat25} from the density of points on a version of
Fig. \ref{simaaran} with many more points and various values for
$\sigma _{12}$.  For each absorber in Table \ref{tababs} we calculate
the probability $P_s$ of absorption at $x,y$ in the spectrum of the
partner QSO.

The probability of one absorber at position $x_i$ in an interval $0 <
x < x_{max}$ of one line of sight is the product of three
probabilities, $P_s=p_lp_ip_r$ where $p_l$ is the probability of
finding no absorbers in $0 < x < x_l$, $p_i$ is the probability of
finding an absorber at $x_i$ and $p_r$ is the probability of finding
no absorbers at $x_r< x < x_{max}$. We have no information on possible
extra absorbers in the interval $x_l < x_i < x_r \sim 0.85$~Mpc,
because we can see a maximum of one absorber per FWHM of the spectra.
From the Poisson distribution, the probability of no absorbers is
$p=e^{-\mu }$ where $\mu $ is the expected number of absorbers. We
have
\begin{equation}
\mu_l = w_i \int^{x_l}_0 p(x,y)\, dx ~~{\rm and} ~~~
\mu_r = w_i \int^{x_{max}}_{x_r} p(x,y)\, dx
\end{equation}
where $p(x,y)$ is probability of an absorber at $(x,y)$ given by the
density of points and $w_i$ is a weighting factor that accounts for
the sensitivity of the spectrum to an absorption system.  Similarly
the probability of the absorber at $x_i$ is $p_i = w_i p(x_i,y)$, and
hence the probability for sight line $s$ is
\begin{equation}
P_s= e^{-\mu_l} w_i p(x_i,y) e^{-\mu_r}.
\end{equation}

The weighting $w_i$ is the fraction of all \wrest\ values that are
larger than the \wpm\ value in Table \ref{tababs}.  If we do not list
a value for either \wrest\ or \wpm\ we assume $w = 0.2$ corresponding
to \wpm $=1.5$~\AA\ in Fig. \ref{figwrcontea}. If \wpm $=-1$, meaning
that the main line could not be seen in the partner spectrum, we use
$w = 0.1$.

The likelihood $\mathcal L$ of the data set for a given model is the
product of the probabilities for each sight line.  We count each
coincidence twice since the \wpm\ values for the two sight lines
differ. We ignore absorbers with \zabs\ $> \Delta z_m +$ \zem\ of the
partner QSO, where $\Delta z_m = (1+z)H(z)x_{max}/c$ is the redshift
equivalent of the distance $x_{max}$.

\begin{figure}
  \includegraphics[width=65mm]{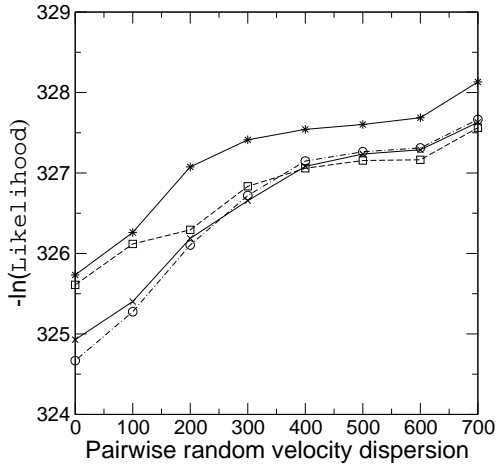}
  \caption{The maximum likelihood estimate of the pair-wise random
    velocity dispersion.  The vertical axis shows the negative of the
    natural likelihood of the absorber-absorber data, both
    coincidences and lack of coincidences. The horizontal axis shows
    the of the random pair-wise velocities dispersion $\sigma _{12}$
    for the exponential distribution in Eqn. \ref{eqncoil}. Reading
    down from the top at $\sigma _{12} < 200$~\kms , the stars show a
    model with no infall, the boxes high mass halos (HiM), the crosses
    medium masses (MidM) and the circles low mass halos (LowM).
}
\label{lnlpair}
\end{figure}

In Fig. \ref{lnlpair}
we see that the likelihood of the data is maximum when the $\sigma
_{12}=0$ with a $1\sigma $ upper limit of 100~\kms\ and a $2\sigma $
limit of 300~\kms . The likelihood is insensitive to larger pair-wise
velocity dispersions because the density of points is then nearly
uniform in space.  These pair-wise velocity dispersion values are very
much at the lower end of the values reported for galaxies.  The
absorber-absorber correlation is more concentrated near the origin
that we expect except for a very ``cold'' population, with low
velocities relative to their neighbours.

In Fig. \ref{lnlpair} we also show the likelihood of the three
other models.  We show a model with no infall velocities, a model
where the probability of absorption in a halo is proportional to
$M^{-2}$ favouring very low mass halos and a model favouring high mass
halos (see Table \ref{tabhalo}).  To first order the models give
similar infall velocities and all are compatible with the data.  In
detail the data are most likely in models with low to medium halo
masses, and less likely by more than $1 \sigma$ in models
using high mass halos or no
infall.  The trends follow because the low and medium mass halo models
both give high density in the caustic region where we have 4 data
points. The high mass halos have excessive velocities giving a
slightly lower density near the origin, while the no-infall model has
the highest density at the origin but a lower density in the caustic
region.

\citet{davis83} estimated line of sight random velocity differences locally
could be  represented as
\begin{equation}
\sigma _{12}(b) = 340 \pm 40 (b/1.4~{\rm Mpc})^{0.13 \pm 0.04}~~ (\kms ),
\end{equation}
for projected separations 14~kpc $< b < $ 1.4~Mpc. Like most authors we will
assume that $\sigma _{12}$ values apply with no $b$ dependence over
a few Mpc. At redshifts $z \sim 0.1
$ \citet{zehavi02a}
find SDSS blue galaxies give $\sigma _{12} \sim 300$ -- 450~\kms\
while red galaxies give $\sigma _{12} \sim 650 - 750$~\kms .
\citet{li06b} find 200 -- 400~\kms\ for blue  and 600 -- 800~\kms\ for red SDSS
galaxies.
For the 2dF Redshift survey  \citet{madgwick03a}
find $\sigma _{12} = 416 \pm 76$\kms\ for active star forming galaxies
and $612 \pm 92$~\kms\ for passive galaxies.
We expect lower velocities at higher redshift but  measured values are only
slightly lower. At $z \sim 1$ \citet{coil07a}
estimate $\sigma _{12} = 240 \pm 20$~\kms\ for blue galaxies and
$530 \pm 50$~\kms\ for red galaxies.

The low pair-wise velocities for the absorbers are marginally
compatible with absorption in some samples of blue galaxies, and
incompatible with red galaxies. This implies that the absorbers tend
to avoid the rare high density regions such as clusters of galaxies
where the red galaxies gain much of their larger pair-wise
velocities. We also believe that our sample is much too small to give
a fair sample of all absorbers and hence we might anticipate a larger
velocity dispersion in a larger sample, as is seen with galaxies.

\subsection{Limits on Wind Outflows}

We can use our upper limit on the random velocities to put a limit on
wind outflow velocities. \citet{adelberger05} recommended this test as
one of the best ways to try to determine how far winds extend from the
centres of galaxies.  Our sample is well suited to this examination
because our redshift errors of 23~\kms\ are small compared to the wind
velocities of hundreds of \kms , and much smaller than the errors
obtained for galaxies at $z=2$ from their optical and UV lines.

Consider absorption in gas flowing radially out from galaxies.  We
assume this gas is transparent so we see either absorption with a
velocity component towards us, or away from us. The result is that the
redshifts of the absorbers are changed by an amount given by the wind
velocities.  Assume pure radial outflow at velocity $v_w$ confined to
a thin spherical shell, and assume that absorption occurs on either
the front or the back of the shell, but not on both sides. The mean
component of the wind velocity along the line of sight is then
$0.5v_w$. If instead we see the same absorption from both sides then
the mean velocity is zero by symmetry. The mean velocity will be less
than $v_w$/2 depending on the frequency of two-sided absorption. Since
we attempt to measure \zabs\ values for the velocity component with
the highest optical depth, we will tend to set the \zabs\ value to one
of the other side rather than the mean of both, except when they give
similar or blended lines.

We will model the component of the wind velocity in the line of sight
as a Gaussian random deviate.  We expect that a Gaussian is a more
realistic distribution than that from a pure radial outflow with a
constant velocity for all galaxies.  We set the standard deviation of
the Gaussian to $\sigma _w = 0.5v_w\sqrt{2}/\sqrt{2/\pi }$.  The last
term is the expected absolute value of a random Gaussian deviate, $
\sqrt{2/\pi }=0.7979$. The first $\sqrt{2}$ term accounts for the wind
from the galaxy that makes the absorber that we place at the origin of
the plot, so we do not need to also add a deviate to the origin. We
choose the absorber that we place at the origin of the plot at random
from each pairing, and we assume equal and uncorrelated winds for both
absorbers. We then add the these Gaussian deviates to each absorber on
the plot, and not to the origin.

\begin{figure}
  \includegraphics[width=65mm]{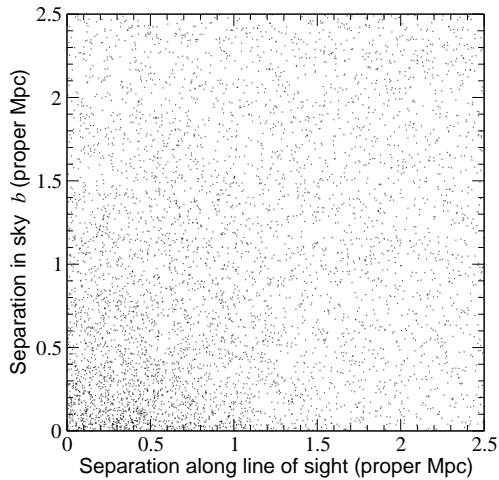}
  \caption{As Fig. \ref{simaa} with observational errors,
  radial infall and random pair-wise velocities from the exponential distribution
  with $\sigma _{12} = 100$~\kms ,
  and now adding  random peculiar velocities from a Gaussian with a
  $\sigma_w = 88$~\kms\ applied to the absorber that is not at the origin.
  This represents radial winds of $v_w = 125$~\kms\
  from each galaxy, including that at the origin.
}
  \label{simgaussian}
\end{figure}

In Fig. \ref{simgaussian}
we add random velocities from a
Gaussian pdf with a dispersion of $\sigma_w = 222$~\kms\ to model a wind with
 $v_w = 250$~\kms . Since the pair-wise random velocities can hardly be zero, we
 use a models with $\sigma_{12} = 100$~\kms . This model has much more dispersion than
 our data.

\citet[Fig. 1]{adelberger05} have estimated the redshifts of LBGs at
$2 < z < 3.5$ from three different spectral features, the \lya\
emission line, the ISM UV absorption lines and nebulae emission lines
such as [OII], H$\alpha$, H$\beta $ and [OIII]. The nebulae lines are
believed to be close to the systemic velocities and give errors of
60~\kms\ for 90 galaxies.  The \lya\ lines and UV lines were seen in
spectra with about 10~\AA\ (600 ~\kms ) resolution and have much
larger errors of approximately 200 -- 275~\kms\ (from the $\sigma_z $
values below Eqn. 3 in \citet{adelberger05} for $z=2.6$) The \lya\
lines give redshifts that are systematically larger by up to
1000~\kms\ and typically 450~\kms . The UV absorption lines give
redshifts systematically smaller than the nebular lines by about
250~\kms\ with wide range from 800~\kms\ smaller to 200~\kms\
larger. The UV absorption lines are direct evidence for outflowing
winds.  Separately, \citet{adelberger05} see strong C~IV lines in
background objects 40~kpc from LBGs and weak lines extending much
farther, but they do not know if this is wind material, and they do not
know if the winds escape from the LBGs. They notice that the mean
absorption within 40~kpc of foreground galaxies is large \wrest\
2.7~\AA\ and implies significant absorption over at least a 260~\kms\
velocity range.

\citet[Fig. 12]{adelberger05} shows the positions of C~IV absorbers
(about 17 systems with 68 components) that they observed with
$<0.4$~proper Mpc sky separation and $< \pm 4$~Mpc redshift difference
from LBGs. For each impact parameter they calculated the redshift
difference out to which the 3D correlation function would place 50\%
of absorbers if there were no peculiar velocities.  They see that most
absorption components have larger velocity differences from their
QSOs, implying several hundred ~\kms\ peculiar velocities.  They
suggest that these peculiar velocities may be winds, but they do not
discuss normal pair-wise random peculiar velocities that can be
comparable in size.

\begin{figure}
  \includegraphics[width=65mm]{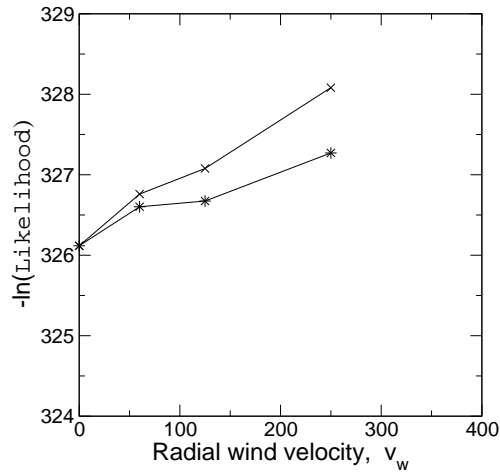}
  \caption{The negative of the natural likelihood of the absorber-absorber
  data, both coincidences and lack of coincidences,
   as a function of the wind velocity $v_w$ (\kms ), shown by the upper
   line marked with crosses.
   We also show (lower line with stars) a case where the wind velocities are restricted
   to 1/3 of the galaxies.
}
  \label{lnlwind}
\end{figure}

We do not see such large peculiar velocities.
In Fig. \ref{lnlwind}
we see the likelihood of the absorber-absorber correlation data
declines significantly with increasing $v _w$ that we use to represent
the random wind velocity. The data prefer zero wind velocity with a
$1\sigma $ limit of $v_w < 45$~\kms\ and a $2\sigma $ limit of
250~\kms .

The \citet{adelberger05} sample differs from our in two obvious ways.
We measure distances between pairs of absorbers while they measure
distances from LBGs to absorbers. We take absorber redshifts from the
main component visible in moderate resolution spectra while they
consider all the components visible in higher resolution
spectra. Hence they are exploring the low column density gas around
LBGs while we are more sensitive to higher column density gas that is
likely near to typical absorbing galaxies that may not be LBGs.

Absorption in gas flowing out from galaxies at a mean velocity of $v_w
=250$~\kms\ would produce much more elongation along the line of sight
than we see. We conclude that the absorbing gas is not in fast winds.
The winds seen by \citet{adelberger05} in LBG spectra are not
representative of absorption systems that we see. Either the winds are
confined to LBGs, or they do not extend to $> 40$~kpc with large
velocities, or they do not produce absorption we can detect. Here
40~kpc is the distance from LBGs at which \citet{adelberger05} see
strong C~IV absorption (not necessarily from winds).  Typical
absorption systems are too common to be confined to smaller distances
from galaxies.

\citet[\S 3.1]{adelberger05} further deduce that the LBGs in their
sample can account for roughly one-third of C~IV 1548 lines with
\wrest $>0.4$~\AA\ that is the typical \wrest\ for our AA absorbers.
We model this assuming that their LBGs account for 1/3 of all \zabs\
that we can detect.  For 1/9 of the points, where both absorbers arise
in fast winds, we set $\sigma _w = 0.5v_w\sqrt{2}/\sqrt{2/\pi }$ as
usual. For 4/9 of points where only one of the two absorbers arises in
a wind we set $\sigma _w = 0.5v_w/\sqrt{2/\pi }$, and for the
remaining 4/9, $\sigma _w = 0$.  The lower curves in
Fig. \ref{lnlwind} show that the likelihood is still lower for higher
wind velocities, with a $1\sigma $ limit of $< 95$~\kms\ but now
250~\kms\ is allowed at the $2\sigma $ level.

Hence our small sample of absorber-absorber coincidences could arise
in fast winds from LBGs alone extending to 40~kpc and making 1/3rd of
strong C~IV or Mg~II lines that we could detect.  However the data
prefer absorption in a cold population, with no extra velocity
dispersion from winds, or otherwise. We are not consistent with most
C~IV or Mg~II coming from such high velocity winds.  If most galaxies
have high velocity absorbing winds, these winds must be confined to
$<< 40$~kpc, the typical radius around a galaxy at which strong metal
lines are seen.  If all galaxies have winds that travel out $>40$~kpc
then the wind material must have low velocities where we see
absorption, or not have the density, metal abundance, ionization and
velocity structure necessary to make metal lines that we can see.

\section{Spatial Distribution of Absorbers around QSOs}
\label{secspatialea}

We now leave the absorber-absorber coincidences and turn to the
QSO-absorber coincidences. The EA pairings share some features with
the AA pairings (3D correlation, infall velocities, pair-wise random
velocities) but they differ in other ways, especially the much larger
errors on the \zem\ values and the possibility of larger halo masses
for the QSOs.  

We will look to see if there are any signs of an asymmetric
distribution of absorbers around the QSOs.  If the UV radiation from
QSOs can change or destroy absorbers, then we will see an asymmetric
distribution of absorbers around QSOs if the QSO UV is confined to a
narrow beam, or alternatively if QSOs emit isotropically but for only
0.3 -- 10~Myr.  We will not use the absorber-QSO correlation to look
for signs of winds from galaxies, because the errors on the \zem\
values are too large.

\begin{figure}
  \includegraphics[width=84mm]{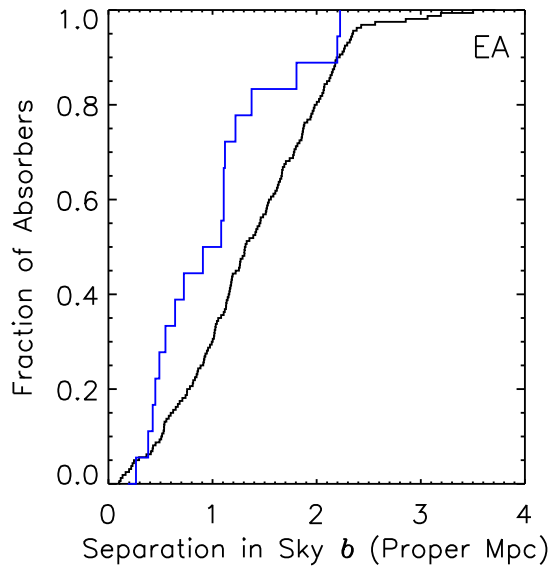}
  \caption{
The cumulative distribution of the separations in the sky of the QSO pairs.
 The smooth black line
shows the distances from all foreground QSOs to the partner
  lines of sight,  plus the distances for the background to the foreground QSOs
  when the \zem\ difference is $< 0.005$.
The 18 EA coincidences are the sub-set shown by the fainter blue stepped line.
}
  \label{figcontmpcsepea}
\end{figure}

In Fig. \ref{figcontmpcsepea}
we show the cumulative distribution of separations of the sight lines
from the foreground QSOs in each pairing.  We show the 18 EA
coincidences and separately the sample from which they were selected,
which includes the associated absorbers.  We see that the EA
coincidences are preferentially seen at impact parameters $b <
1.2$~Mpc.  There is a 2\% chance of seeing a larger difference between
the two cumulative distributions in a random sample of 18 pairs.
Preference for small separations is not as strong as for the AA
coincidences seen in Fig. \ref{figcontmpcsepaa}.

\begin{figure}
  \includegraphics[width=84mm]{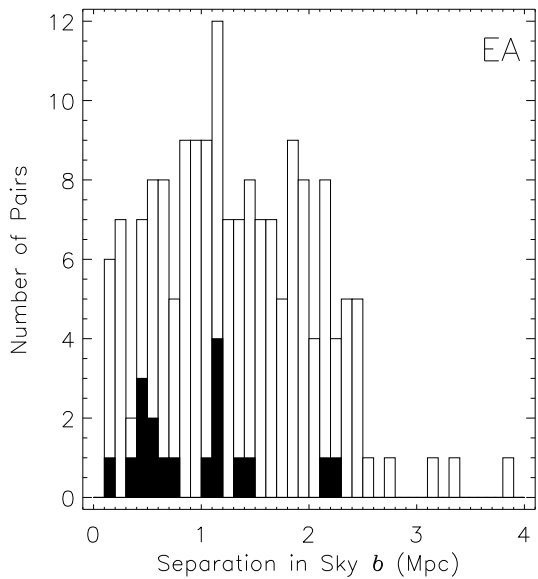}
  \caption{The distribution of impact parameters between the 18
  EA coincidences (black) and the sample from which they are drawn (clear).
  We show the distances between the sight lines measured in the plane of the sky.
  The clear histogram shows the distances from all foreground QSOs to the partner
  lines of sight, plus the distances from the background to the foreground QSOs
  when the \zem\ difference is $< 0.005$ and including all EA pairings.
}
  \label{figMpc_separationEA}
\end{figure}

In Fig. \ref{figMpc_separationEA}
we show the distribution of the separations
 of the EA systems compared to all QSO separations, now as a histogram.
 As for the AA coincidences in Fig. \ref{figMpc_separation}, we divide the
 histograms to estimate the probability of seeing absorption as a function of
 impact parameter. The probability of seeing an EA coincident absorption when a
 sight line passes a QSO is
1/6 (17\%)  at impact parameters of $b=100$ -- 200~kpc,
4/16 (25\%) at 200 -- 500~kpc,
4/40 (10\%) at 0.5 -- 1~Mpc,
7/43 (21\%) at 1 -- 1.5~Mpc and
2/62 (3\%)  at 1.5 -- 2.5~Mpc.
As for the AA coincidences,
these ratios are very much lower limits because we do not detect most weak metal
lines. For example, if we add the EAV
coincidences that extend out to $\dvea \pm 1000$~\kms\ we are less likely
to miss a coincidence that has a large \zem\ error, and the probabilities
rise by about 30/18.
On scales below 100~kpc where we have no sight lines,
\citet{bowen06a} found Mg~II in 4/4 sight lines at impact parameters of
26 -- 97~kpc,  while \citet[Fig. 1]{hennawi06a} see LLS in 3/3 cases at $<100$~kpc.
At 100 -- 200~kpc our probability of 1/6 for EA coincidences is not much lower
than the 3/8 from \citet[Fig. 1]{hennawi06a}, for their LLS absorbers.

We are surprised that the probability of detecting an
EA or EAV  coincidence is approximately constant from about 0.1 out to 1.5~Mpc.
We expect the probability to rise rapidly as the $b$-value decreases,
especially at 100 -- 300~kpc, because of galaxy clustering, as we saw for the
AA coincidences.
The distribution of absorbers around QSOs
(Fig. \ref{figMpc_separationEA}) is less centrally concentrated than the
 distribution of absorbers around absorbers (Fig. \ref{figMpc_separation}).

Why are absorbers be more concentrated around
absorbers than around QSOs? If anything we might expect more concentration around the
QSO host galaxies than around random galaxies. The difference is not
caused the the \zem\ errors because we are discussing a distribution in the
plane of the sky, not in redshift.
For the absorber-QSO correlation the \zem\ errors can cause systems to
have a larger \dzea\ or \dvea\ than we consider for an EA or EAV coincidence,
but this will not be a function of the impact parameter.
Fig. \ref{figMpc_separationEA} looks similar when drawn for both the EA + EAV
coincidences that extend out to 1000~\kms , but with a bit less concentration
around the QSO, presumably because we pick up more associated absorbers 
that are not strongly influenced by clustering.

We propose that the probability
of seeing an EA or EAV coincidence does not rise rapidly at low impact
parameters because the  absorbers nearest to the QSOs
are often destroyed by the UV from the QSOs, perhaps by photoevaporation
 \citep{hennawi07}.

\subsection{Redshift-space distribution of absorbers around QSOs}

\begin{figure*}
  \includegraphics[width=180mm]{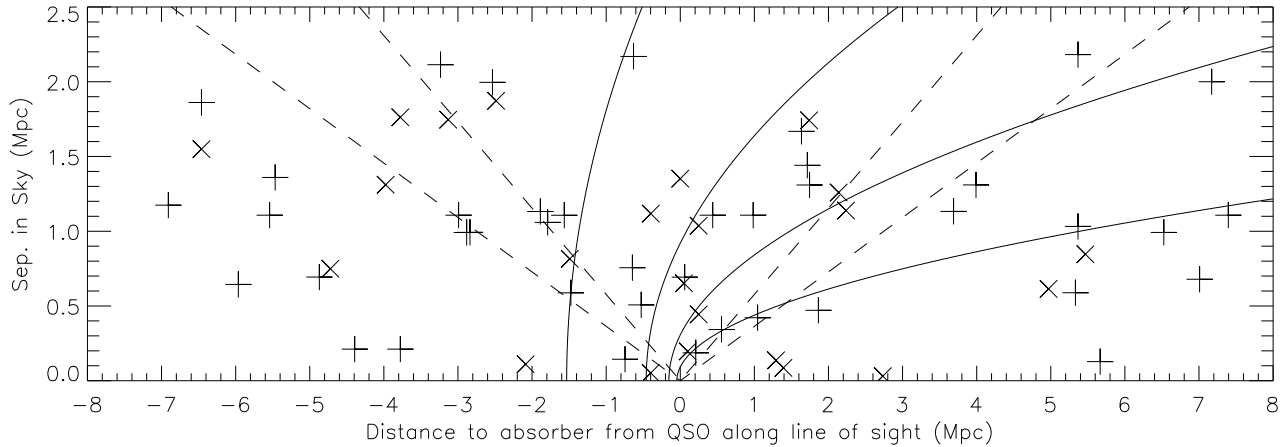}
  \caption{The distribution of absorbers near to QSOs that we place at the
  origin of the plot.
  Rays from the QSOs travelling towards us are horizontal lines
  extending to the right. The horizontal axis
  is derived from the velocity difference \dvea $ \sim c(z_{em}-z_{abs})/(1+z)$
  which is
  positive for absorbers that have a smaller redshift than the emission
  redshift of the partner QSO. We have converted the \dvea\ values into
  proper Mpc using the Hubble flow alone, ignoring peculiar velocities.
  The vertical axis shows the proper distance between the two QSOs in the
  plane of the sky, coming from their angular separation.
   We truncate the plot at $b=2.5$~Mpc because
  Fig. \ref{figMpc_separationEA} shows that the density of impact parameters
  for possible EA coincidences is nearly constant up to this value, but then
  declines by  a factor of several.
  We show the 41 absorbers (+ symbols) that we saw
  around 167 QSOs (the sum of the open histograms at $b < 2.5$~Mpc on Fig.
  \ref{figMpc_separationEA})  , and 23 absorbers (x)
   towards 146 QSOs from \citet{hennawi06a}. We use our \zabs\ for 3 cases where we
   report the same absorber in the same QSO,
    and 2 additional \zabs\ from their Table 1 are outside the area of this plot.
    The dashed lines show angles of 20 and 40 degrees from the horizontal axis, while
    the parabolas show the maximum distance the light travels from a QSO if that
    QSO switched on 0.3, 1, 3 or 10~Myr ago. Points below the dashed lines
    and to the right of the parabolas are more likely to be illuminated by the
    UV flux that we see from each QSO.Errors in the \zem\ values
    move points horizontally 2 - 4 or more Mpc.
  }
  \label{EAskysepMpc}

\end{figure*}

In Fig. \ref{EAskysepMpc}
we show how the absorption systems are distributed around the QSOs in two
dimensions. This Fig. is like Fig. \ref{figmpcsepscat}
 but  we now place the QSOs at the origin of the plot.
The $x$-axis is distance from the QSO to the absorber along the line-of-sight.
We obtain this from \zem\ -- \zabs , giving positive values when \zabs\
$<$ \zem . If there were no redshift errors or peculiar velocities, the
right hand of the plot would contain space nearer to us than the QSOs.
The $y$-axis is the
 impact parameter in the plane of the sky.
 As we discussed for the EA values in \S  \ref{secea}
 we include both foreground and background QSOs since many pairs have similar \zem\
 values and \zem\ errors can be large.
We show an 8~Mpc range for the $x$-axis because
the UV from the QSOs is expected to a factor of a few larger than the UVB
out to about 4~Mpc for this sample of QSOs \citep{kirkman07a}.

We include on the plot the absorption systems from Table 1 of
\citet{hennawi06a}.  The systems are Lyman limits and damped \lya\
lines within 1500~\kms\ of the \zem\ values of 149 foreground QSOs at
projected separations of 0.031 -- 2.4~Mpc.  We are interested in the
distribution of the absorbers in velocity, and not in the absolute
rate of detection, and hence we are not concerned that some of their
systems might not have high H~I column densities.

Errors in the \zem\ values will often be several hundred \kms\ and
hence the $x$-axis location of many points will have large random
errors.  As we discussed in \S \ref{secea} the narrow width of \dvea\
values for the coincident EA systems suggests that some QSOs have \zem\
errors of $<525$~\kms . However, we also expect larger
errors for many QSOs.  \citet{hennawi06a} list errors of 300, 500,
1000 and 1500~\kms , depending on the emission lines seen, for the
\zem\ values that they provide and we use.

The interpretation of Fig. \ref{EAskysepMpc} is complicated by the
presence of background QSOs on the left hand side of the plot. The
spectra of these QSOs sample only part of the line-of-sight past the
foreground QSO, the part to the right of the QSOs position. We have 5
QSOs near zero, 3 at $-$0.5 to $-$4~Mpc and 8 at $-$4 to $-$8~Mpc.
These QSOs are all from our sample, since we allow EA systems to come
from QSOs with similar or identical \zem\ values.  \citet{hennawi06a}
did not consider QSOs within 8~Mpc of their foreground QSOs.
There few QSOs have a small effect on the total sample of 313 QSOs.

We discussed in \S \ref{secinfall} the effects of systematic infall
velocities for the absorber-absorber correlation. That discussion also
applies to absorbers around QSOs, with the difference that the infall
velocities may be different, though not necessarily higher \citep{slosar06},
if the QSOs reside in more massive halos than typical galaxies.
\citet{croom05a} estimated QSO halo masses of $4.2 \pm
2.3 \times 10^{12}$ solar masses in the 2QZ sample at all
redshifts. Less directly, \citet{kim07a} use the distribution of H~I
absorption seen in background QSOs to estimate the masses of nearby
foreground QSO halos.  They find a mean mass of $\log M =
12.48^{+0.53}_{-0.89}$ in solar units for QSOs at $z=3$ with absolute
G-band magnitude $-27.5$, a factor of 20 above the mass of LBGs.

\citet[Fig. 16]{adelberger04a} shows corrections for systematic infall
towards QSOs plus random peculiar velocities. These corrections are
typically 0.2 -- 0.4~Mpc at $z=3$ for the distances out to 4 Mpc,
implying (Hubble flow) velocities of 40 -- 80~\kms . Systematic
corrections of this size have little effect on our plots. As we
mentioned in \S \ref{secinfall} \citet[Fig. 2]{kim07a} use simulations to
estimate how the mean radial infall velocities increase with the mass
of the halo.  These infall velocities are also modest, except for
their most massive halo with $ 5.1 \times 10^{12}$ solar masses that
gave peak infall velocities of 270~\kms\ at 1.2~Mpc.

\subsection{Is the UV emission from QSOs is beamed?}

The first popular explanation for why we see few transverse absorbers
along the los to individual QSOs is that the UV from QSOs illuminates
the los but not the transverse directions. Absorbers in the UV beam
and near to the QSO are destroyed or photo-evaporated
\citep{hennawi07} so they no longer make absorption lines that we can
see.

Let us put aside the \zem\ errors, peculiar velocities, and
complications of QSOs near to QSOs and look for evidence that the QSO
UV is beamed. Let us assume that the UV luminosities of the QSOs have
not changed for $>17$~Myr so they can illuminate all of
Fig. \ref{EAskysepMpc}.  Assume that the UV from all QSOs is emitted
only inside a pair of coaxial cones, one opening towards the Earth and
the other away, and with their vertices at the QSO.  If the QSO UV
destroys the transverse absorbers that we do not see in the los, then
we expect absorbers will be less common inside the cones of UV
radiation.

We show on Fig. \ref{EAskysepMpc} dashed lines for cone apex
half-angles of $\theta =20$ and 30 degrees measured up from the
horizontal axis, the line-of-sight to the Earth.
We see numerous absorbers between the horizontal axis and the dashed lines, and
to first order the distribution of absorbers around the QSOs looks isotropic.

However, three issues persuade us that the plot is compatible with
radiation confined to cones where these is a lower density of absorbers.
First, the UV may
destroy absorbers out to at most 4~Mpc from the QSO at the origin. The
precise distance will depend on the absorber density and structure,
the ions we see, the sensitivity of the spectrum, and the QSO
luminosity.  The density of points is lower inside the 20 degree cone
out to about 4~Mpc.  Second, \zem\ errors will more often move points
into the cone than outside it when the impact parameter is small. This
can account for the absorbers near the $x$-axis. Third, line-of-sight
to the Earth can be anywhere in the UV illuminated cone, including at
the edge of the cone.  The probability of looking into a cone at a
some angle to the cone axis is proportional to that angle. If a
line-of-sight is along one edge of the cone, then we expect absorbers
below the dashed line down to the horizontal axis, on both sides of
the origin, because the figure ignores the position angle on the sky
from the QSO at the origin to the absorber.

\subsection{QSO lifetimes}

The second popular explanation for why we see no transverse absorbers
along the los to individual QSOs is that the QSOs have a short life
time of $\sim 1$~Myr. By life time we mean how long has a QSO has had
a UV luminosity similar to that seen today.

If the QSOs emit isotropically but were less luminous in the past,
then we expect a lack of absorbers confined to regions illuminated
while the QSOs are luminous. These regions are bounded by parabolic
surfaces with the QSO at the focus and the apex of the parabola
farther away from us by half the distance that light can travel in the
QSO life. All rays leaving a QSO at a given time and reflecting on
this parabola reach us at the same time.
\citet[Fig. 3]{adelberger04a} and \citet[Fig. 1]{visbal07a} show these
parabolic surfaces that have the equation
\begin{equation}
y=\sqrt{2d(x+d/2)},
\end{equation}
where $x$ is the horizontal axis of our plots, along the line-of-sight
to Earth and $y$ is the impact parameter in the plane of the sky.
With $x$ and $y$ in proper Mpc, the $d$ in the equation is the delay
measured in Mpc which corresponds to a time delay t in Myr of $t ({\rm
  Myr})=3.2617d ({\rm Mpc})$. We show delays of $t = 0.3$, 1, 3 and 10
Myr, corresponding to $d = 0.09198, 0.3066$, 0.9198 and 3.066 Mpc.
All points in space that are to the right of the parabola for a given
life time will have been exposed to the QSO flux.

A key point is that these parabolas are strongly asymmetric about the
QSO while the bi-cone beaming hypothesis gives a symmetric pattern.
We should then be able to distinguish beaming from QSO life time by
examining the distribution of absorbers about the QSOs. We again
assume that the QSOs UV is only able to destroy absorbers out to some
distance, and we ignore the infall velocities that are probably much
smaller than the errors on the \zem\ values.

There are many points to the right of the parabolas where we would
expect none if the QSOs have short lives, and their UV destroys all
nearby absorbers.  We expect some absorbers inside the parabolas for
two of the three reasons already mentioned for the beamed bi-cone
model.  The QSO UV will only destroy absorbers out to at most 4~Mpc,
depending on the QSO and absorber, and the \zem\ errors will move many
points horizontally.  Hence the plot could be compatible with life
times of about 0.3~Myr, leading to the low density of points to the
right of the smallest parabola and extending out to 5~Mpc. We do not
advocate this explanation because a lifetime of 0.3~Myr is at the
lower limit of other estimates.  \citet{croom05a} estimate QSO active
life times of $4 \times 10^6$ -- $6 \times 10^8$ yr at $z \sim 2$.
\citet{adelberger05a} argue that the lack of dependence of the
QSO-galaxy correlation length on QSO luminosity implies that lower
luminosity AGN live longer.

\section{BAL Systems}

So far we have ignored the BAL systems. We now present two figures
that explore potential coincidences in BAL systems between the paired
lines-of-sight. We do not expect to see any statistically significant
excess coincidences over the random distributions, because the BAL
winds are believed to extend much less than a Mpc from the QSOs.

\begin{figure}
  \includegraphics[width=84mm]{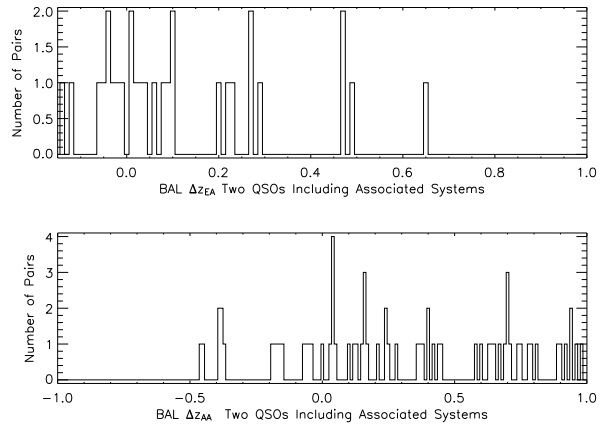}
  \caption{The \dzaa\ and \dzea\ distributions for the 34 BAL systems, with \dz\
   bins of 0.01.
   }
  \label{figbals}
\end{figure}

In Fig. \ref{figbals}
we show the distribution of BAL system redshifts relative to the
redshifts of the partner QSO. We see that about 50\% of the
differences are confined to $0.1 < \dzea < 0.1$.  In comparison
\ref{figangdistsep} showed that the non-BAL absorbers show an excess
that is more confined around $\dzea = 0$. The \dzea\ values for the
BAL systems may have a wider dispersion in part from the difficulty in
selecting a single \zabs\ for a BAL system that shows many components
spread over a large velocity range. in addition the BAL systems might
ignore the partner QSO and the excess at $0.1 < \dzea < 0.1$ may come
from the similarity of the two \zem\ values combined with the tendency
of BAL systems to arise at \vabs $\sim 0$.  To make progress we would
prefer a sample in which the background QSO \zem\ was much larger than
the foreground value.

In the lower panel of Fig. \ref{figbals} we look for coincidences
between BAL systems in one sight line with BAL systems in the other
sight line.  There is no excess near to \dzaa $\sim 0$, except to the
occurrence of 4 pairings at \dzaa $ \sim 0.03$.

\section{Discussion}
\label{sdiscn}

In this paper we report the first sample of absorption systems showing
metal lines in a large sample of pairs of QSOs that are close in the
sky. The QSO pairs are separated by 0.1 -- 2.5~Mpc at $0.2 < z < 4$
and typically 1~Mpc at $z = 2$.  We found 691 absorption systems in
the spectra of 310 QSOs in 170 pairings.  We used medium resolution
spectra (FWHM $\sim 170$~\kms ) and saw lines with typical rest frame
equivalent widths \wrest = 0.5~\AA , with a 90\% range of 0.2 --
1.3~\AA . We summarise this work under several headings as follows.

\subsection{Line-of-sight Associated Absorption}

The sample of absorption systems that we use does not conform to some
minimum \wrest\ limit.  The top panel of Fig. \ref{figcloseup} shows
the strong tendency of the absorbers in a given spectrum to have
\zabs\ $\simeq $ \zem . We see many absorption systems out to $\vabs =
3000$~\kms\ and continuing to approximately 10,000~\kms . This
distribution is largely determined by the ease of finding absorption
near to the emission lines where the SNR is high.  While our
distribution is very similar to that of \citet{weymann81} for a sample
like ours that does not employ an equivalent width cutoff, it is also
possible that the lower luminosity QSOs presented here have different
intrinsic absorption than that seen in the samples of high
luminosity QSOs analysed in the 1980s.

In Fig. \ref{figsepzem} we showed that many of the QSOs in our
sample have similar emission redshifts to their partners. When we compare the absorption
redshifts that we see in one spectrum to those in the partner, we
often find accidental associations favoured by the interplay of the
similar \zem\ values with the excess of absorbers with \vabs $\sim $
\zem\ in each sight line.

\subsection{AA associations of absorbers with absorbers in the paired sight lines}

The correlation of absorbers about absorbers describes the
distribution of metals around galaxies and the clustering of those
galaxies.  We learn about the distribution of the pair-wise velocities
of the halos that cause the absorbers and we obtain limits on the
peculiar velocities of the gas relative to the halos.  Our absorber
redshift errors of $\sim 23$~\kms\ are about ten times smaller than
the typical errors for galaxies. This allows us to study the redshift
space distortion on rather small scales in a sample that is tiny by
galaxy standards.

In the bottom two panels of Fig. \ref{figangdistsep} and
\ref{figcloseup} we see the highly significant excess of absorbers at
the same redshift as an absorber in a second sight line.  We see 17
cases where absorption in one line-of-sight is within 200~\kms\ of
absorption in the second line-of-sight.  Twelve of these 17
coincidences are both $> 3000$~\kms\ or approximately 15~Mpc, from
their QSOs.  This is the first secure detection of coincident
absorption in two sight lines separated by about 1~Mpc.

In Fig. \ref{figMpc_separation} we saw that the incidence of AA
coincidences is very high for small impact parameters and drops
rapidly with increasing separation.  In \S \ref{secaaimpact} we found
that the probability of finding an absorber in the partner spectrum is
at least $\approx 50$\% at $<100$~kpc, declining rapidly from 23\% at
100 -- 200~kpc to 0.7\% by 1 -- 2~Mpc. These probabilities are very
much lower limits because high resolution spectra with high SNR and
full wavelength coverage will show factors of several times more
absorbers than do our moderate resolution spectra with partial
wavelength coverage.

The rapid drop in the rate of coincidences with separation explains
why we are able to detect a significant number of coincidences in 170
pairings of QSOs while \citet{coppolani06} did not see any significant
correlation in 139 C~IV systems towards 32 pairs of QSOs, except for
an over-density of C~IV in front of a group of 4 QSOs.  Their QSOs had
a mean separation of $>2$ arcmin which is too large.

We can explain the distribution of absorbers around absorbers if each
arises in a typical galaxy. Galaxy clustering accounts for the rapid
rise in the number of coincidences with small separations. The
distribution of points in Fig. \ref{simaa} shows the expected
distribution of absorbers around absorbers from galaxy clustering, and
including our errors on the \zabs\ values that are $\sim 23$~\kms .

When we add the expected systematic infall velocities of $\sim 70 -
100$~\kms\ we move the points towards a caustic shape with
concentrations near both the $x$ and $y$-axes.  We assume that the
infall velocities are given by the simulations of \citet{kim07a} for a
distribution of halo masses $\propto 1/M$.  This distribution in
Fig. \ref{simaainfall} looks more like our data.  Other halo
mass distribution give similar results because the infall velocities
are not much different.

\subsection{Absorbers arise in a cold population: blue galaxies, not in winds}

The absorber-absorber correlation is sensitive to the redshift-space
distortions caused by random peculiar velocities that make the
distribution of absorber-absorber separations elongated along the
line-of-sight. We see a hint in Figs. \ref{figmpcsepscat} and
\ref{figmpcsepscat25} that points are more strongly clustered about
the origin in the sky ($y$) direction but less so in the redshift
($x$) direction. This is the usual appearance of redshift-space
distortion.  However, when we calculated the likelihood of obtaining
the data as a function of $\sigma _{12}$ and find that the preferred
value is zero, with a $1 \sigma $ upper limit of 100~\kms\ and a
$2\sigma $ limit of 300~\kms .  The pair wise velocities are expected
to be at least $\sim 240$~\kms\ from galaxy data, and hence our
absorbers arise in a very ``cold'' population with minimal random
velocities.  We conclude that the absorbers arise in the blue rather
than the red samples of galaxies discussed in the references of \S
\ref{secest}.

We expect that the distribution of halo masses sets both the infall
velocity field and the pair-wise random velocities $\sigma _{12}$. We
expect a consistent pairing on infall velocities and pair-wise
velocities for a given halo mass distribution, with smaller velocities
for smaller masses.  In Fig \ref{lnlpair} we saw that changing the
halo mass distribution from $\propto M^{-1}$ to $\propto M^{-2}$ had
no effect on the allowed $\sigma _{12}$ values because low mass halos
dominate in both cases.  The \citet{kim07a} simulations show only a
small decrease in the infall velocities for decreasing halo mass below
$\sim 2.9 \times 10^{10}$ solar masses. Hence it is unclear if even
much lower halo mass would allow the data to be consistent with
significantly larger velocity dispersions, from either pair-wise random
motions or from winds.  We also found that a model with more high mass
halos gave similar results to a model with no infall, and that both
give lower likelihoods for the data.

Our data set is not compatible with most absorbers arising in winds
that flow quickly out from galaxies.  Absorption in gas flowing out
from galaxies at a mean velocity of 250~\kms\ would produce more
elongation than we see. We conclude that the absorbing gas does not
arise in outflowing winds.  The winds seen by \citet{adelberger05} in
LBG spectra are not representative of absorption systems that we
see. Either the winds are confined to LBGs, or they do not extend to
$>40$~kpc with large velocities, or they do not produce absorption we
can detect.  We are compatible with 1/3 of strong metal line systems
coming from fast winds from LBGs.

In a similar vein, \citet{rauch01a} found the the \lya\ absorption in
two pairs of QSOs separated by 86~pc at $z \sim 3.3$ was very similar,
implying that strong winds blowing for a substantial fraction of the
Hubble time fill less than 20\% of the volume of the universe.

\subsection{Size and Effect of \zem\ errors}

Before we discuss the correlation of absorbers around QSOs, we should
summarise what we learnt about the distribution of \zem\ errors.

We know that \zem\ values are often in error by many hundreds of \kms
.  The SDSS QSOs in our sample have \zem\ values that are intended to
correct for the typical systematic errors in \zem\ values obtained
from the common emission lines to give systemic redshifts. We see from
Fig. \ref{figcloseup}(a) that our sample contains about 15 QSOs that
probably have \zem\ value too small by $\sim 1000$~\kms\ or more,
while others will also have such errors and by chance do not show
\zabs\ at negative \vabs\ velocities.

However, we also see in Fig. \ref{figcloseup}(b) that about 12 of our
QSOs have \zem\ values with errors of $< 400$~\kms . In
Fig. \ref{figcloseup}(b) we saw that the excess could be represented
by a gaussian with a mean $-150$~\kms\ and a $\sigma \sim $525~\kms
. The small absolute value for the mean implies that the redshifts
that we use do not have a large systematic error relative to systemic
values. The small $\sigma $ implies that the \zem\ values have small
random differences from the systemic redshifts, especially because
much of the measured $\sigma $ will come from the pair-wise random
velocities of the absorbing galaxies relative to the QSO hosts.  Since
EA coincidences are rare, we can conclude that many of our QSOs have
such small \zem\ errors. This conclusion is not very secure because
there are few relevant pairings and hence a larger sample could show a
wider or displaced peak.

We are lead to speculate that overall the \zem\ errors sample at least
two populations; small errors for some QSOs with favourable emission
lines and large errors in other cases. The absorption redshifts in the
spectra of the partner QSOs may often be better measurements of the
systemic velocities of the QSO hosts than are the \zem\ values
themselves!

\subsection{QSO host population}

We can use the distribution of the velocity differences in the
QSO-absorber pairings to comment on the QSO host population. The
velocity differences include terms from the \zem\ errors and the
pair-wise velocity dispersions of QSOs relative to absorbers.  We know
that both terms can be as large or larger than the small dispersion of
$\sigma \sim 525$~\kms\ that we see.  This suggests that the QSOs,
like the absorbers, are in hosts with a low pair-wise velocity
dispersion, favouring blue over red galaxies.  With the current data
and analysis, this conclusion is speculation on what might be done
in the near future.  The 525~\kms\ is not well determined because it
is based on few pairing. The value may also be biased because the
distribution of absorbers around QSOs appears to be anisotropic
because the QSOs destroy some absorbers.

\subsection{EA association of absorbers with QSOs in the paired sight lines}

We see excess absorption at redshifts similar to the \zem\ values of
QSOs close in the sky. We see these absorbers in the bins near zero 
in Fig. \ref{figangdistsep}(b).  We suspect that some
of this excess in not caused by ordinary line-of-sight associated
absorbers that are selected by chance because the two QSOs have
similar \zem\ values. We suspect this because the width of the peak in
Fig.  \ref{figcloseup}(b) may be narrower than the distribution of los
associated absorbers in Fig.  \ref{figcloseup}(a), and because it
would be consistent with the findings of \citet{bowen06a} and
\citet{hennawi06a}.  The EA pairings select absorbers that are
more concentrated near the QSO host velocities than are the los
associated absorbers as a whole.  Fig. \ref{figcloseup}(b) shows that
the excess of absorbers around QSOs can be represented by a gaussian
with a $\sigma \sim 525$~\kms\ that is a much wider distribution than
we saw for the clustering of absorbers about absorbers seen in
Fig. \ref{figcloseup}(d) and (e), because of the \zem\ errors are much
larger than the \zabs\ errors and because the QSOs may destroy the
nearest absorbers.

In \S 6 we saw that the EA absorbers have absorption lines with
\wrest\ values that are typical for our sample as a whole. Most of the
systems show C~IV lines and we see no excess of N~V, though we did not
explicitly search for N~V in each case.

We discussed the spatial distribution of the absorbers around the QSOs
in \S \ref{secspatialea}. In Fig. \ref{figcontmpcsepea} we saw that
the EA coincidences are more concentrated to pairings with small
impact parameters than are random absorbers. Only 2\% of random
samples of absorbers would have a stronger preference for small impact
parameters. On the other hand, Fig. 23 shows that the probability of
seeing an EA coincidence is relatively constant from 0.1 --
1.5~Mpc. This is a surprise because we know that galaxy clustering
should give a rapid increase in probability of a coincidence for
smaller impact parameters, as we saw for AA coincidences. We proposed
that we do not see the rapid rise in the probability of a coincidence
at small impact parameters because QSOs destroy some near by
absorbers.

\subsection{Is the distribution of absorbers around QSOs anisotropic?}

We see signs of anisotropy in the distribution of absorbers around
QSOs, both in redshift differences, in the distribution of impact
parameters and in prior data. The arguments include the following.

1. \citet{bowen06a} found 4 Mg~II systems that established the
existence of transverse associated absorbers that are not seen in the
line of sight. \citet{hennawi06a} reached a similar conclusion in
their sample of 149 background QSOs that showed 27 LLS and DLAs near
to foreground QSOs.

2. We see transverse absorption that is centred on the \zem\ of the
partner QSOs (Fig. \ref{figcloseup}b) and more tightly centred than
are the usual los associated absorbers.  Fig. \ref{figcontmpcsepea}
showed the same in the plane of the sky.  In redshift and separately
 in the plane of the sky, the probability of 
seeing at least this concentration around the QSOs by accident is 2\%.

3. We saw that the distribution of absorbers around QSOs is much less
concentrated to low impact parameters, than are absorbers around
absorbers. This is not due to \zem\ errors, but might be because QSOs
destroy the nearest absorbers.  This does not mean there is
anisotropy, but it implies a process that can cause anisotropy.

4. In a parallel study \citet{kirkman07b} using a subset of this
sample of QSOs, see no sign of the H~I transverse proximity effect.
The metal lines studied here provide a very sparse sampling of the
space around the QSOs. In contrast, the H~I provides more samples
along a given line of sight.  However, they also do not detect the los
proximity effect. The lack of the los proximity might be caused by
some combination of several effects; systematic \zem\ errors, large
random \zem\ errors, higher density of gas near to the QSOs cancelling
the QSO UV, and cancellation by extra absorption from the \lya\ lines
of the associated C~IV systems that we see in
Fig. \ref{figcloseup}(a).  If the UV is the cause, the UV flux in the
transverse direction is less than that along the los, by about a
factor of 10 -- 100. The \lyaf\ data sampling is dense enough to rule out
UV emission restricted to
irregular patches on the sky, including the line of sight that we see,
but without any single well defined beam.

5. The distribution of absorbers around QSOs (Fig. \ref{EAskysepMpc})
is to first order isotropic, with only a hint of the pattern expected
if QSO radiation were beamed.

If the absorber distribution is anisotropic, then the obvious
explanations include beamed UV emission \citep{barthel90} and short
life times.  We can not yet make a distinction with the data presented
here, though improved \zem\ values will help. If the QSO UV were
anisotropic the total emissivity in UV photons per Mpc$^3$ is
unchanged, since this is set by the number of QSOs observed in any
random direction.  However there will be more QSOs luminous at a given
time, and their radiation is concentrated in a way that will change
the standard picture of ionization spreading out from QSOs in roughly
of spherical bubbles that expand and overlap over time.  

\section*{Acknowledgments}

The data presented herein were obtained using the Kast spectrograph on
the Lick Observatory 3m-Shane telescope, and LRIS spectrograph on the
Keck-I telescope.  The W.M. Keck Observatory is operated as a
scientific partnership among the California Institute of Technology,
the University of California and the National Aeronautics and Space
Administration and was made possible by the generous financial support
of the W.M. Keck Foundation. We are exceedingly grateful for the help
we receive from the staff at both observatories. We recognise and
acknowledge the very significant cultural role and reverence that the
summit of Mauna Kea has always had within the indigenous Hawaiian
community.  We are extremely grateful to have the opportunity to
conduct observations from this mountain. Former UCSD students John
O'Meara and Nao Suzuki helped to obtain some of the spectra used in
this paper. We thank Don Schneider, Chuck Steidel, Jason Prochaska,
Jeffrey Newman, Jeff Cooke, Mat Malkan and Peter Bartel for important
discussions. Our work was inspired by the
pioneering work of Chuck Steidel and his collaborators.
Angela Chapman was funded by the NSF REU program grant
to the Physics Dept. at UCSD.  This work was funded in part by NSF
grants AST-0098731 and 0507717 and by NASA grants NAG5-13113.

\bibliographystyle{mn2e}
\bibliography{archive}

\begin{thebibliography}{}

\bibitem[\protect\citeauthoryear{{Adelberger}}{{Adelberger}}{2004}]{adelberger%
04a}
{Adelberger} K.~L.,  2004, \apj, 612, 706

\bibitem[\protect\citeauthoryear{{Adelberger}}{{Adelberger}}{2005}]{adelberger%
05b}
{Adelberger} K.~L.,  2005, in {Williams} P.,  {Shu} C.-G.,   {Menard} B.,  eds,
  IAU Colloq. 199: Probing Galaxies through Quasar Absorption Lines {Galaxies,
  intergalactic absorption lines, and feedback at high redshift}.
pp 341--348

\bibitem[\protect\citeauthoryear{{Adelberger}, {Shapley}, {Steidel}, {Pettini},
  {Erb} \& {Reddy}}{{Adelberger} et~al.}{2005}]{adelberger05}
{Adelberger} K.~L.,  {Shapley} A.~E.,  {Steidel} C.~C.,  {Pettini} M.,  {Erb}
  D.~K.,    {Reddy} N.~A.,  2005, \apj, 629, 636

\bibitem[\protect\citeauthoryear{{Adelberger} \& {Steidel}}{{Adelberger} \&
  {Steidel}}{2005}]{adelberger05a}
{Adelberger} K.~L.,  {Steidel} C.~C.,  2005, \apj, 630, 50

\bibitem[\protect\citeauthoryear{{Adelberger}, {Steidel}, {Kollmeier} \&
  {Reddy}}{{Adelberger} et~al.}{2006}]{adelberger06a}
{Adelberger} K.~L.,  {Steidel} C.~C.,  {Kollmeier} J.~A.,    {Reddy} N.~A.,
  2006, \apj, 637, 74

\bibitem[\protect\citeauthoryear{{Adelberger}, {Steidel}, {Pettini}, {Shapley},
  {Reddy} \& {Erb}}{{Adelberger} et~al.}{2005}]{adelberger05c}
{Adelberger} K.~L.,  {Steidel} C.~C.,  {Pettini} M.,  {Shapley} A.~E.,  {Reddy}
  N.~A.,    {Erb} D.~K.,  2005, \apj, 619, 697

\bibitem[\protect\citeauthoryear{{Alcock} \& {Paczynski}}{{Alcock} \&
  {Paczynski}}{1979}]{alcock79}
{Alcock} C.,  {Paczynski} B.,  1979, Nature, 281, 358

\bibitem[\protect\citeauthoryear{{Antonucci}}{{Antonucci}}{1993}]{antonucci93}
{Antonucci} R.,  1993, AAp, 31, 473

\bibitem[\protect\citeauthoryear{{Barthel}}{{Barthel}}{1989}]{barthel89}
{Barthel} P.~D.,  1989, \apj, 336, 606

\bibitem[\protect\citeauthoryear{{Barthel}, {Tytler} \& {Thomson}}{{Barthel}
  et~al.}{1990}]{barthel90}
{Barthel} P.~D.,  {Tytler} D.~R.,    {Thomson} B.,  1990, \aap, 82, 339

\bibitem[\protect\citeauthoryear{{Best}}{{Best}}{2007}]{best07a}
{Best} P.~N.,  2007, New Astronomy Review, 51, 168

\bibitem[\protect\citeauthoryear{{Bowen}, {Hennawi}, {M{\'e}nard}, {Chelouche},
  {Inada}, {Oguri}, {Richards}, {Strauss}, {Vanden Berk} \& {York}}{{Bowen}
  et~al.}{2006}]{bowen06a}
{Bowen} D.~V.,  {Hennawi} J.~F.,  {M{\'e}nard} B.,  {Chelouche} D.,  {Inada}
  N.,  {Oguri} M.,  {Richards} G.~T.,  {Strauss} M.~A.,  {Vanden Berk} D.~E.,
   {York} D.~G.,  2006, \apj, 645, L105

\bibitem[\protect\citeauthoryear{{Boyle}, {Smith}, {Shanks}, {Croom} \&
  {Miller}}{{Boyle} et~al.}{1997}]{boyle97a}
{Boyle} {Smith} R.~J.,  {Shanks} T.,  {Croom} S.~M.,    {Miller} L.,  1997,
  Proc. IAU Symp. 183, `Fundamental Cosmological Parameters in Kyoto' (Kyoto
  1997)

\bibitem[\protect\citeauthoryear{{Broadhurst}, {Ellis}, {Koo} \&
  {Szalay}}{{Broadhurst} et~al.}{1990}]{broadhurst90a}
{Broadhurst} T.~J.,  {Ellis} R.~S.,  {Koo} D.~C.,    {Szalay} A.~S.,  1990,
  Nature, 343, 726

\bibitem[\protect\citeauthoryear{{Chelouche}, {M{\'e}nard}, {Bowen} \&
  {Gnat}}{{Chelouche} et~al.}{2007}]{chelouche07a}
{Chelouche} D.,  {M{\'e}nard} B.,  {Bowen} D.~V.,    {Gnat} O.,  2007, ArXiv
  e-prints 0706.4336, 706

\bibitem[\protect\citeauthoryear{{Chen}, {Lanzetta} \& {Webb}}{{Chen}
  et~al.}{2001}]{chen01b}
{Chen} H.,  {Lanzetta} K.~M.,    {Webb} J.~K.,  2001, \apj, 556, 158

\bibitem[\protect\citeauthoryear{{Churchill}, {Kacprzak}, {Steidel} \&
  {Evans}}{{Churchill} et~al.}{2007}]{churchill07a}
{Churchill} C.~W.,  {Kacprzak} G.~G.,  {Steidel} C.~C.,    {Evans} J.~L.,
  2007, \apj, 661, 714

\bibitem[\protect\citeauthoryear{{Coil}, {Newman}, {Croton}, {Cooper}, {Davis},
  {Faber}, {Gerke}, {Koo}, {Padmanabhan}, {Wechsler} \& {Weiner}}{{Coil}
  et~al.}{2007}]{coil07a}
{Coil} A.~L.,  {Newman} J.~A.,  {Croton} D.,  {Cooper} M.~C.,  {Davis} M.,
  {Faber} S.~M.,  {Gerke} B.~F.,  {Koo} D.~C.,  {Padmanabhan} N.,  {Wechsler}
  R.~H.,    {Weiner} B.~J.,  2007, ArXiv e-prints, 2007arXiv0708.0004C

\bibitem[\protect\citeauthoryear{{Cooke}, {Wolfe}, {Gawiser} \&
  {Prochaska}}{{Cooke} et~al.}{2006}]{cooke06a}
{Cooke} J.,  {Wolfe} A.~M.,  {Gawiser} E.,    {Prochaska} J.~X.,  2006, \apj,
  652, 994

\bibitem[\protect\citeauthoryear{{Coppolani}, {Petitjean}, {Stoehr},
  {Rollinde}, {Pichon}, {Colombi}, {Haehnelt}, {Carswell} \&
  {Teyssier}}{{Coppolani} et~al.}{2006}]{coppolani06}
{Coppolani} F.,  {Petitjean} P.,  {Stoehr} F.,  {Rollinde} E.,  {Pichon} C.,
  {Colombi} S.,  {Haehnelt} M.~G.,  {Carswell} B.,    {Teyssier} R.,  2006,
  \mnras, 370, 1804

\bibitem[\protect\citeauthoryear{{Cristiani}, {D'Odorico}, {D'Odorico},
  {Fontana}, {Giallongo} \& {Savaglio}}{{Cristiani}
  et~al.}{1997}]{cristiani97b}
{Cristiani} S.,  {D'Odorico} S.,  {D'Odorico} V.,  {Fontana} A.,  {Giallongo}
  E.,    {Savaglio} S.,  1997, \mnras, 285, 209

\bibitem[\protect\citeauthoryear{{Croft}}{{Croft}}{2004}]{croft04a}
{Croft} R.~A.~C.,  2004, \apj, 610, 642

\bibitem[\protect\citeauthoryear{{Croom}, {Boyle}, {Shanks}, {Smith}, {Miller},
  {Outram}, {Loaring}, {Hoyle} \& {da {\^ A}ngela}}{{Croom}
  et~al.}{2005}]{croom05a}
{Croom} S.~M.,  {Boyle} B.~J.,  {Shanks} T.,  {Smith} R.~J.,  {Miller} L.,
  {Outram} P.~J.,  {Loaring} N.~S.,  {Hoyle} F.,    {da {\^ A}ngela} J.,  2005,
  \mnras, 356, 415

\bibitem[\protect\citeauthoryear{{Crotts}}{{Crotts}}{1989}]{crotts89}
{Crotts} A.~P.~S.,  1989, \apj, 336, 550

\bibitem[\protect\citeauthoryear{{Crotts}, {Bechtold}, {Fang} \&
  {Duncan}}{{Crotts} et~al.}{1994}]{crotts94}
{Crotts} A.~P.~S.,  {Bechtold} J.,  {Fang} Y.,    {Duncan} R.~C.,  1994, \apj,
  437, L79

\bibitem[\protect\citeauthoryear{{Crotts}, {Burles} \& {Tytler}}{{Crotts}
  et~al.}{1997}]{crotts97a}
{Crotts} A.~P.~S.,  {Burles} S.,    {Tytler} D.,  1997, \apj, 489, L7

\bibitem[\protect\citeauthoryear{{Crotts} \& {Fang}}{{Crotts} \&
  {Fang}}{1998}]{crotts98}
{Crotts} A.~P.~S.,  {Fang} Y.,  1998, \apj, 502, 16

\bibitem[\protect\citeauthoryear{{Davis} \& {Peebles}}{{Davis} \&
  {Peebles}}{1983}]{davis83}
{Davis} M.,  {Peebles} P.~J.~E.,  1983, \apj, 267, 465

\bibitem[\protect\citeauthoryear{{Di Matteo}, {Springel} \& {Hernquist}}{{Di
  Matteo} et~al.}{2005}]{dimatteo05a}
{Di Matteo} T.,  {Springel} V.,    {Hernquist} L.,  2005, Nature, 433, 604

\bibitem[\protect\citeauthoryear{{Dobrzycki} \& {Bechtold}}{{Dobrzycki} \&
  {Bechtold}}{1991a}]{dobrzycki91a}
{Dobrzycki} A.,  {Bechtold} J.,  1991a, \apj, 377, L69

\bibitem[\protect\citeauthoryear{{Dobrzycki} \& {Bechtold}}{{Dobrzycki} \&
  {Bechtold}}{1991b}]{dobrzycki91b}
{Dobrzycki} A.,  {Bechtold} J.,  1991b, in {Crampton} D.,  ed., The Space
  Distribution of Quasars Vol.~21 of Astronomical Society of the Pacific
  Conference Series, {The influence of quasars on the Lyman-alpha forest}.
pp 272--279

\bibitem[\protect\citeauthoryear{{D'Odorico}, {Petitjean} \&
  {Cristiani}}{{D'Odorico} et~al.}{2002}]{dodorico02}
{D'Odorico} V.,  {Petitjean} P.,    {Cristiani} S.,  2002, \aap, 390, 13

\bibitem[\protect\citeauthoryear{{Faucher-Giguere}, {Lidz}, {Zaldarriaga} \&
  {Hernquist}}{{Faucher-Giguere} et~al.}{2007}]{faucher07a}
{Faucher-Giguere} C.~.,  {Lidz} A.,  {Zaldarriaga} M.,    {Hernquist} L.,
  2007, Astrophysics e-prints, astro-ph/0701042

\bibitem[\protect\citeauthoryear{{Fernandez-Soto}, {Barcons}, {Carballo} \&
  {Webb}}{{Fernandez-Soto} et~al.}{1995}]{fernandez95}
{Fernandez-Soto} A.,  {Barcons} X.,  {Carballo} R.,    {Webb} J.~K.,  1995,
  \mnras, 277, 235

\bibitem[\protect\citeauthoryear{{Ganguly} \& {Brotherton}}{{Ganguly} \&
  {Brotherton}}{2007}]{ganguly07a}
{Ganguly} R.,  {Brotherton} M.~S.,  2007, ArXiv e-prints, 710

\bibitem[\protect\citeauthoryear{{Gaskell}}{{Gaskell}}{1982}]{gaskell82}
{Gaskell} C.~M.,  1982, \apj, 263, 79

\bibitem[\protect\citeauthoryear{{Gaskell}}{{Gaskell}}{1983}]{gaskell83}
{Gaskell} C.~M.,  1983, \apj, 267, L1

\bibitem[\protect\citeauthoryear{{Guimar{\~a}es}, {Petitjean}, {Rollinde}, {de
  Carvalho}, {Djorgovski}, {Srianand}, {Aghaee} \& {Castro}}{{Guimar{\~a}es}
  et~al.}{2007}]{guimaraes07a}
{Guimar{\~a}es} R.,  {Petitjean} P.,  {Rollinde} E.,  {de Carvalho} R.~R.,
  {Djorgovski} S.~G.,  {Srianand} R.,  {Aghaee} A.,    {Castro} S.,  2007,
  \mnras, 377, 657

\bibitem[\protect\citeauthoryear{{Hamilton}, {Casertano} \&
  {Turnshek}}{{Hamilton} et~al.}{2002}]{hamilton02a}
{Hamilton} T.~S.,  {Casertano} S.,    {Turnshek} D.~A.,  2002, \apj, 576, 61

\bibitem[\protect\citeauthoryear{{Hennawi} \& {Prochaska}}{{Hennawi} \&
  {Prochaska}}{2007}]{hennawi07}
{Hennawi} J.~F.,  {Prochaska} J.~X.,  2007, \apj, 655, 735

\bibitem[\protect\citeauthoryear{{Hennawi}, {Prochaska}, {Burles}, {Strauss},
  {Richards}, {Schlegel}, {Fan}, {Schneider}, {Zakamska}, {Oguri}, {Gunn},
  {Lupton} \& {Brinkmann}}{{Hennawi} et~al.}{2006}]{hennawi06a}
{Hennawi} J.~F.,  {Prochaska} J.~X.,  {Burles} S.,  {Strauss} M.~A.,
  {Richards} G.~T.,  {Schlegel} D.~J.,  {Fan} X.,  {Schneider} D.~P.,
  {Zakamska} N.~L.,  {Oguri} M.,  {Gunn} J.~E.,  {Lupton} R.~H.,    {Brinkmann}
  J.,  2006, \apj, 651, 61

\bibitem[\protect\citeauthoryear{{Hennawi}, {Strauss}, {Oguri}, {Inada},
  {Richards}, {Pindor}, {Schneider}, {Becker}, {Gregg}, {Hall}, {Johnston},
  {Fan}, {Burles}, {Schlegel}, {Gunn}, {Lupton}, {Bahcall}, {Brunner} \&
  {Brinkmann}}{{Hennawi} et~al.}{2006}]{hennawi06b}
{Hennawi} J.~F.,  {Strauss} M.~A.,  {Oguri} M.,  {Inada} N.,  {Richards} G.~T.,
   {Pindor} B.,  {Schneider} D.~P.,  {Becker} R.~H.,  {Gregg} M.~D.,  {Hall}
  P.~B.,  {Johnston} D.~E.,  {Fan} X.,  {Burles} S.,  {Schlegel} D.~J.,  {Gunn}
  J.~E.,  {Lupton} R.~H.,  {Bahcall} N.~A.,  {Brunner} R.~J.,    {Brinkmann}
  J.,  2006, \aj, 131, 1

\bibitem[\protect\citeauthoryear{{Hui}, {Stebbins} \& {Burles}}{{Hui}
  et~al.}{1999}]{hui99a}
{Hui} L.,  {Stebbins} A.,    {Burles} S.,  1999, \apj, 511, L5

\bibitem[\protect\citeauthoryear{{Jakobsen}, {Jansen}, {Wagner} \&
  {Reimers}}{{Jakobsen} et~al.}{2003}]{jakobsen03a}
{Jakobsen} P.,  {Jansen} R.~A.,  {Wagner} S.,    {Reimers} D.,  2003, \aap,
  397, 891

\bibitem[\protect\citeauthoryear{{Jakobsen} \& {Perryman}}{{Jakobsen} \&
  {Perryman}}{1992}]{jakobsen92a}
{Jakobsen} P.,  {Perryman} M.~A.~C.,  1992, \apj, 392, 432

\bibitem[\protect\citeauthoryear{{Jakobsen}, {Perryman}, {di Serego Alighieri},
  {Ulrich} \& {Macchetto}}{{Jakobsen} et~al.}{1986}]{jakobsen86a}
{Jakobsen} P.,  {Perryman} M.~A.~C.,  {di Serego Alighieri} S.,  {Ulrich}
  M.~H.,    {Macchetto} F.,  1986, \apj, 303, L27

\bibitem[\protect\citeauthoryear{{Kaiser}}{{Kaiser}}{1987}]{kaiser87a}
{Kaiser} N.,  1987, \mnras, 227, 1

\bibitem[\protect\citeauthoryear{{Kaiser}}{{Kaiser}}{1988}]{kaiser88}
{Kaiser} N.,  1988, \mnras, 231, 149

\bibitem[\protect\citeauthoryear{{Kim} \& {Croft}}{{Kim} \&
  {Croft}}{2007}]{kim07a}
{Kim} Y.-R.,  {Croft} R.,  2007, ArXiv e-prints, astro-ph/0701012

\bibitem[\protect\citeauthoryear{{Kirkman}, {Tytler} \& {Gleed}}{{Kirkman}
  et~al.}{2007}]{kirkman07b}
{Kirkman} D.,  {Tytler} D.,    {Gleed} M.,  2007, ArXiv e-prints

\bibitem[\protect\citeauthoryear{Kirkman, Tytler, Lubin \& Charlton}{Kirkman
  et~al.}{2007}]{kirkman07a}
Kirkman D.,  Tytler D.,  Lubin D.,    Charlton J.,  2007, \mnras, 376, 1227

\bibitem[\protect\citeauthoryear{{Lanzetta} \& {Bowen}}{{Lanzetta} \&
  {Bowen}}{1990}]{lanzetta90}
{Lanzetta} K.~M.,  {Bowen} D.,  1990, \apj, 357, 321

\bibitem[\protect\citeauthoryear{{Li}, {Jing}, {Kauffmann}, {B{\"o}rner},
  {White} \& {Cheng}}{{Li} et~al.}{2006}]{li06b}
{Li} C.,  {Jing} Y.~P.,  {Kauffmann} G.,  {B{\"o}rner} G.,  {White} S.~D.~M.,
   {Cheng} F.~Z.,  2006, \mnras, 368, 37

\bibitem[\protect\citeauthoryear{{Liske}}{{Liske}}{2000}]{liske00a}
{Liske} J.,  2000, \mnras, 319, 557

\bibitem[\protect\citeauthoryear{{Liske} \& {Williger}}{{Liske} \&
  {Williger}}{2001}]{liske01}
{Liske} J.,  {Williger} G.~M.,  2001, \mnras, 328, 653

\bibitem[\protect\citeauthoryear{{Loeb} \& {Eisenstein}}{{Loeb} \&
  {Eisenstein}}{1995}]{loeb95}
{Loeb} A.,  {Eisenstein} D.~J.,  1995, \apj, 448, 17

\bibitem[\protect\citeauthoryear{{Loh}, {Quashnock} \& {Stein}}{{Loh}
  et~al.}{2001}]{loh01a}
{Loh} J.-M.,  {Quashnock} J.~M.,    {Stein} M.~L.,  2001, \apj, 560, 606

\bibitem[\protect\citeauthoryear{{Madgwick}, S. \& {and others}}{{Madgwick}
  et~al.}{2003}]{madgwick03a}
{Madgwick} D.,  S. H.~D.,    {and others} 2003, \mnras, 344, 847

\bibitem[\protect\citeauthoryear{{McCarthy}, {Cohen}, {Butcher}, {Cromer},
  {Croner}, {Douglas}, {Goeden}, {Grewal}, {Lu}, {Petrie}, {Weng}, {Weber},
  {Koch} \& {Rodgers}}{{McCarthy} et~al.}{1998}]{mccarthy98}
{McCarthy} J.~K.,  {Cohen} J.~G.,  {Butcher} B.,  {Cromer} J.,  {Croner} E.,
  {Douglas} W.~R.,  {Goeden} R.~M.,  {Grewal} T.,  {Lu} B.,  {Petrie} H.~L.,
  {Weng} T.,  {Weber} B.,  {Koch} D.~G.,    {Rodgers} J.~M.,  1998, in Proc.
  SPIE Vol. 3355, p. 81-92, Optical Astronomical Instrumentation, Sandro
  D'Odorico; Ed. {Blue channel of the Keck low-resolution imaging
  spectrometer}.
pp 81--92

\bibitem[\protect\citeauthoryear{{McDonald}}{{McDonald}}{2003}]{mcdonald03}
{McDonald} P.,  2003, \apj, 585, 34

\bibitem[\protect\citeauthoryear{{McDonald} \& {Miralda-Escud{\'
  e}}}{{McDonald} \& {Miralda-Escud{\' e}}}{1999}]{mcdonald99a}
{McDonald} P.,  {Miralda-Escud{\' e}} J.,  1999, \apj, 518, 24

\bibitem[\protect\citeauthoryear{{Miller}, {Lopes}, {Smith}, {Croom}, {Boyle},
  {Shanks} \& {Outram}}{{Miller} et~al.}{2004}]{miller04a}
{Miller} L.,  {Lopes} A.~M.,  {Smith} R.~J.,  {Croom} S.~M.,  {Boyle} B.~J.,
  {Shanks} T.,    {Outram} P.,  2004, \mnras, 348, 395

\bibitem[\protect\citeauthoryear{{Milutinovic}, {Misawa}, {Lynch}, {Masiero},
  {Palma}, {Charlton}, {Kirkman}, {Bockenhauer} \& {Tytler}}{{Milutinovic}
  et~al.}{2007}]{milutinovic07a}
{Milutinovic} N.,  {Misawa} T.,  {Lynch} R.~S.,  {Masiero} J.~R.,  {Palma} C.,
  {Charlton} J.~C.,  {Kirkman} D.,  {Bockenhauer} S.,    {Tytler} D.,  2007,
  ArXiv e-prints, 2007arXiv0705.1353M

\bibitem[\protect\citeauthoryear{{Misawa}, {Charlton}, {Eracleous}, {Ganguly},
  {Tytler}, {Kirkman}, {Suzuki} \& {Lubin}}{{Misawa} et~al.}{2007}]{misawa07b}
{Misawa} T.,  {Charlton} J.~C.,  {Eracleous} M.,  {Ganguly} R.,  {Tytler} D.,
  {Kirkman} D.,  {Suzuki} N.,    {Lubin} D.,  2007, \apjs, 171, 1

\bibitem[\protect\citeauthoryear{{Misawa}, {Eracleous}, {Charlton} \&
  {Tajitsu}}{{Misawa} et~al.}{2005}]{misawa05b}
{Misawa} T.,  {Eracleous} M.,  {Charlton} J.~C.,    {Tajitsu} A.,  2005, \apj,
  629, 115

\bibitem[\protect\citeauthoryear{{Moller} \& {Kjaergaard}}{{Moller} \&
  {Kjaergaard}}{1992}]{moller92}
{Moller} P.,  {Kjaergaard} P.,  1992, \aap, 258, 234

\bibitem[\protect\citeauthoryear{{Narayanan}, {Hamann}, {Barlow}, {Burbidge},
  {Cohen}, {Junkkarinen} \& {Lyons}}{{Narayanan} et~al.}{2004}]{narayanan04}
{Narayanan} D.,  {Hamann} F.,  {Barlow} T.,  {Burbidge} E.~M.,  {Cohen} R.~D.,
  {Junkkarinen} V.,    {Lyons} R.,  2004, \apj, 601, 715

\bibitem[\protect\citeauthoryear{{Oke}, {Cohen}, {Carr}, {Cromer}, {Dingizian},
  {Harris}, {Labrecque}, {Lucinio}, {Schaal}, {Epps} \& {Miller}}{{Oke}
  et~al.}{1995}]{oke95}
{Oke} J.~B.,  {Cohen} J.~G.,  {Carr} M.,  {Cromer} J.,  {Dingizian} A.,
  {Harris} F.~H.,  {Labrecque} S.,  {Lucinio} R.,  {Schaal} W.,  {Epps} H.,
  {Miller} J.,  1995, \pasp, 107, 375

\bibitem[\protect\citeauthoryear{{Pen}}{{Pen}}{1999}]{pen99}
{Pen} U.-L.,  1999, \apjs, 120, 49

\bibitem[\protect\citeauthoryear{{Petitjean} \& {Bergeron}}{{Petitjean} \&
  {Bergeron}}{1990}]{petitjean90a}
{Petitjean} P.,  {Bergeron} J.,  1990, \aap, 231, 309

\bibitem[\protect\citeauthoryear{{Petitjean} \& {Bergeron}}{{Petitjean} \&
  {Bergeron}}{1994}]{petitjean94a}
{Petitjean} P.,  {Bergeron} J.,  1994, \aap, 283, 759

\bibitem[\protect\citeauthoryear{{Petitjean}, {Rauch} \&
  {Carswell}}{{Petitjean} et~al.}{1994}]{petitjean94b}
{Petitjean} P.,  {Rauch} M.,    {Carswell} R.~F.,  1994, \aap, 291, 29

\bibitem[\protect\citeauthoryear{{Pichon}, {Scannapieco}, {Aracil},
  {Petitjean}, {Aubert}, {Bergeron} \& {Colombi}}{{Pichon}
  et~al.}{2003}]{pichon03a}
{Pichon} C.,  {Scannapieco} E.,  {Aracil} B.,  {Petitjean} P.,  {Aubert} D.,
  {Bergeron} J.,    {Colombi} S.,  2003, \apj, 597, L97

\bibitem[\protect\citeauthoryear{{Pieri}, {Schaye} \& {Aguirre}}{{Pieri}
  et~al.}{2006}]{pieri06a}
{Pieri} M.~M.,  {Schaye} J.,    {Aguirre} A.,  2006, \apj, 638, 45

\bibitem[\protect\citeauthoryear{{Prochaska}, {Hennawi} \&
  {Herbert-Fort}}{{Prochaska} et~al.}{2007}]{prochaska07a}
{Prochaska} J.~X.,  {Hennawi} J.~F.,    {Herbert-Fort} S.,  2007, ArXiv
  e-prints, astro-ph/0703594

\bibitem[\protect\citeauthoryear{{Quashnock}, {vanden Berk} \&
  {York}}{{Quashnock} et~al.}{1996}]{quashnock96a}
{Quashnock} J.~M.,  {vanden Berk} D.~E.,    {York} D.~G.,  1996, \apj, 472, L69

\bibitem[\protect\citeauthoryear{{Rauch}, {Sargent}, {Barlow} \&
  {Carswell}}{{Rauch} et~al.}{2001}]{rauch01a}
{Rauch} M.,  {Sargent} W.~L.~W.,  {Barlow} T.~A.,    {Carswell} R.~F.,  2001,
  \apj, 562, 76

\bibitem[\protect\citeauthoryear{{Rauch}, {Sargent}, {Barlow} \&
  {Simcoe}}{{Rauch} et~al.}{2002}]{rauch02}
{Rauch} M.,  {Sargent} W.~L.~W.,  {Barlow} T.~A.,    {Simcoe} R.~A.,  2002,
  \apj, 576, 45

\bibitem[\protect\citeauthoryear{{Rauch}, {Sargent}, {Womble} \&
  {Barlow}}{{Rauch} et~al.}{1996}]{rauch96a}
{Rauch} M.,  {Sargent} W.~L.~W.,  {Womble} D.~S.,    {Barlow} T.~A.,  1996,
  \apj, 467, L5

\bibitem[\protect\citeauthoryear{{Richards}, {Vanden Berk}, {Reichard}, {Hall},
  {Schneider}, {SubbaRao}, {Thakar} \& {York}}{{Richards}
  et~al.}{2002}]{richards02a}
{Richards} G.~T.,  {Vanden Berk} D.~E.,  {Reichard} T.~A.,  {Hall} P.~B.,
  {Schneider} D.~P.,  {SubbaRao} M.,  {Thakar} A.~R.,    {York} D.~G.,  2002,
  \aj, 124, 1

\bibitem[\protect\citeauthoryear{{Richards}, {York}, {Yanny}, {Kollgaard},
  {Laurent-Muehleisen} \& {vanden Berk}}{{Richards} et~al.}{1999}]{richards99}
{Richards} G.~T.,  {York} D.~G.,  {Yanny} B.,  {Kollgaard} R.~I.,
  {Laurent-Muehleisen} S.~A.,    {vanden Berk} D.~E.,  1999, \apj, 513, 576

\bibitem[\protect\citeauthoryear{{Rollinde}, {Petitjean}, {Pichon}, {Colombi},
  {Aracil}, {D'Odorico} \& {Haehnelt}}{{Rollinde} et~al.}{2003}]{rollinde03}
{Rollinde} E.,  {Petitjean} P.,  {Pichon} C.,  {Colombi} S.,  {Aracil} B.,
  {D'Odorico} V.,    {Haehnelt} M.~G.,  2003, \mnras, 341, 1279

\bibitem[\protect\citeauthoryear{{Rollinde}, {Srianand}, {Theuns}, {Petitjean}
  \& {Chand}}{{Rollinde} et~al.}{2005}]{rollinde05a}
{Rollinde} E.,  {Srianand} R.,  {Theuns} T.,  {Petitjean} P.,    {Chand} H.,
  2005, \mnras, 361, 1015

\bibitem[\protect\citeauthoryear{{Romani}, {Filippenko} \& {Steidel}}{{Romani}
  et~al.}{1991}]{romani91a}
{Romani} R.~W.,  {Filippenko} A.~V.,    {Steidel} C.~C.,  1991, \pasp, 103, 154

\bibitem[\protect\citeauthoryear{{Russell}, {Ellison} \& {Benn}}{{Russell}
  et~al.}{2006}]{russell06a}
{Russell} D.~M.,  {Ellison} S.~L.,    {Benn} C.~R.,  2006, \mnras, 367, 412

\bibitem[\protect\citeauthoryear{Sargent, Young, Boksenberg \& Tytler}{Sargent
  et~al.}{1980}]{sargent80}
Sargent W.,  Young P.,  Boksenberg A.,    Tytler D.,  1980, \apjs, 42, 41

\bibitem[\protect\citeauthoryear{{Sargent}, {Boksenberg} \&
  {Steidel}}{{Sargent} et~al.}{1988}]{sargent88a}
{Sargent} W.~L.~W.,  {Boksenberg} A.,    {Steidel} C.~C.,  1988, \apjs, 68, 539

\bibitem[\protect\citeauthoryear{{Sargent}, {Boksenberg} \& {Young}}{{Sargent}
  et~al.}{1982}]{sargent82a}
{Sargent} W.~L.~W.,  {Boksenberg} A.,    {Young} P.,  1982, \apj, 252, 54

\bibitem[\protect\citeauthoryear{{Sargent} \& {Steidel}}{{Sargent} \&
  {Steidel}}{1987}]{sargent87a}
{Sargent} W.~L.~W.,  {Steidel} C.~C.,  1987, \apj, 322, 142

\bibitem[\protect\citeauthoryear{{Scannapieco}}{{Scannapieco}}{2005}]{scannapi%
eco05a}
{Scannapieco} E.,  2005, \apj, 624, L1

\bibitem[\protect\citeauthoryear{{Scannapieco}, {Pichon}, {Aracil},
  {Petitjean}, {Thacker}, {Pogosyan}, {Bergeron} \& {Couchman}}{{Scannapieco}
  et~al.}{2006}]{scannapieco06a}
{Scannapieco} E.,  {Pichon} C.,  {Aracil} B.,  {Petitjean} P.,  {Thacker}
  R.~J.,  {Pogosyan} D.,  {Bergeron} J.,    {Couchman} H.~M.~P.,  2006, \mnras,
  365, 615

\bibitem[\protect\citeauthoryear{{Schirber}, {Miralda-Escud{\' e}} \&
  {McDonald}}{{Schirber} et~al.}{2004}]{schirber04a}
{Schirber} M.,  {Miralda-Escud{\' e}} J.,    {McDonald} P.,  2004, \apj, 610,
  105

\bibitem[\protect\citeauthoryear{{Schirber}, {Miralda-Escud{\'e}} \&
  {McDonald}}{{Schirber} et~al.}{2004}]{schriber04a}
{Schirber} M.,  {Miralda-Escud{\'e}} J.,    {McDonald} P.,  2004, \apj, 610,
  105

\bibitem[\protect\citeauthoryear{{Scoccimarro}}{{Scoccimarro}}{2004}]{scoccima%
rro04}
{Scoccimarro} R.,  2004, \prd, 70, 083007

\bibitem[\protect\citeauthoryear{{Shanks}, {Boyle}, {Croom}, {Loaring},
  {Miller} \& {Smith}}{{Shanks} et~al.}{2000}]{shanks00a}
{Shanks} T.,  {Boyle} B.~J.,  {Croom} S.,  {Loaring} N.,  {Miller} L.,
  {Smith} R.~J.,  2000, in {Mazure} A.,  {Le F{\`e}vre} O.,   {Le Brun} V.,
  eds, Clustering at High Redshift Vol.~200 of Astronomical Society of the
  Pacific Conference Series, {The 2dF QSO Redshift Survey}.
pp 57--+

\bibitem[\protect\citeauthoryear{{Shaver}, {Boksenberg} \&
  {Robertson}}{{Shaver} et~al.}{1982}]{shaver82a}
{Shaver} P.~A.,  {Boksenberg} A.,    {Robertson} J.~G.,  1982, \apj, 261, L7

\bibitem[\protect\citeauthoryear{{Shaver} \& {Robertson}}{{Shaver} \&
  {Robertson}}{1983}]{shaver83a}
{Shaver} P.~A.,  {Robertson} J.~G.,  1983, \apj, 268, L57

\bibitem[\protect\citeauthoryear{{Shaver} \& {Robertson}}{{Shaver} \&
  {Robertson}}{1985}]{shaver85a}
{Shaver} P.~A.,  {Robertson} J.~G.,  1985, \mnras, 212, 15P

\bibitem[\protect\citeauthoryear{{Shen}, {Strauss}, {Oguri}, {Hennawi}, {Fan},
  {Richards}, {Hall}, {Gunn}, {Schneider}, {Szalay}, {Thakar}, {Vanden Berk},
  {Anderson}, {Bahcall}, {Connolly} \& {Knapp}}{{Shen} et~al.}{2007}]{shen07a}
{Shen} Y.,  {Strauss} M.~A.,  {Oguri} M.,  {Hennawi} J.~F.,  {Fan} X.,
  {Richards} G.~T.,  {Hall} P.~B.,  {Gunn} J.~E.,  {Schneider} D.~P.,  {Szalay}
  A.~S.,  {Thakar} A.~R.,  {Vanden Berk} D.~E.,  {Anderson} S.~F.,  {Bahcall}
  N.~A.,  {Connolly} A.~J.,    {Knapp} G.~R.,  2007, \aj, 133, 2222

\bibitem[\protect\citeauthoryear{{Simcoe}, {Sargent}, {Rauch} \&
  {Becker}}{{Simcoe} et~al.}{2006}]{simcoe06}
{Simcoe} R.~A.,  {Sargent} W.~L.~W.,  {Rauch} M.,    {Becker} G.,  2006, \apj,
  637, 648

\bibitem[\protect\citeauthoryear{{Slosar}, {Seljak} \& {Tasitsiomi}}{{Slosar}
  et~al.}{2006}]{slosar06}
{Slosar} A.,  {Seljak} U.,    {Tasitsiomi} A.,  2006, \mnras, 366, 1455

\bibitem[\protect\citeauthoryear{{Srianand}}{{Srianand}}{1997}]{srianand97a}
{Srianand} R.,  1997, \apj, 478, 511

\bibitem[\protect\citeauthoryear{{Steidel} \& {Sargent}}{{Steidel} \&
  {Sargent}}{1991}]{steidel91}
{Steidel} C.~C.,  {Sargent} W.~L.~W.,  1991, \aj, 102, 1610

\bibitem[\protect\citeauthoryear{{Stoughton} \& {et al.}}{{Stoughton} \& {et
  al.}}{2002}]{stoughton02}
{Stoughton} C.,  {et al.} 2002, \aj, 123, 485

\bibitem[\protect\citeauthoryear{{Suzuki}, {Tytler}, {Kirkman}, {O'Meara} \&
  {Lubin}}{{Suzuki} et~al.}{2003}]{suzuki03b}
{Suzuki} N.,  {Tytler} D.,  {Kirkman} D.,  {O'Meara} J.~M.,    {Lubin} D.,
  2003, \pasp, 115, 1050

\bibitem[\protect\citeauthoryear{{Tonry}}{{Tonry}}{1998}]{tonry98}
{Tonry} J.~L.,  1998, \aj, 115, 1

\bibitem[\protect\citeauthoryear{{Tytler}}{{Tytler}}{1982}]{tytler82}
{Tytler} D.,  1982, Nature, 298, 427

\bibitem[\protect\citeauthoryear{{Tytler}}{{Tytler}}{1987}]{tytler87b}
{Tytler} D.,  1987, \apj, 321, 49

\bibitem[\protect\citeauthoryear{{Tytler} \& {Fan}}{{Tytler} \&
  {Fan}}{1992}]{tytler92}
{Tytler} D.,  {Fan} X.-M.,  1992, \apjs, 79, 1

\bibitem[\protect\citeauthoryear{{Tytler}, {O'Meara}, {Suzuki}, {Kirkman},
  {Lubin} \& {Orin}}{{Tytler} et~al.}{2004}]{tytler04a}
{Tytler} D.,  {O'Meara} J.~M.,  {Suzuki} N.,  {Kirkman} D.,  {Lubin} D.,
  {Orin} A.,  2004, \aj, 128, 1058

\bibitem[\protect\citeauthoryear{{Tytler}, {Sandoval} \& {Fan}}{{Tytler}
  et~al.}{1993}]{tytler93a}
{Tytler} D.,  {Sandoval} J.,    {Fan} X.-M.,  1993, \apj, 405, 57

\bibitem[\protect\citeauthoryear{{Urry} \& {Padovani}}{{Urry} \&
  {Padovani}}{1995}]{urry95}
{Urry} C.~M.,  {Padovani} P.,  1995, \pasp, 107, 803

\bibitem[\protect\citeauthoryear{{Vale} \& {Ostriker}}{{Vale} \&
  {Ostriker}}{2004}]{vale04}
{Vale} A.,  {Ostriker} J.~P.,  2004, \mnras, 353, 189

\bibitem[\protect\citeauthoryear{{Vanden Berk} \& {et al.}}{{Vanden Berk} \&
  {et al.}}{2001}]{vandenberk01}
{Vanden Berk} D.~E.,  {et al.} 2001, \aj, 122, 549

\bibitem[\protect\citeauthoryear{{Vanden Berk}, {Stoughton}, {Crotts}, {Tytler}
  \& {Kirkman}}{{Vanden Berk} et~al.}{2000}]{vandenberk00}
{Vanden Berk} D.~E.,  {Stoughton} C.,  {Crotts} A.~P.~S.,  {Tytler} D.,
  {Kirkman} D.,  2000, \aj, 119, 2571

\bibitem[\protect\citeauthoryear{{Visbal} \& {Croft}}{{Visbal} \&
  {Croft}}{2007}]{visbal07a}
{Visbal} E.,  {Croft} R.~A.~C.,  2007, ArXiv e-prints, 2007arXiv0709.2364V

\bibitem[\protect\citeauthoryear{{Weymann}, {Carswell} \& {Smith}}{{Weymann}
  et~al.}{1981}]{weymann81}
{Weymann} R.~J.,  {Carswell} R.~F.,    {Smith} M.~G.,  1981, AnnRAAp, 19, 41

\bibitem[\protect\citeauthoryear{{Williger}, {Hazard}, {Baldwin} \&
  {McMahon}}{{Williger} et~al.}{1996}]{williger96}
{Williger} G.~M.,  {Hazard} C.,  {Baldwin} J.~A.,    {McMahon} R.~G.,  1996,
  \apjs, 104, 145

\bibitem[\protect\citeauthoryear{{Wise}, {Eracleous}, {Charlton} \&
  {Ganguly}}{{Wise} et~al.}{2004}]{wise04a}
{Wise} J.~H.,  {Eracleous} M.,  {Charlton} J.~C.,    {Ganguly} R.,  2004, \apj,
  613, 129

\bibitem[\protect\citeauthoryear{{Worseck}, {Fechner}, {Wisotzki} \&
  {Dall'Aglio}}{{Worseck} et~al.}{2007}]{worseck07a}
{Worseck} G.,  {Fechner} C.,  {Wisotzki} L.,    {Dall'Aglio} A.,  2007, ArXiv
  e-prints, 2007arXiv0704.0187W

\bibitem[\protect\citeauthoryear{{Worseck} \& {Wisotzki}}{{Worseck} \&
  {Wisotzki}}{2006a}]{worseck06a}
{Worseck} G.,  {Wisotzki} L.,  2006a, \aap, 450, 495

\bibitem[\protect\citeauthoryear{{Worseck} \& {Wisotzki}}{{Worseck} \&
  {Wisotzki}}{2006b}]{worseck06b}
{Worseck} G.,  {Wisotzki} L.,  2006b, ArXiv Astrophysics e-prints,
  2006astro.ph.10895W

\bibitem[\protect\citeauthoryear{{Wright}}{{Wright}}{2006}]{wright06a}
{Wright} E.~L.,  2006, \pasp, 118, 1711

\bibitem[\protect\citeauthoryear{{York} \& {et al.}}{{York} \& {et
  al.}}{2000}]{york00}
{York} D.~G.,  {et al.} 2000, \aj, 120, 1579

\bibitem[\protect\citeauthoryear{{Young}, {Sargent} \& {Boksenberg}}{{Young}
  et~al.}{1982}]{young82a}
{Young} P.,  {Sargent} W.~L.~W.,    {Boksenberg} A.,  1982, \apjs, 48, 455

\bibitem[\protect\citeauthoryear{{Zehavi} \& {et al.}}{{Zehavi} \& {et
  al.}}{2002}]{zehavi02a}
{Zehavi} I.,  {et al.} 2002, \apj, 571, 172

\end{thebibliography}

\section{Appendix: Comments on Individual systems}

\subsubsection{P1}

EA and AAA coincidences. For EA1 an absorption system in P1a is coincident
with the \zem\ of P1b. The C~IV and N~V are slightly to the blue of
of their emission
lines. Both absorbers are single velocity components. The N~V $\lambda 1242.80$
transition is blended with unidentified absorption. For AAA21 the same system
in P1a is coincident with a system showing a single velocity component in C~IV
and Si~IV.

\subsubsection{P3}

An AA coincidence. Both P3a and P3b show strong Mg~II and
Fe~II absorption. There are two velocity components in the Mg~II absorption of
P3b. The Mg~I in P3b shows a single velocity component. The Mg~II and Fe~II
in P3a are also single velocity components.

\subsubsection{P5}

EA, AA, AAA, and AAV coincidences. For AA2 both P5a and P5b show single velocity
components in C~IV absorption. For EA2 an absorption system in P5a is
coincident
with the \zem\ of P5b. AAA22 consists of this same absorption system in P5a
and a system in P5b that shows single velocity components in C~IV and Si~IV.
AAV17 consists of an absorption system in P5a
containing
single velocity components in C~IV and Si~IV and an absorption system in P5b
containing
 a single velocity component in C~IV.

\subsubsection{P6}

Both P6a and P6b show Mg~II absorption comprising the AA3 coincidence.
The Mg~II in
P6a is in a blend but seems reliable. The Mg~II in P6b is less secure.
Its Mg~II 2803 is the blue side
of a blend with N~V 1238. We use the redshift $z=2.5593$ for the N~V from
its Si~III 1206.

The EA3 absorption system in P6a is coincident with
the \zem\ of P6b. The C~IV has multiple velocity components that are blended.
The N~V is a single velocity component. Both the N~V and C~IV are slightly
blueward of their emission lines.

The same absorption system in P6a that comprises EA3 is coincident with an
absorption system in P6b, which is AAA23. The system in P6b shows a single
velocity component in C~IV, and it is coincident with the \zem\ of P6a to
create EAV19.

\subsubsection{P7}

An AA coincidence. The C~IV in P7a is very strong and is a single velocity
component. The Al~III is a multiple velocity component. The C~IV in P7b is a
 single velocity component.

\subsubsection{P8}

EA, EAV, and AAA coincidences. For EA4 an absorption system in P8a is
coincident with the
\zem\ of P8c. The C~IV is a single velocity component and slightly blueward
of
its emission line. The P8a absorption system of EA4 is also coincident with an
associated system of P8c to form AAA24 and an associated system of P8b to form
AAA25. The AAA24 absorption system in P8c shows a single velocity
component in C~IV.

The C~IV in the system in P8b is strong and its $\lambda 1550.77$
transition is blended with another C~IV at a different redshift.
For EA5 the \zem\ of P8a is coincident with the AAA25 absorption system in P8b.
For EAV20 the \zem\ of P8c is also coincident with the AAA25 absorption system
in P8b.
For EA6 the \zem\ of P8a is coincident with another absorption system
in P8b. The C~IV in this system in P8b is strong and its $\lambda 1548.19$
transition is blended with the C~IV in the system mentioned above. We find a
Ca~II in P8a at $z = 0.4215$.

\subsubsection{P22}

EA, AA, AAA, EAV, and AAV coincidences. An absorption system in P22a is coincident
with the
\zem of P22b. The same absorption system in P22a is coincident with an
absorption system in P22c and two different absorption systems in P22b.
The C~IV in P22a
is strong and has multiple velocity components that are blended. The Si~IV has
multiple velocity components.  The N~V in the system in P22c is a single
velocity component and is in the \lya\ emission line. The two coincidences with
P22b are AAA28 and AAA29. For P22b the $\lambda 1550.77$ transition of the C~IV
in AAA28 is blended with the $\lambda 1548.19$ transition of the C~IV in AAA29.
In addition, the system in P22c mentioned above is also coincident with the two
close proximity systems of P22b to form AAA26 and AAA27 and the \zem of P22b to
form EAV21.
For the AAV coincidence,
in P22a the C~IV and Si~IV are both show a large single velocity component.
In P22b
the C~IV and Si~IV  both show a small single velocity component.

\subsubsection{P23}

An EA coincidence. An absorption system in P23a is coincident with the \zem\
of
P23b. The C~IV is at the peak of the C~IV emission line and is a single velocity
component.

\subsubsection{P25}

Two AA coincidences. For P25a the C~IV in both coincident systems is a
single velocity component.  For P25b the C~IV in both coincident systems
show multiple velocity components.

\subsubsection{P31}

Two AA coincidences and an AAV coincidence. For AA8 an Mg~II absorption system
in P31a
is coincident with a Mg~II absorption system in P31b.  The Mg~II in P31b is a
single velocity component,
and the Mg~II in P31a is weak but is also a single velocity component. AA9
consists of an absorption system in P31a that
shows strong multiple velocity components.
C~IV and an absorption system in P31b with C~IV that is a single velocity
component. AAV19 consists of the C~IV absorption of P31a of AA9 and C~IV
absorption in P31b that is a single velocity component.

\subsubsection{P36}

EA and AAA coincidences. An absorption system in P36b is coincident with the
\zem\ of P36a and also with an associated system of P36a. The system in P36b
consists of a single velocity component in Si~II,Al~II and Al~III. AAA30 is
this system plus a single velocity component N~V absorption in P36a.

\subsubsection{P38}

Two AA coincidences. An absorption system in P38a with large C~IV that has
multiple velocity components is coincident with an absorption system in P38b
with large C~IV that has multiple velocity components.  Another absorption
system in P38a with weak C~IV that has multiple velocity components is
coincident with an absorption system in P38b with weak C~IV that has multiple
velocity components.

\subsubsection{P42}

Two AA coincidences. An absorption system in P42a with Mg~II that has multiple
velocity components is coincident to a system in P42b with Mg~II that is a
single velocity component. An absorption system in P42a is coincident with
an absorption system in P42b.  The C~IV in both systems has strong multiple velocity
components.

\subsubsection{P44}

An AA coincidence. An absorption system in P44a with C~IV,Mg~II,Al~II and
Fe~II identified is coincident with an absorption system in P44b that has
C~IV identified.  The C~IV in both systems are large and appear in the \lyaf .

\subsubsection{P70}

An AA coincidence. An absorption system in P70a with Mg~II and Fe~II is
coincident with an absorption system in P70b with C~IV.

\subsubsection{P83}

An AAA coincidence. An absorption system in P83a with C~IV is coincident with
and absorption system in P83b with Si~III and H~I. The C~IV is a single
velocity component.

\subsubsection{P125}

An AAV coincidence. Both of the absorption systems contain C~IV. The C~IV in
P125a is a large single velocity component and the C~IV in P125b is a small
single velocity component.

\subsubsection{P147}

An AAA and two EAV coincidences. EAV29 is the coincidence of an absorption
system in P147a and the \zem\ of P147b. This same absorption system is also
coincident with an absorption system in P147b with a single velocity component
in C~IV identified. A system in P147b with a single velocity component in C~IV
identified slightly to the red of the emission line is coincident with the
\zem\ of P147a.

\subsubsection{P153}

EA, AAA and EAV coincidences. One absorption system in P153a and one in P153b
are coincident with each other, and with the opposite \zem\. For the system in
P153a only H~I and Si~III were identified. For the system in P153b, weak
single velocity components in C~IV and Si~IV are identified.
\subsubsection{P155}

AA and AAA coincidences. An absorption system in P155a is coincident with an
absorption system in P155b. Also, another absorption system in P155a is
coincident with a separate absorption system in P155b.

\end{document}